\documentclass[prbl,aps,amssym,nofootinbib,floatfix,twocolumn,notitlepage]{revtex4-2} 

\usepackage{hyperref}
\hypersetup{colorlinks,linkcolor=blue,urlcolor=blue,citecolor=blue}

 \usepackage{xcolor}
\usepackage{amsmath,amssymb,bbold,bm}
\usepackage{graphicx}

\newcommand{\be}{\begin{equation}}
\newcommand{\ben}{\begin{equation*}}
\newcommand{\ee}{\end{equation}}
\newcommand{\een}{\end{equation*}}
\newcommand{\bs}{\begin{split}}
\newcommand{\es}{\end{split}}
\newcommand{\bmx}{\begin{array}}
\newcommand{\emx}{\end{array}}
\newcommand{\bea}{\begin{eqnarray}}
\newcommand{\bean}{\begin{eqnarray*}}
\newcommand{\eea}{\end{eqnarray}}
\newcommand{\eean}{\end{eqnarray*}}
\newcommand{\dg}{^{\dagger}}
\newcommand{\dn}{^{\vphantom{\dagger}}}

\newcommand{\ra}{\rightarrow}

\newcommand{\ua}{\uparrow}
\newcommand{\da}{\downarrow}
\newcommand{\bb}[1]{\mathbb{#1}}
\newcommand{\qqquad}{\qquad\qquad\qquad}
\newcommand{\so}{\qquad\rightarrow\qquad}

\newcommand{\andd}{\qquad\text{and}\qquad}

\newcommand{\eps}{\epsilon}

\newcommand{\sgn}[1]{{\rm sign}{#1}}
\newcommand{\pref}[1]{(\ref{#1})}

\newcommand{\intob}[1]{\int_{0}^{\beta}{#1}}

\newcommand{\re}[1]{{\rm Re}\left[ #1 \right]}

\newcommand{\abs}[1]{\left\vert #1 \right\vert}
\newcommand{\bra}[1]{\left\langle #1 \right\vert}
\newcommand{\ket}[1]{\left\vert #1\right\rangle}
\newcommand{\braket}[1]{\left\langle #1\right\rangle}
\newcommand{\sbraket}[1]{\langle #1\rangle}

\newcommand{\mat}[1]{\left(\bmx{cc}#1\emx\right)}
\newcommand{\matc}[2]{\left(\bmx{#1}#2\emx\right)}
\newcommand{\matn}[1]{\bmx{cc}#1\emx}

\newcommand{\bw}[1]{\begin{widetext}}
\newcommand{\ew}[1]{\end{widetext}}

\newcommand{\gray}[1]{}

\begin{document}
\title{Triplet pairing, orbital selectivity and correlations in iron-based superconductors}
\author{Yashar Komijani$^{1 *}$}
\author{Elio K\"onig$^{2}$}
\author{Piers Coleman$^{3,4}$}
 \affiliation{ $^1$Department of Physics, University of Cincinnati, Ohio 45221, USA}
 \affiliation{ $^2$Max-Planck-Institut f\"ur Festk\"orperforschung, 70569 Stuttgart, Germany}
 \affiliation{ $^3$Department of Physics and Astronomy, Rutgers University, Piscataway, NJ 08854, USA}
 \affiliation{ $^4$Department of Physics, Royal Holloway, University of London, Egham, Surrey TW20 0EX, UK}

\date{\today}
\begin{abstract}
We use {a} slave-boson approach to study {the} band renormalization and pair susceptibility in the normal state of iron-based superconductors in presence of strong Coulomb repulsion and Hund's interaction. Our results show orbital selectivity toward localization of $xy$ orbitals and its interplay with superconductivity. We also compare the recently proposed triplet resonating valence bond theory of superconductivity in Iron based superconductors with the more conventional $s_\pm$ pairing and show that both favor a superconductivity when the $xy$ orbital is delocalized.
\end{abstract}
\maketitle

\section{Introduction}
{A major} 
open question in the theory of strongly correlated electronic systems is the origin of the superconductivity 
in iron-based superconductors (FeSC) \cite{Hosono2018}. Since their
discovery more than a decade ago, a large body of data has been
collected which has largely been interpreted in the framework of
$s_\pm$ singlet superconductivity \cite{Mazin2008,Hirschfeld2011}. A
widely accepted paradigm for the pairing in these materials, based on 
itinerant 
electrons in {the} presence of an attractive pairing glue
\cite{Chubukov2012,Chubukov2015a} that binds them into Cooper pairs,
has been successful in explaining a subset of this data. There is, however, less consensus on the origin of {the pairing glue.} 
Various {exchange bosons} have been suggested, from spin, nematic and
orbital fluctuations to a combination
thereof\,\cite{Si08,Fernandes2016}. {At the same time, as we expand
upon below, } 
a variety of new experimental data has shown signatures of an
interplay between complex orbital and local moment physics
\cite{Xu2020} with superconductivity. {The} 
consistency {of these findings} with the 
paradigm of electrons subject to a fluctuation-mediated pairing glue
is currently unclear.

Considering the complexity of this multi-orbital system and presence
of many competing orders and putative critical points, the pursuit
of a single origin for the
superconductivity may be fraught with difficulty.  A more modest goal 
is to seek 
the minimal ingredients for superconductivity in FeSC. In a recent paper \cite{tRVB} we suggested
that Hund's coupling is a the driver \cite{VafekChubukov2017} of the
superconductivity, giving rise to an orbitally-odd
spin-triplet superconductor. Although itinerant triplet pairing in
iron-based SC was discussed early on by Wen and Lee
\cite{LeeWen2008} this idea was quickly dismissed 
due to the observation of a Knight shift and the  absence of nodes on the
Fermi surface (FS). Triplet pairing in
iron-based superconductors  has recently been re-visited
\cite{PuetterKee2012,Hu13,Hao14,Hao14b} in the context of a possible 
 interplay between pairing and the topological bandstructure.

Given the pre-dominant support for an $s_\pm$ singlet pairing
scenario,  it is
instructive to re-iterate the motivation for a triplet state.
Dynamical mean-field theory (DMFT) \cite{Georges96} has long shown
that the Hund's interaction plays a dominant role in the correlations
of the normal state
\cite{Haule08,Haule09,Yin11,Yin12,Georges13,Deng2019}, a picture that is also confirmed in slave-spin mean-field studies \cite{Georges13}. The basic idea is that coherent metallicity
in DMFT can be understood  within a self-consistent impurity
framework, where the underlying physics is akin to
Kondo screening of an equivalent impurity problem. Hund's interactions
tend to force electrons into high-spin configurations which are 
hard to screen, thus effectively suppressing the Kondo
coherence to low temperatures
\cite{Schrieffer1967,Nevidomskyy2009}. At the same time, the orbital degrees
of freedom can be quenched at higher temperature creating a
spin-orbit{al} separation (SOS) regime
\cite{Stadler15,DrouinTouchette2021}. A related phenomen{on} is the
spin freezing effect \cite{Werner2008}, where under orbital symmetry
the spin correlation functions do not decay in time. Therefore, unhindered by the
hybridization and down to lowest temperatures, the Hund's interaction is
the dominant interaction 
whose pair-hopping effects can give rise to superconductivity. Indeed, the
authors in \cite{Hoshino15} proposed spatially isotropic triplet
pairing for SrRuO$_3$ in the spin-freezing regime.

{An important experimental} 
development is the realization that FeSCs {appear to display a
universal yet} non-BCS {ratio of maximal zero-temperature gap and
transition temperature} \cite{MiaoDing2018} $2\Delta/T_C\sim
7.2$. {Given the wide variety of different Fermi surface topologies in
the large FeSC family, this apparent  universality suggests that
superconductivity may be intimately linked to the only common feature
they share - namely, the local crystalline environment of the Fe ion. Indeed the universal $2\Delta/T_C$ ratio} 
can be explained \cite{Lee18} {semi-phenomenologically} 
by local critical spin-fluctuations of the form $-\chi''/\omega\sim
\omega^{-\gamma}$ {and $\gamma = 1.2$}. {Pairing by 
power-law-spin fluctuations is a problem of substantial present-day interest \cite{Classen2021}, partly due to its interconnection to Sachdev-Ye-Kitaev models \cite{InkofSchmalian2022}. The value of $\gamma = 1.2$ relevant to the observed $\Delta/T_c$ in FeSC occurs in DMFT solutions of Hund's metals \cite{Yin12,Stadler15},  and (approximately) in} the {aforementioned} SOS regime of {multi-orbital \cite{WaltervonDelft2020} and} mixed-valent Hund's impurities.
\cite{DrouinTouchette2022} {Combining experimental observations with the above-mentioned Hund's phenomenology,} 
naturally leads to the question of investigating the possibility of triplet superconductivity in FeSC.

Another pertinent aspect of these materials  concerns the role of
local atomic orbitals in the development of 
orbital selective Mott phases (OSMP) \cite{Medici14}, orbital selective pairing \cite{SprauDavis2017}, orbital loop currents \cite{Klug2018} and the pairing enhanced by orbital fluctuations \cite{Yue2021}. In particular, the angle-resolved photo-emission spectroscopy (ARPES) has enabled orbitally resolved study of the band-structure. The strong orbital selectivity, first predicted by DMFT \cite{Yin12} has been confirmed by ARPES \cite{Yi13,Yi17}. Certainly, by focusing on the Fermi surface the {entirely} 
itinerant {approach} picture 
misses the orbital degrees of freedom of the local physics. 

{A further question of importance in the context of
superconductivity is the local Coulomb repulsion which tends to enforce
the anomalous Green's function to vanish at equal points in space
and time. Phonon-mediated superconductivity satisfies this Coulomb
constraint by exploiting the strong retardation of the electron phonon
interaction, to build in nodes in
frequency space \cite{BogolyubovShirkov1959,McMillan1968}. By
contrast, non-BCS
pairing of strongly correlated fermions typically
involves a finite angular momentum of Cooper pair, which is symmetry protected
against onsite repulsion \cite{AndersonMorel,Coleman2015}.
The theory of electron-interaction-mediated $s_\pm$ in FeSC is
criticized \cite{KoenigColeman2019} for realizing neither of these
established escape-routes.}

A conceptually different paradigm which may be easier to reconcile
with the orbital complexity of this material is the picture of
pre-formed pairs \cite{Anderson87,Anderson1987,Baskaran87} as in the
theory of resonating valence bonds in a doped Mott insulator
\cite{Lee06}. The idea of Hund's driven spin-triplet
superconductivity in strongly correlated materials was first discussed by Anderson
\cite{Anderson84a,Anderson84b,Anderson85} who, in the context of
heavy-fermions, argued that an odd-parity of the Cooper pair requires at
least two atoms per unit cell with a center of inversion between the
two, an ingredient that is satisfied in FeSC. Recently, the idea of
triplet pairing resurfaced in the context of ferromagnetic systems, 
with the observation of strange metal phase in a pure
heavy-fermion CeRh$_6$Ge$_4$ \cite{Cerge}. In this material, a 
magnetic easy-plane anisotropy was
argued to lead to presence of triplet resonating valence bonds (tRVB)
in the ground state and a highly entangled ordered phase. It was
subsequently shown that doping such a tRVB host can lead to odd-parity
spin-triplet superconductivity \cite{Koenig2022,
LopezPowellMerino2022}.

Recently, we proposed \cite{tRVB} that in FeSC the Hund's induced
electron triplets resonate between various orbitals of each atom. {To stabilize of a} 
non-zero  {superconducting} parameter order, 
Anderson's {above-mentioned argument about the} two atoms per unit cell {proves} critical. 
 We argued that a tRVB state is consistent with the body of available
experimental data, from Knight shift \cite{Carretta2020},
quasi-particle interference \cite{HanaguriTakagi2010,ChiPennec2014}
and neutron spin-resonance measurements \cite{tRVB}. Moreover, a 
tRVB order parameter is relatively
robust against disorder and predicts staggered component to the pair
wavefunction that, we argued, will be discernible
using scanning Josephson spectroscopy. 
The goal of the present paper is to extend these early studies, 
contrasting the orbital complexity and the effects of strong intra-orbital Coulomb repulsion in
the tRVB and $s_\pm$ states within these two perspectives. 

To this end, we study the {(orbital selective) band} renormalization and superconducting instability of the normal state of FeSC under strong on-site Coulomb interaction. In the trade-off between complexity and tractability, we choose the simplest possible model, i.e. a two-dimensional three-band model of layered FeSC, which seem to capture the main physics. We also assume {absent} nematicity and use slave-bosons \cite{Barnes1976,Coleman1984} to represent the Hubbard operators in the 
limit {of infinite intra-orbital repulsion}, which are treated via mean-field theory. Our approach connects previous slave-boson works of \cite{Ruckenstein87,KotliarLiu1988} and more recent slave-spin approaches \cite{Yu13}.

The content of the paper is as follows: in section \ref{sec:model} we
introduce the model and the decoupling of the Hund's
interaction. Section \ref{sec:U}  describes
how the slave-bosons are used to treat the intra-orbital Coulomb
repulsion and
the Hund's induced renormalization of the inter-orbital Coulomb interaction. Section \ref{sec:mf} contains the mean-field analysis of the interacting Hamiltonian and an analysis of band-renormalization, orbital selectivity and pair susceptibility within the mean-field theory. 
\section{Model}\label{sec:model}

The model is described by the Hamiltonian
\be
\hspace{-.1cm}H_0=\hspace{-.25cm}\sum_{ij,\mu\nu,\alpha\beta} c\dg_{i\mu\alpha}{\cal H}_0^{i\mu\alpha,j\beta\nu}c_{j\nu\beta}+H_{\rm int},
\ee
where the non-interacting Hamiltonian matrix is
\be
{\cal H}_0^{i\mu\alpha,j\beta\nu}=t^{i\mu,j\nu}\delta^{\alpha\beta}+\delta^{ij}(\eps_\mu\delta^{\mu\nu}\delta^{\alpha\beta}-\lambda_{S}\vec L_{\mu\nu}\cdot\vec \sigma_{\alpha\beta}).
\ee
Here, $\alpha,\beta=\ua,\da$ are spin, {$i,j$ site,} and
$\mu,\nu=xy,xz,yz$ are the orbital {indices} ({the ladder with}in the
d-shell of Fe).  Furthermore, $\sigma^a_{\alpha\beta}$ are Pauli
matrices in spin space and $L^a_{\mu\nu}=-i\eps_{a\mu\nu}$ are three
totally anti-symmetric matrices in orbital space. Throughout this
paper, the upper/lower
position of indices is equivalent. We use a three band model
\cite{DaghoferDagotto2010} expressed in the original two atom per unit
cell basis ({cf.} Appendix \ref{sec:appmodel}).  The atomic spin-orbit
coupling $\lambda_S$ can be regarded as a spin-dependent inter-orbital
hopping. The interaction is on-site and using the notation
$\bar\delta_{\mu\nu}=1-\delta_{\mu\nu}$, can be written as
\be
\hspace{-.1cm}H_{\rm int}=\hspace{-.1cm}\frac{1}{2}\sum_{j,\mu\nu}\Big[n_{i\mu\sigma}(U\delta_{\mu\nu}+U'\bar\delta_{\mu\nu})n_{i\nu\sigma'}-J_H\vec S_{j\mu}\cdot\vec S_{j\nu}\Big],\label{eq3}
\ee
in terms of $n_{j\mu\sigma}=c\dg_{j\mu\sigma}c\dn_{j\mu\sigma}$ and
$\vec S_{j\mu}=\frac{1}{2}c\dg_{j\mu\alpha}\vec\sigma_{\alpha\beta}
c\dn_{j\mu\beta}$.  The interaction $H_{\rm int}$ contains an intra-orbital Coulomb
interaction $U$, an inter-orbital part $U '$ as well as a local Hund's
interaction $J_H>0$, which is an intra-atomic spin-spin
interaction which favors higher-spin states.

The largest energy scale in $H_{\rm int}$ 
is the intra-orbital Coulomb repulsion $U\sim$1-5eV.
In terms of $U$, the remaining parameters have the typical hierarchy
$U'\sim U/4$, $t\sim J_H\sim U/10$ and $\lambda_{S}\sim U/100$, to be
compared with a typical iron-based superconducting transition temperature $T_c\sim U/1000\sim$ 10-50K.
Our strategy is to study a simplified limit of the problem in which the intra-orbital $U$ is sent to infinity.\\

After the onsite Coulomb interaction, the most important interaction 
term is the Hund's
interaction.  We now re-write this term in terms of the inter-orbital 
triplet interactions.
By using a modified Fierz identity (Appendix \ref{sec:eq4}):
\be
\vec\sigma_{\alpha\beta}\cdot\vec\sigma_{\alpha'\beta'}=(\vec\sigma\sigma^y)_{\alpha\alpha'}\cdot(\sigma^y\vec\sigma)_{\beta'\beta}-\vec\sigma_{\alpha\beta'}\cdot\vec\sigma_{\alpha'\beta}\label{eq4}
\ee
we can decouple the Hund's interaction in the triplet channel.
Restoring the fermions by multiplying this identity by 
$c\dg_{j\mu\alpha}c\dg_{j\nu\alpha'}$ on the left and 
$c_{j\nu\beta'}c_{j\mu\beta}$ on the right, the Hund's interaction can
be written as 
\bea\label{eq5}
\vec S_{j\mu}\cdot\vec S_{j\nu}&=&
\Big[ 
(c\dg_{j\mu} \vec\sigma (- i\sigma_{y}) c^{*} _{j\nu})\cdot(c^{T}_{j\nu} (i\sigma_{2})\vec\sigma c_{j\mu})\nonumber\\
&&\hspace{3cm}+(c\dg_{j\mu}\vec\sigma c\dn_{j\nu})\cdot(c\dg_{j\nu}\vec\sigma c\dn_{j\mu})
\Big]\nonumber\\
&=& \left[\vec{\Psi}\dg_{\mu\nu} (j)\cdot \vec{\Psi}_{\mu\nu} (j)+\vec{\Phi}\dg _{\mu\nu } (j)\cdot\vec\Phi_{\mu\nu}(j)
 \right].\nonumber
\eea
Here we use the notation $c^{*}_{j\nu}\equiv (c\dg _{j\nu})^{T}$ to
denote the transpose of the creation operator and we have rewritten
the Fermion bi-linears in terms of 
the triplet pair and particle-hole operators 
\begin{equation}\label{}
\vec\Psi_{\mu\nu} (j) = c_{j\nu} (i\sigma_{y})\vec{\sigma }c_{j\mu},
\qquad 
\vec\Phi_{\mu\nu} (j) = c\dg _{j\nu} \vec{\sigma }c_{j\mu}.
\end{equation}
We note that the anti-commutation properties of the fermion operators
enforce an orbital antisymmetry on the local triplet pair operators
$\vec{\Psi}_{\mu\nu} (j)=-\vec{\Psi}_{\nu\mu} (j)$, whereas the
particle-hole operators are Hermitian, $(\vec{\Phi }_{\mu\nu})\dg = \vec{\Phi }_{\nu\mu}$.
Both interaction
channels in Eq.\,\pref{eq5} are attractive and can acquire an
expectation value for ferromagnetic Hund's coupling.
In a bulk
system, thee triplet operators can condense into a ferromagnetic ground state
\cite{Koenig2022} similar to the way short-range singlet RVBs can
exhibit long-range AFM order \cite{Albuquerque12}. 

It is convenient to decompose the triplet pair and particle-hole
operators as follows:
\begin{eqnarray}\label{}
\vec{\Phi }_{\mu\nu}&=& \frac{1}{2}\sum_{A=1,8}\lambda^{A}_{\mu\nu}\vec{\Phi }_{A}, \cr
\vec{\Psi}_{\mu\nu}&=& \frac{1}{2}\sum_{a=1,3}L^{a}_{\mu\nu}\vec{\Psi }_{A}.
\end{eqnarray}
Here the $\lambda^{A}_{\mu\nu}$ are the eight Gell-Mann matrices.  The
orbital-antisymmetry of the pair operators means that only the three
antisymmetric Gell-Mann matrices appear in the triplet pair 
operators $\vec{\Psi}_{\mu\nu}$,  denoted by 
the angular momentum operators
$L^{a}_{\mu\nu}= -i \epsilon_{a\mu\nu}\equiv
(\lambda^{7},-\lambda^{5},\lambda^{2})$. 
We can rewrite 
\begin{eqnarray}\label{}
\vec{\Phi }_{A} (j) &=& c\dg_{j\nu} \vec{\sigma
}\lambda^{A}_{\nu\mu}c_{j\mu}, \qquad \qquad (A = 1,8),\cr
\vec{\Psi}_{a} (j) &=& c_{j\nu} ( i\sigma_{y} \vec{\sigma
})L^{a}_{\nu\mu}c_{j\mu}, \qquad  (a=1,3),
\end{eqnarray}
where the magnetic vectors $\vec{\Phi }_{A}=\vec{\Phi }\dg_{A}$ are
real. 
If we combine the three vectors $\vec{\Psi}_{a} $ into a 
three-dimensional matrix $\Psi_{ab}=(\vec{\Psi}_{a})_{b}
$, 
and similarly, denote $(\vec{\Phi}_{A})_{b}= \Phi_{Ab}$, then 
the Hund's interaction can be written as 
\begin{equation}\label{eqdef}
H_{\rm H} = - 
\frac{J_{H}}{2}
\sum_{j,ab}{\Psi}\dg_{ab} (j){\Psi }_{ab} (j)-\frac{J_{H}}{2}
\sum_{j,Ab}
[{\Phi } _{Ab} (j)]^{2}.
\end{equation}

Carrying out a Hubbard Stratonovich transformation, we can decouple the Hund's interaction in terms of
magnetic order parameters $\Lambda_{Ab}= (\vec{\Lambda}_{A})_{b}$
and a triplet gap matrix $\Delta_{ab}= (\vec{\Delta }_{a})_{b}$ as follows
\bea
H_{{\rm H}} &\ra&
\sum_{j, Ab}\Big[\frac{\vert{\Gamma_{Ab} (j)}\vert^2 }{2g}+\Phi_{Ab} (j)\Gamma_{Ab} (j)\Big]\nonumber\\
&&+
\sum_{j, ab}\Big[\frac{\vert{\Delta_{ab} (j)}\vert^2}{g}+(\Psi\dg_{ab}\Delta_{ab} (j) +{\rm H.c})\Big],
\quad\label{eqdec}
\eea
where $g=J_{H}/2$ is the bare coupling constant. 

Under renormalization, the above interaction is expected to develop
anisotropies. Moreover, the magnetic and the triplet channels
will behave differently, since the particle-hole triplet
channel associated with $\Phi$ couples to the spin-orbit interaction 
(SOI) and the particle-hole channels are un-nested, 
whereas the presence of center of symmetry between the two
iron-atoms per unit cell in the iron-based superconductors 
means that the triplet Cooper channel will couple to the Fermi
surface, and will undergo a logarithmic renormalization. 
In 
\cite{tRVB}, it was shown that the effects of spin-orbit coupling
at an atomic level create anisotropies in the above pairing
interactions, favoring diagonal gap functions, 
$\Gamma_{ab}\sim {\rm diag}(1,1,1)$, and $\Delta_{ab}\sim{\rm
diag}(1,1,-2)$.  The latter is an example of a triplet resonating
valence bond (tRVB) state. In the next section,
we introduce slave-bosons to study this problem in the $U\to\infty$
limit, but first we comment on the use of the reduced three band model.

\subsection*{Three vs. five band model and occupancy}

In a single ion, the $e_g$ orbitals are filled with four electrons and
$t_{2g}$ orbitals are doubly occupied. In a five-orbital model of
FeSC \cite{Eschrig2009}, the $e_g$ orbitals disperse strongly,
crossing and hybridizing with $t_{2g}$ orbital. A faithful
representation of the Coulomb interaction then would require a
finite-U slave-spin representation of all the five orbitals, which is
a rather heavy calculation. To simplify the procedure, we note that
three-band models of FeSC only in terms of $t_{2g}$ orbitals, agree
qualitatively with ARPES, assuming an occupation of 4 electrons per
site. One can justify the latter as a result of a formal integrating
out of the $e_g$ orbital in five-band orbital. However, due to their
crossing of the Fermi energy, the integrated-out $e_g$ orbitals
introduce poles and zeros into the Green's function. In Appendix
\ref{sec:reduction} we have integrated out the $e_g$ orbitals and
shown that this enlargement of FS is compensated by the FS of the
integrated-out orbitals. In the following, we use a three-band model
where the occupation of the $t_{2g}$ orbital is four electrons per
site.

\section{Slave-boson representation of the $U \rightarrow \infty$ limit}\label{sec:U}

Since the reduced band in the parent compound has the occupation of
$n_{e}=4$ among three orbitals, at least one orbital is doubly
occupied. The electron annihilation operator is represented by
$c_{j\mu\sigma}$ where $\mu=xz,yz,xy$ is the orbital index,
$\sigma=\ua,\da$ is the spin index and $j$ is the site index. In order
to study doping of the parent compound in an infinite-$U$ problem, we
choose to preserve the doubly occupied states of the $\mu=xz,yz$
orbitals (``doublons'') and the empty state (``holon'') of 
the $xy$ orbital. In other words, we discard empty $xz/yz$ orbitals
and doubly occupied $xy$ orbitals, so that the fermionic Hubbard operators are represented as
\bea
\mu&=&xz,yz,\quad c\dg_{j\mu\sigma}=\tilde\sigma\ket{2_{j\mu}}\bra{\bar\sigma_{j\mu}}, \quad c\dg_{j\mu\sigma}\to b\dg_{j\mu}\tilde\sigma f\dn_{j\mu\bar\sigma},\nonumber\\
\mu&=&xy,\quad c\dg_{j,\mu,\sigma}=\ket{\sigma_{j\mu}}\bra{0_{j\mu}}, \quad c\dg_{j,\mu,\sigma}\to b\dn_{j,\mu}f\dg_{j,\mu,\sigma},\qquad\quad\label{eq7b}
\eea
where $\tilde\sigma=\sgn(\sigma)$. These representations are subject to the constraints
\be
n^f_{j\mu}+n^b_{j\mu}=1, \qquad \forall j,\mu, \label{eqcons1}
\ee
where $n_{j\mu}^b=b_{j\mu}\dg b\dn_{j\mu}$ and $n^f_{j\mu}=\sum_\sigma
f\dg_{j\mu\sigma}f\dn_{j\mu\sigma}$. The number of physical electrons
is $n_\mu=n^f_\mu+2n^b_\mu$ for $\mu=xz,yz$ and $n_\mu=n^f_{\mu}$ for
$\mu=xy$, so that the total number of electrons per site is then 
\begin{eqnarray}\label{filling}
n_{e}&=&n^{f}+ 2 (n^{b}_{xz}+n^{b}_{yz})\cr
&=& 3 + (n^{b}_{xz}+ n^{b}_{yz}- n^{b}_{xy}),
\end{eqnarray}
where we have imposed the constraint \pref{eqcons1} in
the last step.
In the mean-field theory, we adjust the overall chemical potential so
that the average number of electrons per site is $n_{e}=4$, i.e. by \eqref{filling},
\be
n^b_{j,xz}+n^b_{j,yz}=1+n^b_{j,xy}, \qquad \forall j.\label{eqcons3}
\ee
The constraints (\ref{eqcons1}-\ref{eqcons3}) are imposed via separate Lagrange multipliers.  A consequence of these constraints is that $n_{xy}^f=n_{xz}^f+n_{yz}^f$. Assuming that xy is dominated by electrons and $xz/yz$ by holes, this would indicate fully compensated electron-hole pockets at the FS.

The inter-orbital Coulomb interaction remains as in Eq.\,\pref{eq3}
but we may use the constraint to express it entirely in terms of
$n^b_{j\mu}$. The $U'$ terms drive various phenomena
\cite{DaghoferDagotto2010} including the nematic phase, which is not
considered in this paper. Within mean-field theory
\be
U'n_{j\mu}^bn_{j\nu}^b\to
U'(n_{j\mu}^b\sbraket{n_{j\nu}^b}+\sbraket{n_{j\mu}^b}n_{j\nu}^b-\sbraket{n_{j\mu}^b}\sbraket{n_{j\nu}^b})\nonumber
\ee
and the $U'$ is completely absorbed by shifting the chemical potential
and the Lagrange multiplier used to impose Eq.\,\pref{eqcons3}. In
other words within mean-field theory, physical quantities expressed in terms of
densities are insensitive to the value of the $U'$ interaction.
Beyond mean-field theory, we will argue in the next section that in
the Hund's dominated regime, inter-orbital repulsion is renormalized
to smaller values by spin-fluctuations.

For future convenience we collect the bosons into $\tilde b$ and spinons into $\tilde f$ such that $c_{j\mu\sigma}=\tilde b\dn_{j\mu}\tilde f\dg_{j\mu\sigma}$  for all $\mu$. 

\subsection*{Charge projectors \& Hund's interaction}
The Hund's interaction takes place entirely in the spin sector. Indeed
if the number of electrons at sites $n_{j\mu}$ and $n_{j\nu}$ differ
from unity, this interaction vanishes. Thus the Hund's interaction
involves a projection into single-electron occupancy which is not faithfully preserved once we decouple the right-hand side of these equations in \pref{eqdec}. Inside a path integral the constraint on these terms is imposed by $\Delta_{ab}$ and $\Gamma_{Ab}$ carrying gauge charges. In (integer-valence) Kondo systems, when a pre-fractionalized pattern for $\vec S^f_{\mu}=\frac{1}{2}f\dg_\mu\vec\sigma f_\mu$ in terms of spinons is used, $\Delta_{\mu\nu}$ and $f_i$ both carry gauge charges. In the present context, however, $c_i$ does not carry a gauge charge. Therefore, in order to extend these decouplings to the mixed-valence regime, we need to include inert charge projectors
\be
J\vec S_{j\mu}\cdot\vec S_{j\nu}\to JP_{j\mu} P_{j\nu}\vec S^f_{j\mu}\cdot\vec S^f_{j\nu},\label{eq10}
\ee
where
$P_{j\mu}=\sum_\sigma\ket{\sigma}_{j\mu}\bra{\sigma}_{j\mu}$. Within
the infinite-$U$ limit, the projectors can be represented as
$P_{j\mu}=\sum_\sigma \tilde f\dg_{j\mu\sigma}\tilde
f\dn_{j\mu\sigma}=1-\tilde b\dg_{j\mu} \tilde b\dn_{j\mu}$ and within
the physical sector, they can be replaced with $P_{j\mu}\to\tilde
b_{j\mu}\tilde b_{j\mu}\dg$ in \pref{eq10}. When we decouple the
interaction we need to decouple the projectors as well. This means
that \pref{eqdef} and \pref{eqdec} can still be used, but with the
$c_{j\mu\alpha}\sim \tilde b\dg_{j\mu}\tilde f_{j\mu\alpha}$
replacement.  This will ensure that each term in the Hamiltonian
commutes with the constraints, indicating that $\Delta_{ab}$ and
$\Gamma_{ab}$ are gauge-invariant, a necessary condition for their
condensation and a safe starting point for a mean-field study. In
other words, $\Psi_{ab}$ and $\Phi_{ab}$ correspond to inter-orbital
pairing and pair-hopping of \emph{physical electrons} rather than
\emph{spinons}. This is somewhat different than the traditional
approach applied to the single-band Hubbard model
\cite{KotliarLiu1988} and we revisit that model in Appendix
\ref{1band}.

The inclusion of charge projectors offers the simplification that we do not really need a gauge theory of fractionalization and the pairing interaction between physical electrons has a finite coupling constant. All that is needed, is to compute the electron pairing susceptibility for the interacting theory whose divergence signals the onset of superconductivity. The downside is that the theory is still interacting and in practice, we have to resort to mean-field theory to compute the susceptibility.

\subsection*{Hund's mediated attraction}

Another implication of charge projectors is to realize that the
spin-fluctuations can produce an attractive charge interaction between
different orbitals: 
\be
-J_H(1-n^b_{j\mu})(1-n^b_{j\nu})\sbraket{\vec S^f_{j\mu}\cdot\vec S^f_{j\nu}}.
\ee
We can understand this by noting that a minimization of the Hund's
energy $-J_H\vec S_\mu\cdot\vec S_\nu$ requires putting one electron
on each orbital (despite $U'$) and effectively producing an attractive
Coulomb interaction between the two orbitals. On the other hand, if
$U'$ wins the competition, the Hund's interaction is reduce by
renormalization, which will typically lead to the nematic phase. Therefore, Hund's and $U'$ (and thus tRVB and nematic phases) are antagonistic.

A similar effect occurs in the single-band t-J model where nearest
neighbor anti-ferromagnetic coupling will produce a reduced charge
repulsion between nearby sites. The competition between RVB and
charge-density wave states could be possibly attributed to this phenomenon.
Moreover, the competition between $U'$ and the Hund's interaction can
also 
be seen in impurity models relevant to DMFT calculations. We have done
a one-loop calculation for an Fe impurity model in Appendix
\ref{sec:beta} and shown that after the decoupling, the Gaussian pair
fluctuations in the disordered normal state do indeed renormalize the
repulsive $U'$ interaction to smaller values, 
\be
\frac{du'_{rr'}}{d\ell}=u'_{rr'}-2g\rho_r\rho_{r'}.
\ee
Here, the dimensionless coupling $u'_{rr'}=U'_{rr'}/D$ and $d\ell=-d\log D$ are expressed in terms of the bandwidth $D$ and $\rho_r$ is the density of states of orbital $r$. Inclusion of the charge projectors ensures that this physics is not lost in subsequent mean-field decouplings.

\section{Mean-field analysis}\label{sec:mf}
A mean-field decoupling of the slave-boson Hamiltonian leads to
\be
\hspace{-.1cm}H_0\to\hspace{-.1cm}\sum_{i,j,\mu\nu,\alpha\beta}\tilde f\dg_{i\mu\alpha}{\cal H}^{i\mu\alpha,j\nu\beta}_f\tilde f\dn_{j\nu\beta}+\sum_{i,j,\mu\nu}\tilde b\dn_{i\mu}{\cal H}_b^{i\mu,j\nu}\tilde b\dg_{j\nu}.
\ee
The coefficients ${\cal H}^b$ and ${\cal H}^f$ are chosen so that
\bea
{\cal H}^{i\mu\alpha,j\nu\beta}_f&=&{\cal H}^{i\mu\alpha,j\nu\beta}_0\sbraket{\tilde b\dn_{i\mu}\tilde b\dg_{j\nu}}, \\
{\cal H}^{i\mu,j\nu}_b&=&\sum_{\alpha\beta}{\cal H}^{i\mu\alpha,j\nu\beta}_0\sbraket{\tilde f\dg_{i\mu\alpha}\tilde f\dn_{j\nu\beta}},\nonumber
\eea
leading to a set of self-consistent equations. Limiting ourselves to the normal state and assuming absence of nematicity, 
we have
self-consistently solved these equations in momentum space, imposing the constraints (\ref{eqcons1},\ref{eqcons3}).

This enables us to study the mean-field Hamiltonian beyond the
single-site approximation used in earlier slave-spin \cite
{Yu13,Yu17,Komijani17} or DMFT approaches, allowing us to address 
the possibility of orbitally selective Mott transitions in the
presence of inter-orbital hopping. Our results for the band renormalizations and pair susceptibility are summarized in the next two sections, while the technical details of the calculation are discussed in Appendix \ref{sec:details}.

\begin{figure}[tp!]
\includegraphics[width=0.32\linewidth]{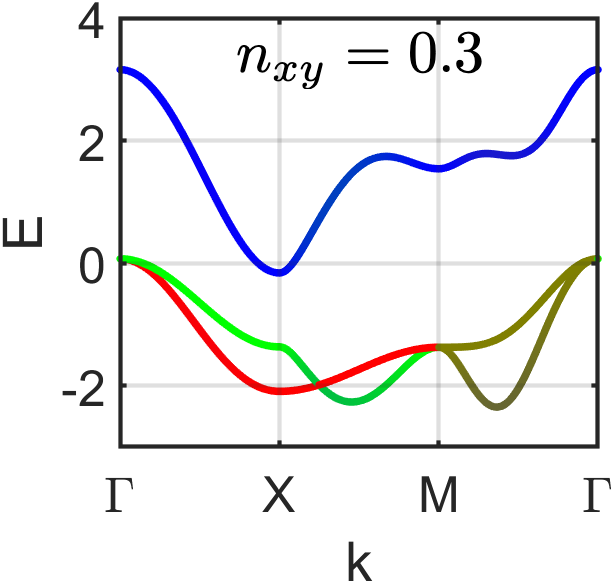}
\includegraphics[width=0.32\linewidth]{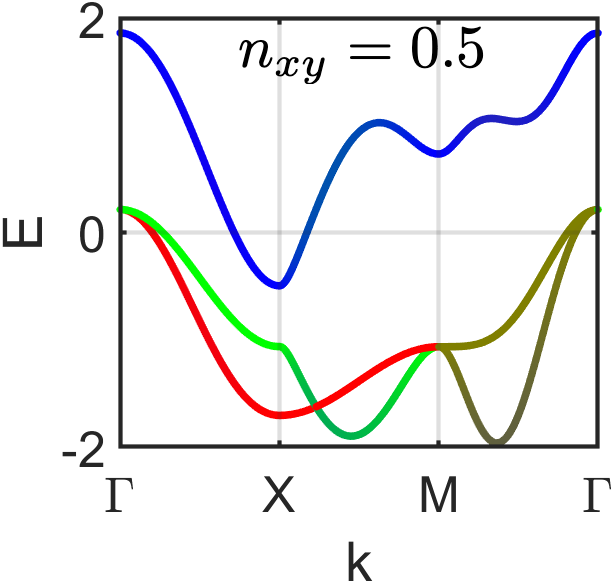}
\includegraphics[width=0.32\linewidth]{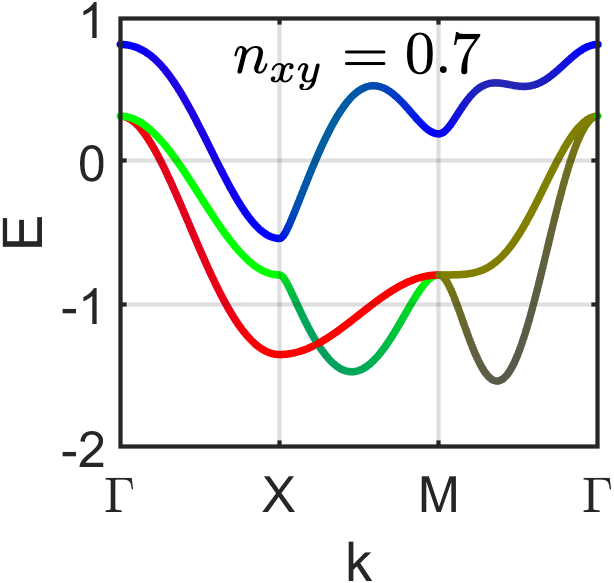}
\includegraphics[width=0.32\linewidth]{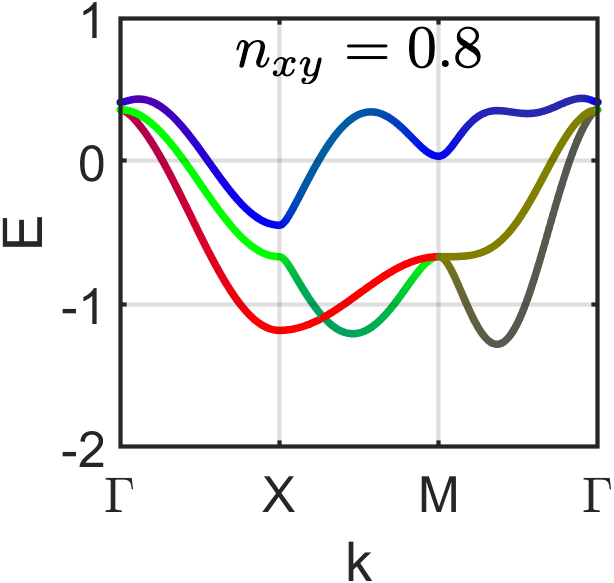}
\includegraphics[width=0.32\linewidth]{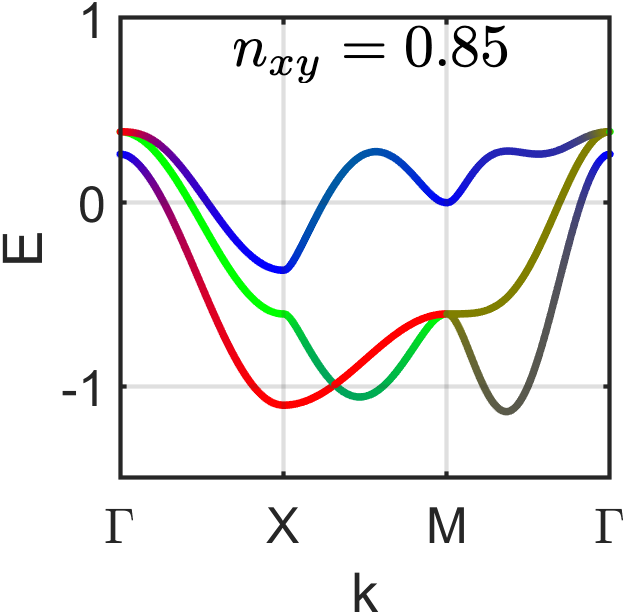}
\includegraphics[width=0.32\linewidth]{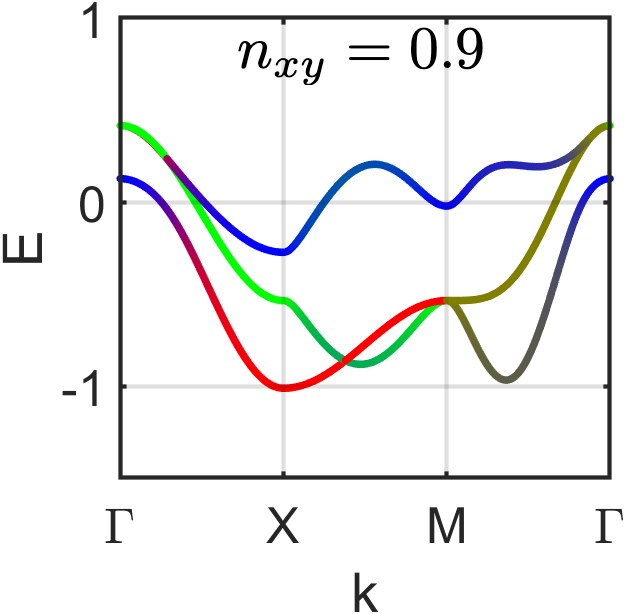}
\includegraphics[width=0.32\linewidth]{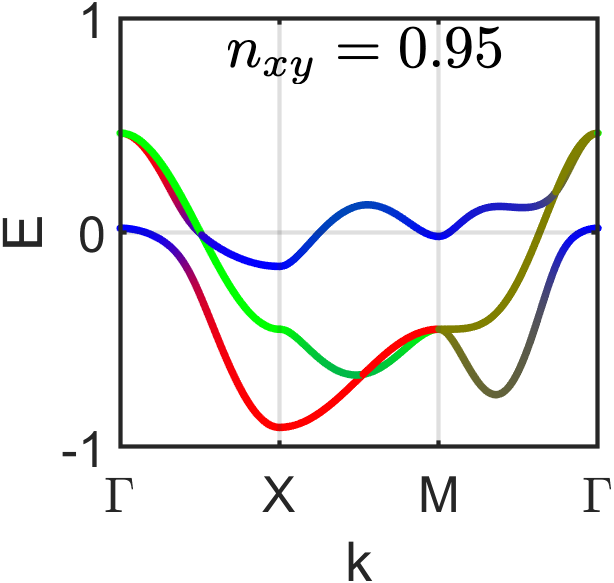}
\includegraphics[width=0.33\linewidth]{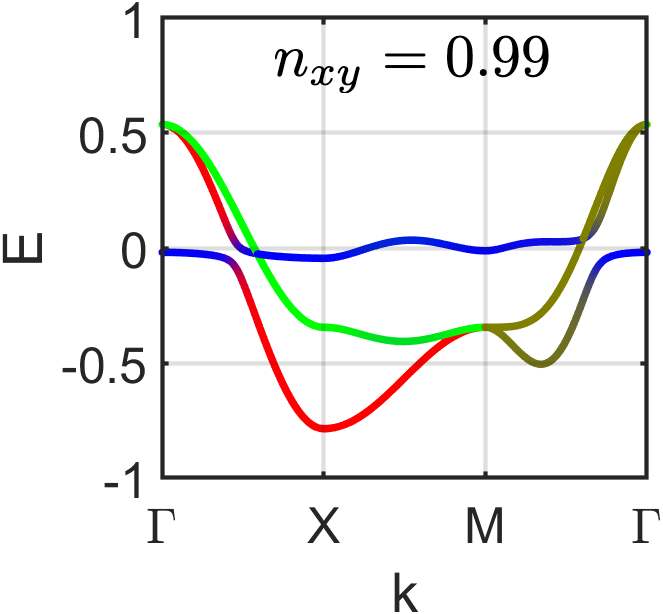}
\includegraphics[width=0.32\linewidth]{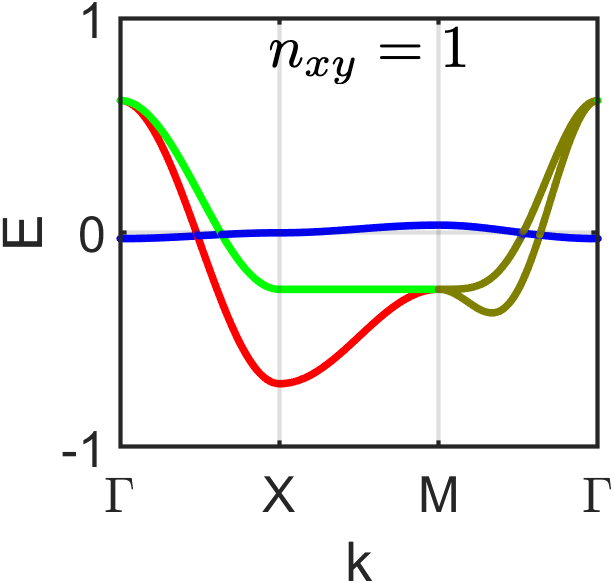}
\caption{\small Renormalized $E(k)$ along a cut through Brillouin zone for various $n_{xy}$ and in absence of spin-orbit interaction. The red/blue/green indicate $xz/yz/xy$ orbital content.}\label{fig:1}
\end{figure}
\begin{figure}[tp!]
\includegraphics[width=0.32\linewidth]{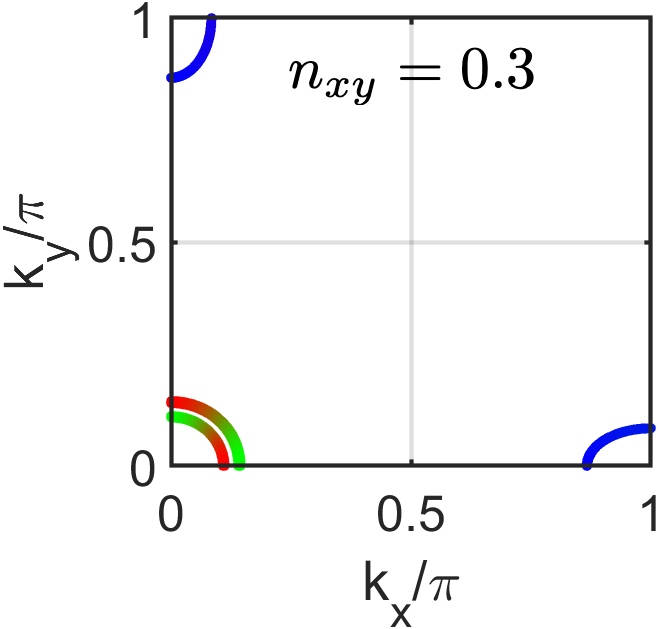}
\includegraphics[width=0.32\linewidth]{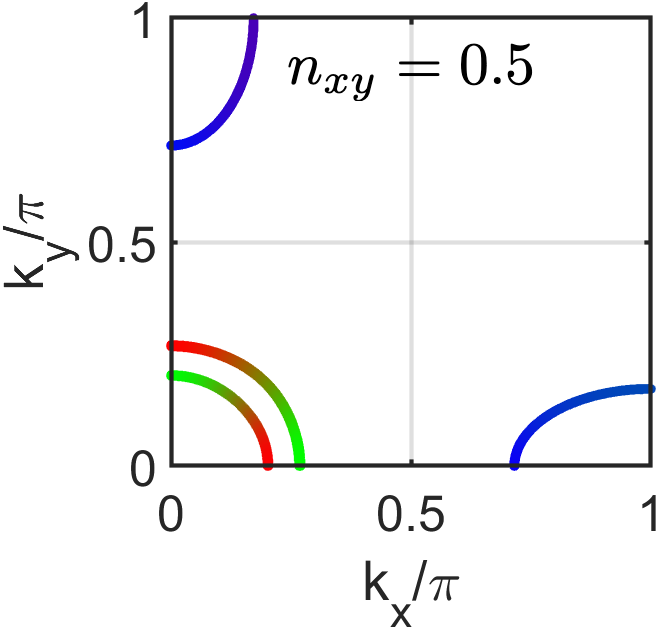}
\includegraphics[width=0.32\linewidth]{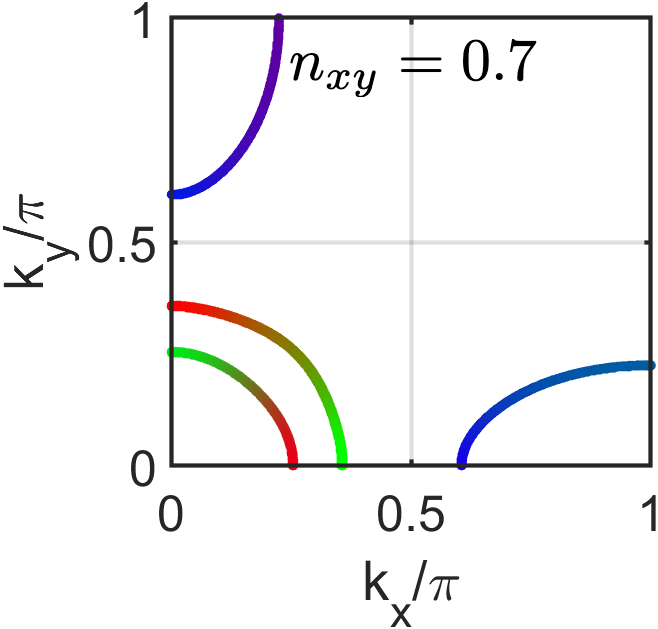}
\includegraphics[width=0.32\linewidth]{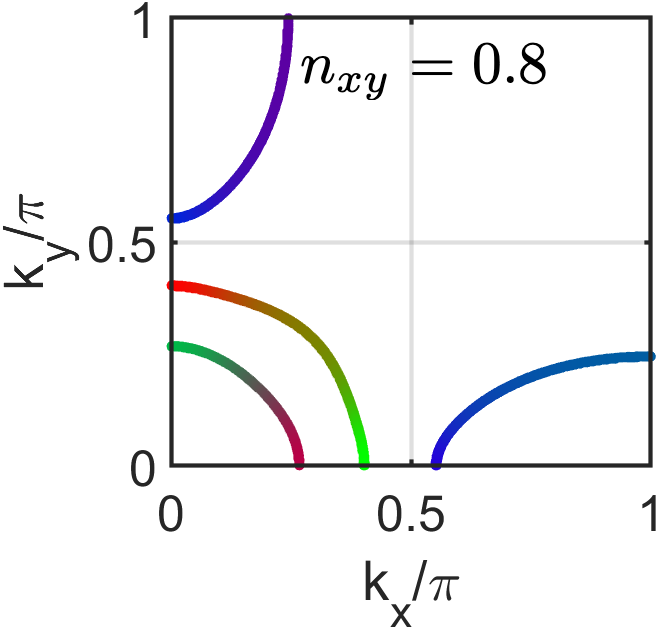}
\includegraphics[width=0.32\linewidth]{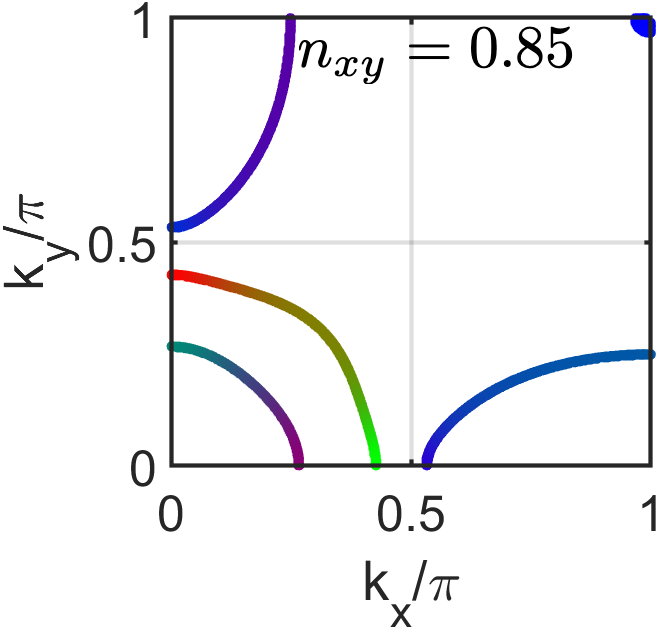}
\includegraphics[width=0.32\linewidth]{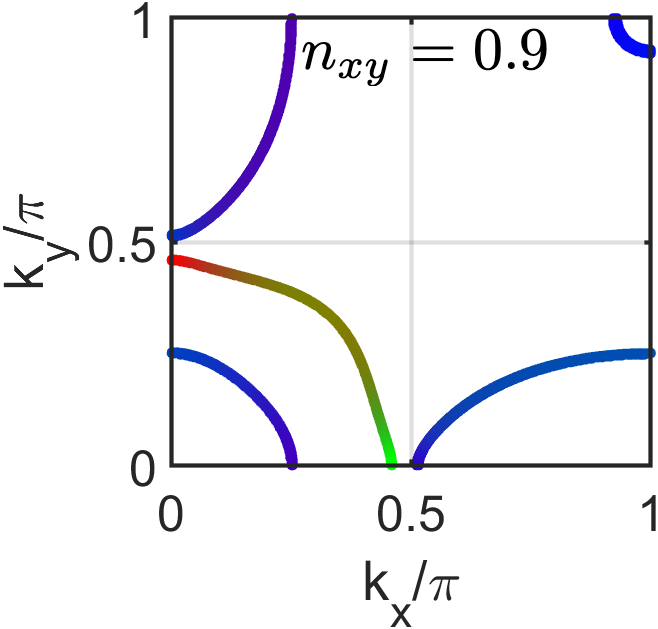}
\includegraphics[width=0.32\linewidth]{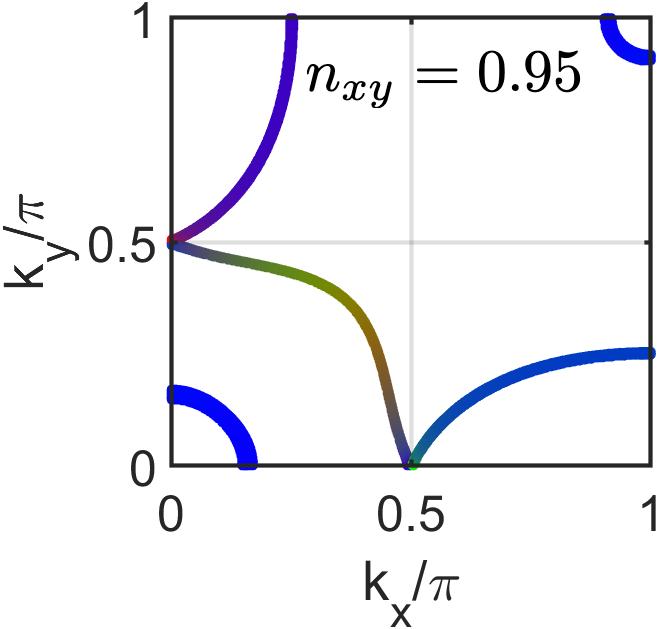}
\includegraphics[width=0.32\linewidth]{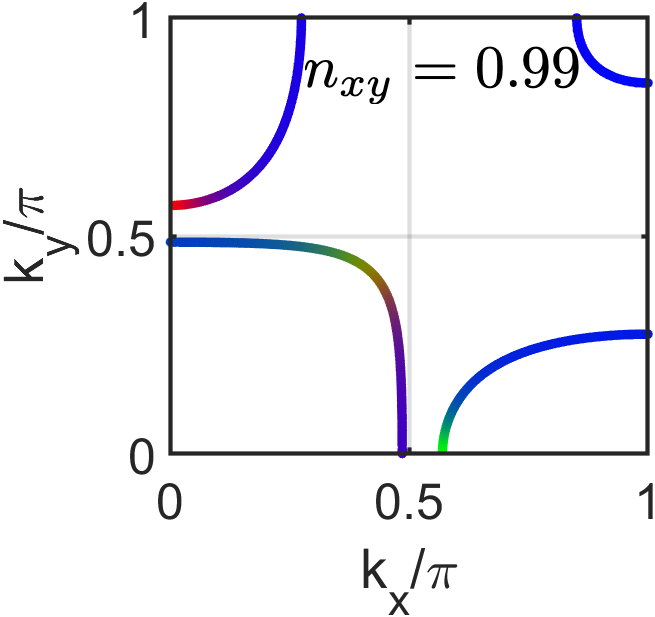}
\includegraphics[width=0.32\linewidth]{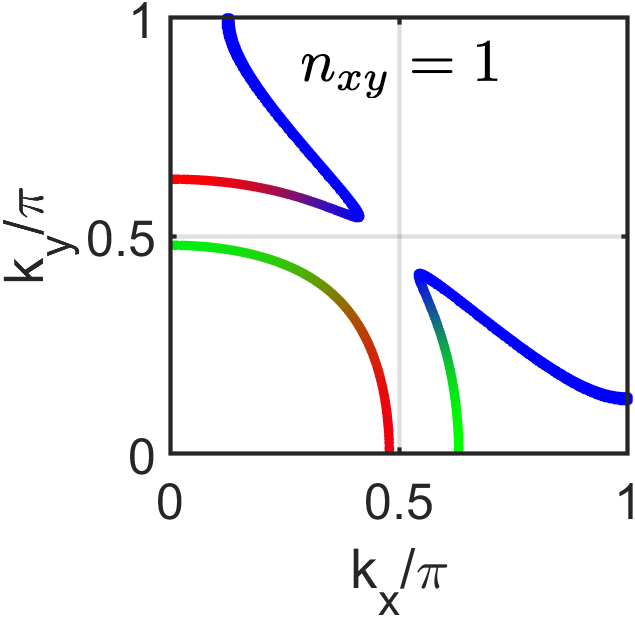}
\caption{\small Renormalized FSs for various $n_{xy}$ and in absence of spin-orbit interaction. The red/blue/green indicate $xz/yz/xy$ orbital content.}\label{fig:2}
\end{figure}


\subsection*{Band renormalization and orbital selectivity}
Fig.\,\ref{fig:1} shows the evolution of the band dispersions as
$n_{xy}$ is varied from {$0.3$ to 1}
while other occupancies are adjusted to maintain the same number of
four electrons in the $t_{2g}$ band. We have found that the spin-orbit
interaction has a small effect on band renormalizations and
therefore, in the following we report the results in the  absence of spin-orbit coupling. Fig.\,\ref{fig:2} shows the evolution of the FS over the same regime. {While for $n_{xy}$ sufficiently smaller than $1$ the standard Fermi surface topology appears, drastic interaction effects become apparent when $n_{xy} \rightarrow 1$.} 

{Indeed, i}n the coherent regime $\sbraket{\tilde b\dn_{i\mu}\tilde
b\dg_{j\nu}}\approx\delta_{\mu\nu}n_{\mu}$ and thus one expects to
have a flat $xy$ band at $n_{xy}=1$, the so-called orbital selective
Mott phase (OSMP). In practice, rather than a total localization a
finite temperature dependent bandwidth remains that goes to zero as
$T\to 0$. The fate of this band upon its hybridization with other
bands has been debated \cite{Yu13,Yu17,Komijani17,Komijani2019}. We
find a finite hybridization that is most manifest in gapping part of
the outer hole pocket in Fig.\,\pref{fig:2}. There are significant FS
re-constructions in the vicinity of OSMP $n_{xy}\to 1$. In this
regime,  additional band renormalizations due to Hund's interaction
\cite{Si2016} are expected to be relevant for the OSMP. As the xy
orbital is doped further, the hybridization between $xy$ orbital and
$xz/yz$ grows and the orbital character of the inner FS is strongly
modified. In the opposite regime of $n_{xy}<0.7$, the $xy$ electron
pockets and $xz/yz$ hole pockets are mostly compensated due to the
condition $n^f_{xy}=n^f_{xz}+n^f_{yz}$ we found earlier.   At
$n_{xy}<0.25$, there is a Lifshitz transition and the system becomes an insulator, in which the finite occupancy of the $xy$ orbital is supported by the finite admixture in the occupied bands. 

There is a window {$0.35 \simeq n_{xy} \simeq 0.9$} where {the xy}
band is partially occupied, with delocalized excitations. We will focus on this regime and study the pair susceptibility of the model in the next section.

\subsection*{Pair susceptibility: tRVB vs. $s_\pm$}

As explained in the introduction, experiments on the iron-based
superconductors point towards an interplay between orbital selectivity
and superconductivity. Motivated by these considerations, here we study the pair susceptibility in this section, providing a comparison between tRVB and $s_\pm$ states.

In the normal state, the free energy has a  Landau-type expansion $F=F_0+\abs{\Delta}^2(1/g-\chi)+u\abs{\Delta}^4+\dots$ in $\Delta$. The quadratic term is controlled by the pair susceptibility
\be
\chi=\int{dx}\int{d\tau}\braket{(\psi\dg{\cal O}\bar\psi)_{x,\tau}(\bar\psi\dg{\cal O}\dg\psi)},
\ee
where ${\cal O}$ contains the matrix structure of the pairing term. Therefore, $\chi(T_c)=1/g$ determines the onset of pairing. Since we do not have access to the renormalized coupling $g$, we plot the susceptibilities vs. temperature, whose divergence appears as co-centric superconductivity domes. These can be directly compared to the onset of superconductivity in FeSC materials.

The bare pair susceptibility can be written as (Appendix \ref{sec:pair})
\be
\chi=-4\sum_{\vec k,nm}\frac{f(\eps_{n,\vec k})-f(-\eps_{m,-\vec k})}{\eps_{n,\vec k}+\eps_{m,-\vec k}}\vert{{\cal M}_{nm}(\vec k)}\vert^2\label{eqchi}
\ee
where $\cal M$ is the matrix elements of ${\cal O}$ in the band basis
\be
{\cal M}_{nm}(\vec k)=\varphi\dg_{n,\vec k}{\cal O}(\vec k)\sigma^y \varphi^*_{m,-\vec k}\label{eqM}
\ee
and the matrix ${\cal O}$ acts in orbital/spin/sublattice space. The
Pauli principle enforces the relation
\be
\sigma^y{\cal O}(\vec k)\sigma^y={\cal O}^T(-\vec k).\label{eqsymmetryk}
\ee 

Assuming ${\cal M}_{nn}(\vec k)\neq 0$ on the FS, the linearly vanishing denominator of \pref{eqchi} leads to a $\chi\sim-\log T$ behavior that eventually becomes dominant at low temperature. This is relevant for infinitesimal coupling $g$, but for generic coupling constants, and in particular, large Hund's coupling, the entire sum in Eq.\,\pref{eqchi} has significance.  For the $s^\pm$ we choose
\be
{\cal O}(\vec k)=\bb 1\cos k_x\cos k_y, \label{eq20}
\ee
where $\bb 1$ acts in orbital, spin and sublattice spaces but the $k$-factor changes sign at $k_{x,y}\sim\pm\pi/2$ between the electron/hole pockets.

The tRVB state is local and odd in orbital and spin
\be
{\cal O}(\vec k)=\sum_{a,b}\Lambda^{ab}L^b\sigma^a, \qquad \Lambda={\rm diag}(\tau^z,\tau^z,-2\tau^0),
\ee
where the Pauli matrix $\tau^z$ acts in the sublattice space, playing
the role of the staggered part of the tRVB order parameter
\cite{tRVB}.  In absence of SOI, ${\cal M}_{nm}(\vec k)=\vec
d_{nm}(\vec k)\cdot\vec \sigma$, where 
\be
d^a_{nm}(\vec
k)=\sum_b\Lambda^{ab}\varphi\dg_{n,\vec k}L^b\varphi^*_{m,-\vec k}]\nonumber
\ee
 is the
spin-triplet d vector which is odd in parity $\vec d_{nm}(\vec
k)=-\vec d_{mn}(-\vec k)$. It was shown in \cite{tRVB} that  $\vec
d_{nn}$ is non-zero on the FS, lying in the x-y plane and vanishing at
eight notes on the outer hole pocket. A finite SOI, rotates $\vec d$
out of the x-y plane and fills the nodes but also adds a singlet admixture to $\cal M$, whose uniform $\tau^0$ part is similar to the superconducting order parameter proposed by Vafek and Chubukov in Ref.\,\cite{VafekChubukov2017}. In addition, there are substantial inter-band and off-resonant contribution to the susceptibility, due to its local nature.

\begin{figure}[t!]
\includegraphics[width=0.49\linewidth]{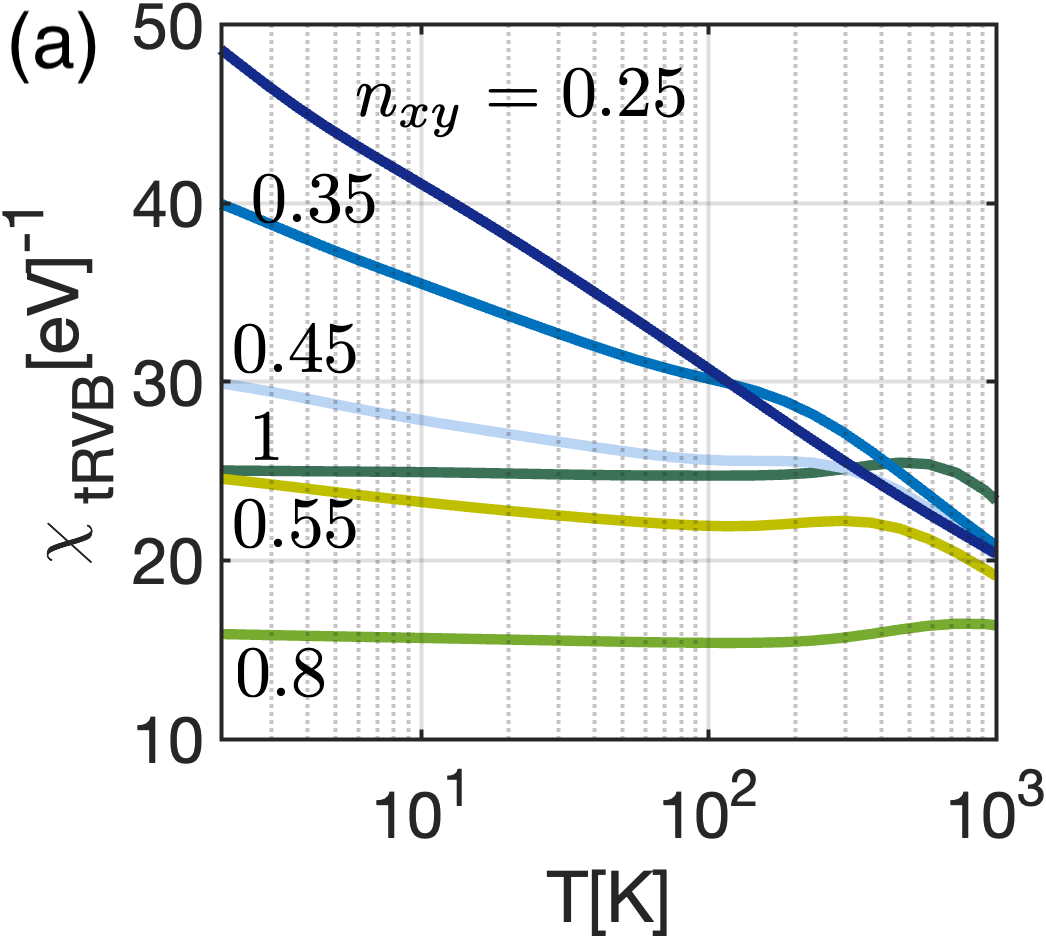}
\includegraphics[width=0.49\linewidth]{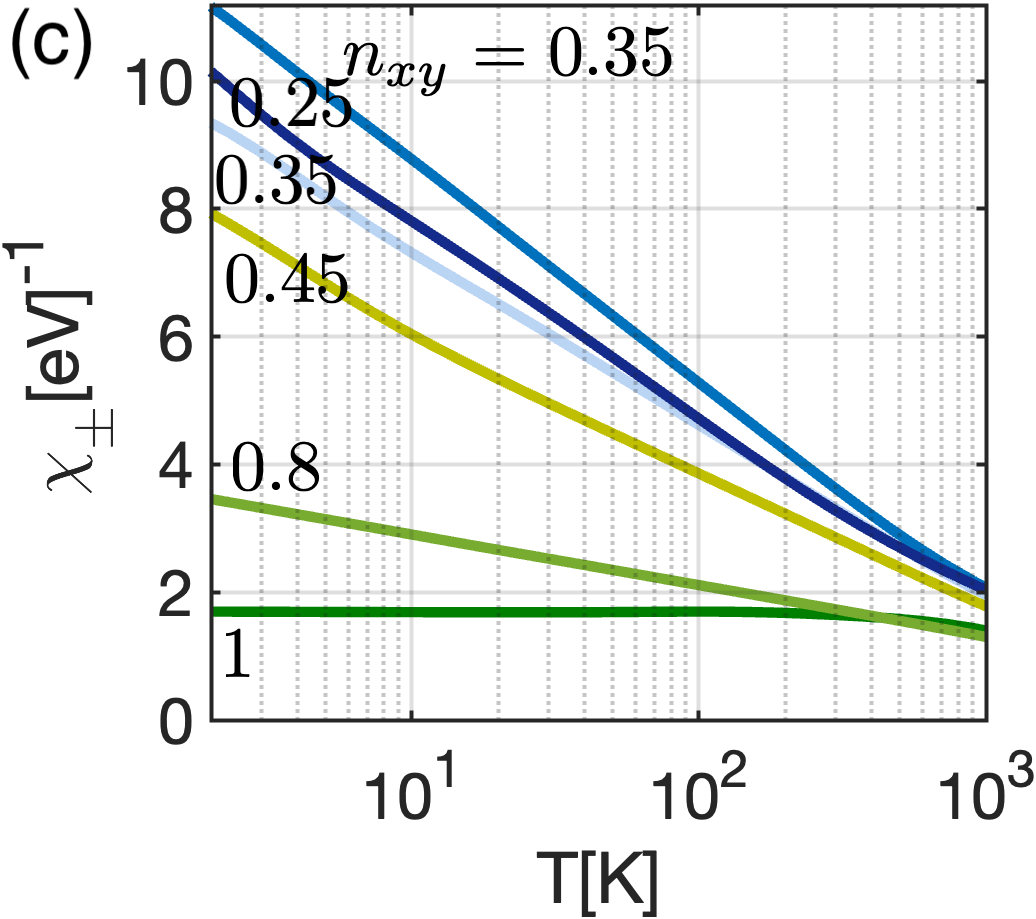}
\includegraphics[width=0.49\linewidth]{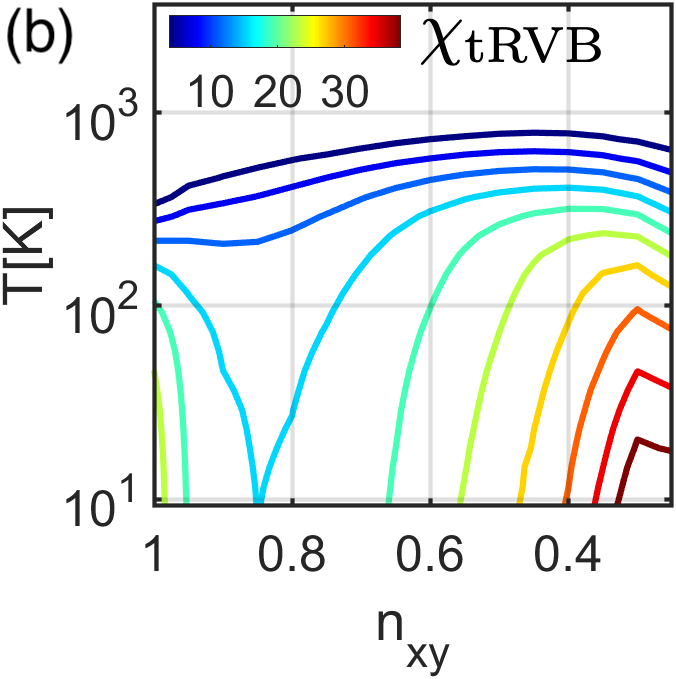}
\includegraphics[width=0.49\linewidth]{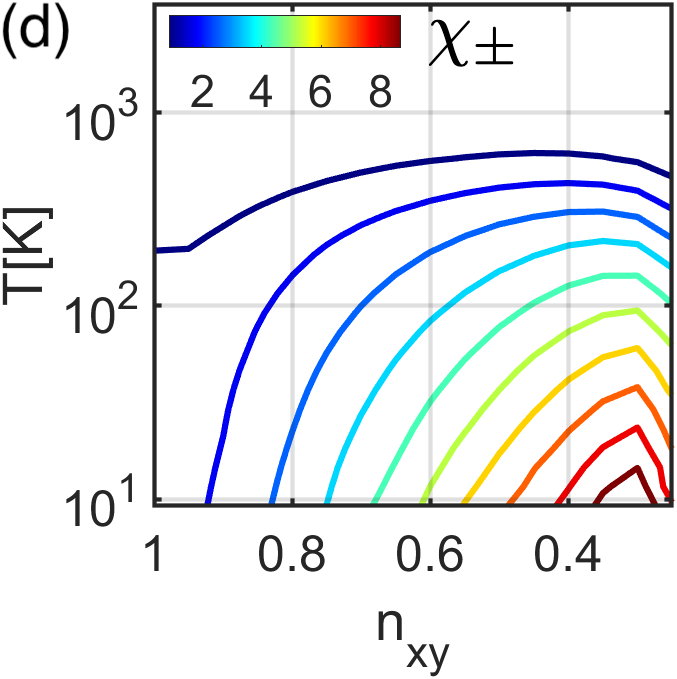}
\caption{Pair susceptibility to (a) tRVB state and (b) $s_\pm$ state, in units of $(eV)^{-1}$, within mean-field theory and under $\lambda_{SO}=20$meV. Only the coherent component of bosons is taken into account. Note that $\chi_{\rm tRVB}$ has much larger magnitude but the dome at small $n_{xy}$ is shared. In addition, tRVB exhibits another superconductivity dome near the OSMP at $n_{xy}\sim 1$.}\label{fig:pair}
\end{figure}

Fig.\,\pref{fig:pair} is the central result of our paper shows a comparison of the pair susceptibility of tRVB and $s_\pm$ as a function of xy doping and the temperature. The absolute magnitude of the two susceptibilities cannot be compared with each other, due to our lack of knowledge about the coupling constants. However, tRVB is driven by the Hund's coupling, which assuming a renormalized value of order 0.1eV, predicts a superconducting dome at $n_{xy}<0.4$ with a $T_c\sim 100 K$.


\section{Discussion and Conclusion}\label{sec:conclusion}

Both tRVB and $s_\pm$ states show a superconducting dome {around the
doping $n_{xy} \approx 0.3$ where} electrons in the  $xy$ orbital are delocalized. {In the presence of realistic spin-orbit coupling, the
mean-field pair susceptibilities of the tRVB state are enhanced by a
factor of about five. Moreover, our mean field theory 
demonstrates a clear correlation between increasing $xy$-orbital localization
and decreasing $T_c$ for both superconducting states,   a feature}  which is
consistent with experimental observations. 

Such correlations between $xy$ occupation and $T_c$ is observed in FeTe$_{1-x}$Se$_x$
\cite{Liu15} which exhibits an anti-ferromagnetic (AFM) order at
$x=0$. As $x$ is varied \cite{Liu15} from FeTe to FeSe, the
antiferromagnetism disappears at about $x=0.1$ and superconductivity
develops at $x>0.25$ with $T_c\sim
10K$ which depends only weakly on doping in the range
$x\in(0.3,0.5)$. A particularly fascinating feature, is that the renormalization of
the $xy$ band, as determined by the effective mass, diverges as 
$x\to 0.2$, so that the $T_c$ and bandwidth of the $xy$ orbital are
correlated at low doping, the latter strongly depending on the
temperature.  It is plausible that as the $xy$ electrons localize, they
produce the AFM order \cite{Huang2022}.

In summary, we have studied the pair susceptibility of the iron-based
superconductors, taking into account the effects of correlations,
orbital selectivity and Hund's interaction. The influence on band
renormalization, evolution of orbitals and Fermi surface reconstruction close to
the orbital selective Mott transition is captured. Away from the OSMT,
a mostly electron-hole compensated Fermi surface develops. We have
argued the importance of including 
charge projectors in the decoupling of the interaction
which enable a study of pair susceptibility in terms of physical
electrons. Furthermore, this provides a Hund's driven
mechanism for renormalizing down the inter-orbital charge repulsion
via spin-fluctuation.

We have
performed this calculation for the tRVB state to study the effects of
orbital localization and compare it to the s$_\pm$ state. We employed
a three-band tight-binding model for this calculation taking into account both
intra-band and inter-band contributions to the susceptibility. Both
states show a superconducting dome around $n_{xy}\sim 0.5$, but the
tRVB state also includes a much weaker superconducting dome close to
the orbitally selective Mott phase at $n_{xy}\sim 1$ due to inter-band contributions.

\acknowledgments
\emph{Acknowledgment} - Discussions with S. Fang are
appreciated. This work was performed in part at Aspen Center for
Physics, which is supported by NSF Grant No. PHY-1607611 and
work was also 
supported by Office of Basic Energy Sciences, Material
Sciences and Engineering Division, U.S. Department of Energy (DOE)
DE-FG02-99ER45790 (PC and EK).

\appendix

\section*{Appendices}
The following appendices contain further details and proof of various statements made in the paper.

\section{The model}\label{sec:appmodel}
We use the notation 
\be
k_x=\frac{k_++k_-}{\sqrt 2}, \qquad k_y=\frac{k_+-k_-}{\sqrt 2}.
\ee
The Hamiltonian is
\be
{\cal H}=\mat{{\cal H}\dn_{AA} & {\cal H}\dn_{AB} \\ {\cal H}_{AB}\dg & {\cal H}\dn_{BB}}\sigma^0-\lambda\tau^0\vec L\cdot\vec\sigma
\ee
in terms of [$c_{x/y}$ and $s_{x/y}$ denote $\cos k_{x/y}$ and $\sin k_{x/y}$.]
\be
{\cal H}_{AA}=\matc{ccc}
{4t_3c_xc_y & 4t_4s_xs_y & 4it_8s_xc_y\\  4t_4s_xs_y  & 4t_3c_xc_y & 4it_8c_xs_y\\  -4it_8s_xc_y& -4it_8c_xs_y& 4t_6c_xc_y+\Delta_{xy}}\nonumber
\ee
and
\be
{\cal H}_{AB}=2\matc{ccc}{-t_1c_y-t_2c_x & & it_7s_x  \\ & -t_1c_x-t_2c_y & it_7s_y \\ it_7s_x  & it_7s_y & t_5(c_x+c_y)}.\nonumber
\ee
We also have ${\cal H}_{BB}=T_z{\cal H}_{AA}T_z$ in terms of $T_z={\rm diag}(-1,-1,1)$. In absence of SOI, it is customary to do a gauge transformation $\psi_B\to T^z\psi_B$. Then, define uniform and staggered components
\be
\mat{\psi_A \\ \psi_B}=\frac{1}{\sqrt 2}\mat{1 & 1 \\ 1 & -1}\mat{\bar\psi \\ \Delta\psi}
\ee
and this reduces the problem to \cite{LeeWen2008} with $\vec Q=(\pi,\pi)$
\be
{\cal H}\to \mat{\tilde {\cal H}(\vec k)\sigma^0-\lambda L^z\sigma^z & -\lambda\vec L_\perp\cdot\vec\sigma_\perp\\ -\lambda\vec L_\perp\cdot\vec\sigma_\perp & \tilde{\cal H}(\vec k+\vec Q)\sigma^0-\lambda L^z\sigma^z}
\ee
where 
\be
\tilde{\cal H}(k)={\cal H}_{AA}(k)+{\cal H}_{AB}(k)T^z.
\ee
The bare parameters are listed in the table \pref{tab:bare}.
\begin{table}
\begin{tabular}{|c|c|c|c|c|c|c|c|c|c|}
\hline
$t_1$ & $t_2$ & $t_3$ & $t_4$ & $t_5$ & $t_6$  & $t_7$ & $t_8$ & $\Delta_{xy}$ & $\mu$\\
\hline
0.06 & 0.02 & 0.03 & -0.01& 0.2 & 0.3 & -0.2 & $-t_7/3$ & 0.4 & 0.212\\
\hline
\end{tabular}
\caption{Model parameters, slightly modified from \cite{DaghoferDagotto2010}.}\label{tab:bare}
\end{table}

\section{Proof of Eq.\,\eqref{eq4}}\label{sec:eq4}

The key equation is the Fierz identity
\be
\vec\sigma_{\alpha_1\beta_1}\cdot\vec\sigma_{\alpha_2\beta_2}+\delta_{\alpha_1\beta_1}\delta_{\alpha_2\beta_2}=2\delta_{\alpha_1\beta_2}\delta_{\beta_1\alpha_2}
\label{eqS0}.
\ee
which we rewrite as 
\be
\vec\sigma_{\alpha_1\beta_1}\cdot\vec\sigma_{\alpha_2\beta_2}=2\delta_{\alpha_1\beta_2}\delta_{\beta_1\alpha_2}-\delta_{\alpha_1\beta_1}\delta_{\alpha_2\beta_2}\label{eqS1}.
\ee
We can contract this with $\sigma^y_{\beta_1\alpha'}$ and $\sigma^y_{\beta'\alpha_2}$ to find
\be
(\vec\sigma\sigma^y)_{\alpha\alpha'}\cdot (\sigma^y\vec\sigma)_{\beta'\beta}=2\delta_{\alpha\beta}\delta_{\alpha'\beta'}-\sigma^y_{\alpha\alpha'}\sigma^y_{\beta'\beta}\label{eqS2}
\ee
after re-labeling $\alpha_1\to\alpha$ and $\beta_2\to\beta$. Eq.\,\pref{eqS1} also gives
\be
\vec\sigma_{\alpha\beta'}\cdot\vec\sigma_{\alpha'\beta}=2\delta_{\alpha\beta}\delta_{\beta'\alpha'}-\delta_{\alpha\beta'}\delta_{\alpha'\beta}\label{eqS3}\\
\ee
\begin{widetext}
Subtracting \pref{eqS3} from \pref{eqS2}, we obtain 
\be
(\vec{\sigma }\sigma^{y})_{\alpha \alpha '}\cdot (\vec{ \sigma }\sigma^{y})_{\beta '\beta }
-\vec{\sigma }_{\alpha \beta '}\cdot \vec{ \sigma }_{\alpha '\beta }
 =\delta_{\alpha\beta'}\delta_{\alpha'\beta}-\sigma^y_{\alpha\alpha'}\sigma^y_{\beta'\beta}\label{eqS4}
\ee
Here, a useful relation is 
\be
\sigma^y_{\alpha\alpha'}\sigma^y_{\beta'\beta}=\delta_{\alpha\beta}\delta_{\alpha'\beta'}-\delta_{\alpha\beta'}\delta_{\beta\alpha'},
\label{eqS5}
\ee
which expresses the fact that the initial states and final states 
are either parallel, or antiparallel, coming in with opposite amplitudes.
Combining \pref{eqS4} and \pref{eqS5} we find
\bea
(\vec{\sigma }\sigma^{y})_{\alpha \alpha '}\cdot (\vec{ \sigma }\sigma^{y})_{\beta '\beta }
-\vec{\sigma }_{\alpha \beta '}\cdot \vec{ \sigma }_{\alpha '\beta }
&=&2\delta_{\alpha\beta'}\delta_{\beta\alpha'}-\delta_{\alpha\beta}\delta_{\alpha'\beta'}\nonumber\\
&=&\vec\sigma_{\alpha\beta}\cdot\vec\sigma_{\alpha'\beta'},\label{eqS6}
\eea
where we have employed the Fierz equality \eqref{eqS1} again in the last step, thus
proving Eq.\,\pref{eq4}. Note that the triplet decoupling \pref{eq4} is very similar (but has opposite in sign) to the singlet decoupling, which can be obtained from \pref{eqS5} and \pref{eqS6}:
\be
\vec\sigma_{\alpha\beta}\cdot\vec\sigma_{\alpha'\beta'}=-\sigma^y_{\alpha\alpha'}\sigma^y_{\beta'\beta}+\delta_{\alpha\beta'}\delta_{\alpha'\beta}\label{eqS7}.
\ee
\end{widetext}

\section{Five-band to three-band reduction}\label{sec:reduction}

\begin{figure}[h!]
\includegraphics[width=\linewidth]{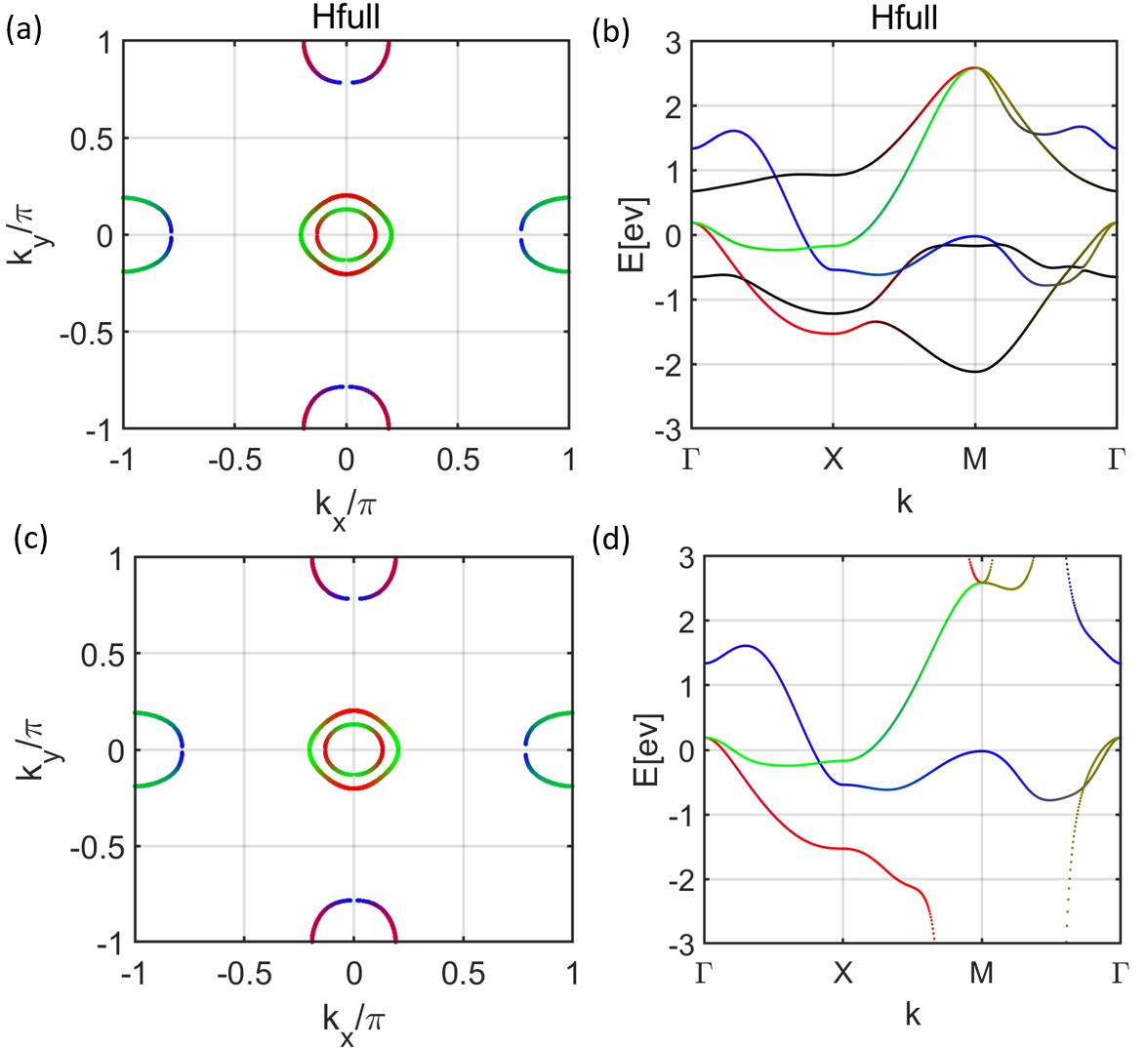}
\caption{\small (a-b) A five band model of FeAs layers in LaOFeAs, reflected in FSs and the dispersion along a cut through the BZ. No spin-orbit interaction is included and the diagrams are unfolded dispersions, with the blue/green/red indicating $xz/yz/xy$ orbital contents. (b-d) An effective three-band model, resulted by integrating out the eg orbitals, with diverging bands, which nevertheless matches the five-band models at low-energies.}\label{fig:3vs5}
\end{figure}

FeSCs are usually described by three-band \cite{DaghoferDagotto2010} or five-band \cite{Graser2009,Eschrig09} models. This extensively discussed in \cite{Fernandes2016,BorisenkoZhigadlo2016}. Since the $e_g$ orbitals mix with $t_{2g}$ orbitals, Fig.\,\ref{fig:3vs5}(a,b), it cannot be ignored and since according to DFT calculations it crosses the chemical potential, it's effect is beyond a merely renormalization of the tight-binding parameters. However, as we show here, an effective 3-band model provides a faithful representation of the material close to the Fermi energy. The Green's function for the 5-band system is
\be
{\cal G}(k,z)=[z{\bb 1}_{5\times 5}-{\cal H}(k)]^{-1}
\ee
where
\be
{\cal H}(k)=\mat{{\cal H}_{tt} & {\cal H}_{te} \\ {\cal H}_{et} & {\cal H}_{ee}}
\ee
Focusing on $t_{2g}$ orbitals, the Green's function is
\be
{\cal G}_{tt}(k,z)=[z{\bb 1}_{3\times 3}-{\cal H}_{tt}(k)-\Sigma_{tt}(k,z)]^{-1}
\ee
where
\be
\Sigma_{tt}(k,z)={\cal H}_{te}(z-{\cal H}_{ee})^{-1}{\cal H}_{et}
\ee
which motivates defining the static Hamiltonian \cite{Wang2012}
\be
{\cal H}_{tt,\rm eff}(k)\equiv{\cal H}_{tt}-{\cal H}_{te}{\cal H}_{ee}^{-1}{\cal H}_{et}.
\ee
Fig.\,\ref{fig:3vs5}(c,d) shows the band structure and FS of ${\cal H}_{tt,\rm eff}(k)$. Clearly, the spectrum diverges at points in the BZ and therefore, a tight-binding representation is not available. However, the FS is captured faithfully and this effective Hamiltonian can use it for practical calculation.

Note that ${\cal H}_{ee}=0$ appears as an infinity of ${\cal H}_{eff}$ or a zero of the effective $t_{2g}$ Green's function. Here, we will discuss the occupancy of the effective Hamiltonian, showing that the $t_{2g}$ orbital will have an occupancy that includes the zeros of the Green's function due to integrated-out $e_g$ states. According to the Luttinger's theorem
\be
n=\frac{1}{(2\pi)^d}{\cal V}_{FS}
\ee
where $d=2$ is the dimensionality and the FS volume is
\be
{\cal V}_{FS}=\frac{1}{\pi}{\rm Im Tr}\log[-{\cal G}^{-1}(k,z)]\Big\vert_{z=0+i\eta},
\ee
so that any place where $\re{{\cal G}^{-1}(k,0)}>0$ is counted as occupied state. Denoting ${\cal G}^{-1}(k,z)=P(k,z)/Q(k,z)$, with $P(k,z)=\prod_{m} (z-p^{m}_k)$, and $Q(z)=\prod_{n}(z-q_k^{n})$ there are two ways the Green's function can change sign: through poles $p^m_k$, covering the total area ${\cal V}_{\rm poles}$ in the BZ, or through zeros $p^n_k$, covering the area ${\cal V}_{\rm zeros}$. This leads to the following relation:
\be
n=\frac{1}{(2\pi)^d}[{\cal V}_{\rm poles}-{\cal V}_{\rm zeros}].
\ee
Therefore, poles alone enclose an enlarged area that contains the FS of the integrated-over orbitals. In the present case, $e_g$ bands have occupancy 2. Therefore, the effective $t_{2g}$ orbital has the occupancy of 4.

\section{Hund's driven attraction, beta function}\label{sec:beta}
In this section, we should that the decoupled Hund's coupling in the normal state of the tRVB order parameter creates an attractive interaction between the orbitals. For simplicity, we consider an infinite-$U$ impurity setting, where a single Fe atom is hybridized with conduction electrons, which realizes a DMFT setting. For simplicity, we assume 

We assume each orbital has its own bath, with a hybridization that can depend on energy. The Hamiltonian is $H=H_0+H_{U'}+H_{\rm Hund}+H_\lambda$ where $H_0$ contains the conduction electron and their hybridization with different orbitals of Fe. The interaction is given by
\bea
H_{U'}&=&\sum_{rr'}U'_{rr'}(b\dn_rb\dg_r)(b\dn_{r'}b\dg_{r'})\\
H_{\rm Hund}&=&\frac{\abs{\Delta}^2}{g}+\sum_{R,R'}\Delta[b_Rf\dg_{R}({\cal O}\sigma^y)_{RR'}f\dg_{R'}b_{R'}+h.c.].\nonumber
\eea
Here we have introduced super-index $R=(r,s)$ and $R'=(r',s')$ where $r,r'$ are orbital and $s,s'$ are spin degrees of freedom and it is understood that $b_R=b_r$. $H_\lambda$ contains Lagrange multipliers that impose the infinite intra-orbital $U$ constraint. For simplicity we have assumed the occupancy of all the orbitals is less than one. This could easily changed by doing $f\to \tilde f$ and $b\to \tilde b$ with the notation of the paper.

\begin{figure}[tp!]
\includegraphics[width=0.45\linewidth]{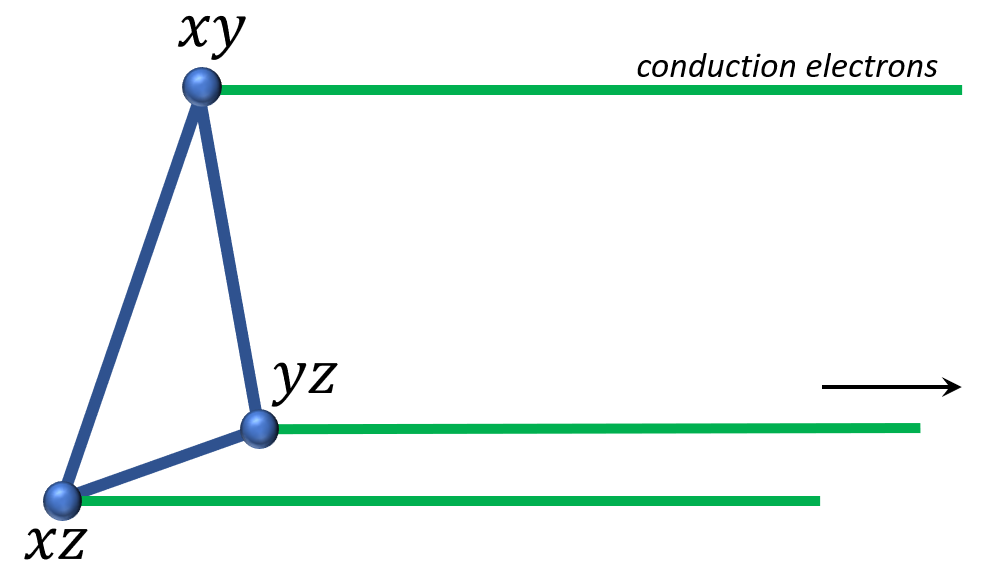}
\qquad
\includegraphics[width=0.35\linewidth]{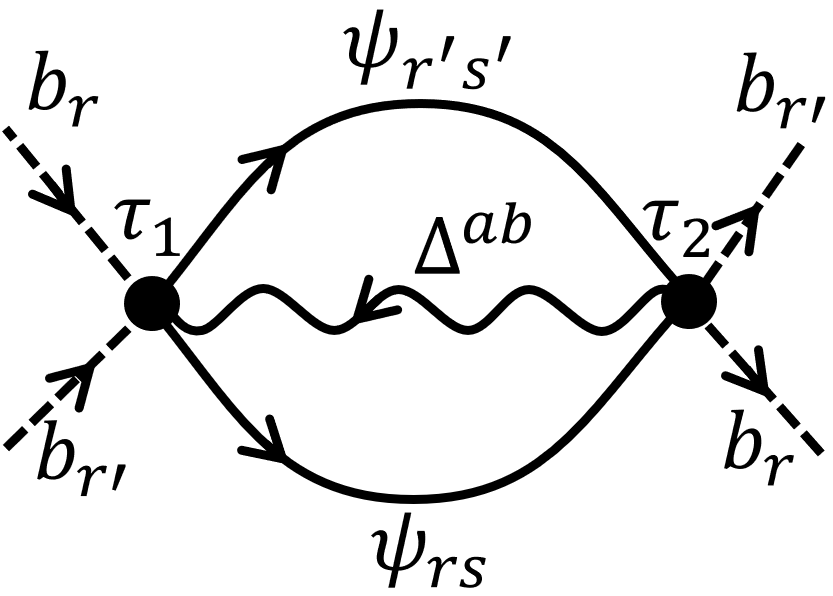}
\caption{\small (a) The impurity setting of a 3-orbital Fe atom, hybridized with conduction electrons, as a first iteration of a dynamical mean-field theory. (b) The basic Feynman diagram we compute here. The wavy $\Delta^{ab}$ line is the propagator of the Hubbard-Stratonovich fields. In the normal phase, the propagator is just the (renormalized) Hund's coupling constant.}\label{fig:rg}
\end{figure}

We would like to do an RG studies of this Hamiltonian. We can write the partition function in the interaction picture w.r.t. $H_0$:
\be
Z/Z_0=\braket{T_\tau e^{-\int_0^\beta d\tau H_{\rm Hund}(\tau)}},
\ee
and compute $Z$ to second order in $w$, representing $\tau_1+\tau_2=2\tau$ and $\tau_1-\tau_2=\tau'$. This is shown by the Feynman diagram \ref{fig:rg}(b). We use the fact that within the time-scale $\tau'\in(\tau_0,\tau_0+\delta\tau_0)$ beside a phase evolution by $b_r(\tau)=e^{-\lambda_r\tau}b_r$, the holons are slowly varying so that
\be
b_r(\tau_1)b\dg_r(\tau_2)\approx e^{-\lambda_r\tau'}b_r(\tau)b\dg_r(\tau).
\ee
For the second order term we find
\bea
Z_2/Z_0&=&\sum_{R,R'}\int_0^{\beta}{d\tau}b_R(\tau)b_{R'}(\tau)b\dg_{M'}(\tau)b\dg_M(\tau)\times\nonumber\\
&&\int_{\tau_0}^{\tau_0+\delta\tau_0}{d\tau'}e^{-(\lambda_r+\lambda_{r'})\tau'}\braket{\Delta(\tau')\Delta\dg}\times\nonumber\\
&&\hspace{-1cm}\braket{f\dg_R(\tau_1)f\dg_{R'}(\tau_1)f_{M'}(\tau_2)f_{M}(\tau_2)}
({\cal O}\sigma^y)_{RR'}(\sigma^y{\cal O}\dg)_{M'M}\nonumber
\eea
At this point we make some simplifying assumption. First we assume that interaction effects can be neglected and we are in the normal phase. This means we can apply the Wick's contraction
\bea
&&\braket{f\dg_Rf\dg_{R'}f_{M'}(\tau)f\dn_{M}(\tau)}\label{eqg79}\\
&&\qquad\qquad=G_{MR}(\tau)G_{M'R'}(\tau)-G_{M'R}(\tau)G_{MR'}(\tau)\quad\nonumber
\eea
Next, we assume we are in the paramagnetic regime and different orbitals are not correlated at high-temperature, meaning that fermion propagators are diagonal in spin and orbital, although orbital asymmetry can be present. The two terms in Eq.\,\pref{eqg79} add up:
\bean
{\cal R}_{rr'}&\equiv&{\rm Tr}_{\rm spin}[({\cal O}\sigma^y)_{RR'}(\sigma^y{\cal O}\dg)_{R'R}]\nonumber \\
&=&-{\rm Tr}_{\rm spin}[({\cal O}\sigma^y)_{RR'}(\sigma^y{\cal O}\dg)_{R'R}]\nonumber \\
&=& {\rm Tr}_{\rm spin}[{\cal O}_{rr'}{\cal O}_{r'r}\dg],
\eean
where $r,r'$ denote the orbital index of $R,R'$ subindices and we have used Eq.\,\pref{eqsymmetryk} to deduce $(\sigma^y{\cal O}\dg)^T=-\sigma^y{\cal O}\dg$.  For the tRVB state ${\cal O}=L^a\sigma^a$, 
\be
{\cal R}_{rr'}=2\sum_aL^a_{rr'}L^a_{r'r}=2(1-\delta_{rr'}).
\ee
Moreover, in the normal phase the $\Delta$ propagator is approximately constant and equal to the inverse (renormalized) coupling constant
\be
g(\tau')=\braket{\Delta(\tau')\Delta\dg}\approx g(\tau_0).
\ee
Then we can write 
\bea
Z_2/Z_0&=&\sum_{rr'}\int{d\tau}(b_rb\dg_r)_\tau(b_{r'}b\dg_{r'})_\tau g(\tau_0) {\cal R}_{rr'}\nonumber\\
&&\qquad\int{d\tau'}e^{-(\lambda_r+\lambda_{r'})\tau'}G_r(0,\tau')G_{r'}(0,\tau').\quad
\eea
At high energies $e^{-\lambda_r\tau'}$ can be dropped out with a similar term with an opposite sign inside $G_r(\tau')\sim -e^{+\lambda_r\tau'}\rho_r/{\tau'} $. We have assumed each orbital is in the Fermi liquid state with a bandwidth governed by $\rho_r$.  Therefore, doing the integral and replacing $w\to 1$ we find
\be
Z_2/Z_0=-\sum_{rr'}\int{d\tau}[(b_rb_r\dg)_\tau(b_{r'}b_{r'}\dg)_\tau]g(\tau_0){\cal R}_{rr'}\rho_r\rho_{r'}\frac{\delta\tau_0}{\tau_0^2}\nonumber
\ee
This has the same form as the $U'$ term and can be absorbed to renormalize its value
\be
U'_{rr'}\to U'_{rr'}+g{\cal R}_{rr'}\rho_r\rho_{r'}\frac{1}{\tau_0}d\log\tau_0
\ee
or using $1/\tau_0=D$ and defining $d\ell=d\log\tau_0=-d\log D$, the dimensionless coupling $u'_{rr'}=U'_{rr'}/D$ is modified to
\be
u'_{rr'}\to u'_{rr'}-g\rho_r\rho_{r'}{\cal R}_{rr'}d\ell
\ee
We add to this a tree-level renormalization of the relevant operator $u'_{rr'}$ which arises due to scaling of $\tau_0\to\tau_0+\delta\tau_0$.
\bea
H_{inter}&=&\sum_{rr'}u'_{rr'}\tau_0 (b\dg_rb\dn_r)(b\dg_{r'}b\dn_{r'})\\
&\to& \sum_{rr'}u'_{rr'}(\tau_0+\delta\tau_0)\frac{\tau_0}{\tau_0+\delta\tau_0} (b\dg_rb\dn_r)(b\dg_{r'}b\dn_{r'})\nonumber
\eea
which means
\be
u'_{rr'}\to u'_{rr'}(1-\delta\tau_0/\tau_0)=u'_{rr'}(1+d\ell)
\ee
So, adding these two contributions we find
\be
\frac{du'_{rr'}}{d\ell}=u'_{rr'}-g\rho_r\rho_{r'}{\cal R}_{rr'}
\ee

\section{Pair susceptibility}\label{sec:pair}
We consider a pairing terms of the type
\be
\hspace{-.12cm}H_\Delta=\Delta\int{dx}\sum_{RR'}\sum_\delta[\psi\dg_R(x)({\cal O}_{\delta}\sigma^y)_{RR'}\psi\dg_{R'}(x+\delta)+h.c.]\nonumber
\ee
where $R,R'$ are super-indices containing orbital/spin/sublattice and $\delta$ denotes the relative position of the two electrons in a Cooper pair. Note that the order parameter has the symmetry
\be
({\cal O}_{\delta}\sigma^y)_{RR'}=-({\cal O}_{-\delta}\sigma^y)_{R'R}\label{eqsymmetry}
\ee
due to Pauli principle. A second-order perturbation theory in $\Delta$ gives a contribution $\Delta F=\Delta^2\chi$ to the Free energy where the susceptibility $\chi$ is given by
\bea
\chi&=&-\sum_{\delta\delta'}\int{d^2x}\intob{d\tau}({\cal O}_\delta \sigma^y)_{RR'}(\sigma^y{\cal O}_{\delta'}\dg)_{M'M}\nonumber\\
&&\qqquad\times[\Pi(x,\tau)+\Pi(x,-\tau)],\quad\label{eq35b}
\eea
expressed in terms of the fermionic bubble
\be
\Pi(x,\tau)=\braket{\psi\dg_R\psi\dg_{R'}(\delta,0)\psi\dn_{M'}(x+\delta',\tau)\psi\dn_{M}(x,\tau)}.\\
\ee\\

\subsection*{Bare Susceptibility}
Using Wick's contraction we find
\bea
\Pi(x,\tau)&=&G_{MR}(x,\tau)G_{M'R'}(x+\delta'-\delta,\tau)\nonumber\\
&&\qqquad-G_{M'R}(x+\delta',\tau)G_{MR'}(x-\delta,\tau). \nonumber
\eea
The Green's functions can be expressed as
\be
G_{MR}(x,\tau)=\frac{1}{\beta}\sum_{n,k} e^{ikx-i\omega_n\tau}\int{\frac{d\omega}{2\pi}}\frac{A_{MR}(k,\omega)}{i\omega_n-\omega}\label{eqG}
\ee
in terms of $A(\omega)\equiv-[G(\omega+i\eta)-G(\omega-i\eta)]$. Doing the imaginary-time integral, and the Matsubara sum and using that the two terms in Eq.\,\pref{eq35b} are equal we find
\bea
\chi&=&\sum_k\int{\frac{d\omega d\omega'}{(2\pi)^2}}A_{MR}(k,\omega)A_{M'R'}(-k,-\omega')\frac{f(\omega)-f(-\omega')}{\omega+\omega'}\nonumber\\
&&\hspace{-.5cm}\times 2\hspace{-.2cm}\sum_{RR'MM'}\sum_\delta\Big[({\cal O}_{\delta}\sigma^y)_{RR'}-({\cal O}_{-\delta}\sigma^y)_{R'R}\Big]e^{-ik\delta}\nonumber\\
&&\qquad\times \sum_{\delta'}(\sigma^y{\cal O}_{\delta'}\dg)_{M'M}e^{ik\delta'}.\nonumber
\eea
Finally, using the anti-symmetry of the order parameter \pref{eqsymmetry} and that the spectral function in the band-basis has a simple form
\be
A(k,\omega)=2\pi\sum_n\varphi_n(k)\delta(\omega+i\eta-\eps_{n,k})\varphi\dg_n(k)\label{eq41}
\ee
we arrive at Eqs.\,\pref{eqchi} and \pref{eqM}.
\subsection*{Susceptibility from mean-field theory}
In this case 
\be
\Pi(x,\tau)=\Pi^f(x,\tau)\Pi^b(x,\tau)\label{eqmfPi}
\ee
where
\bea
\Pi^f(x,\tau)=\sbraket{\tilde f\dg_R\tilde f\dg_{R'}(\delta,0)\tilde f\dn_{M'}(x+\delta',\tau)\tilde f\dn_{M}(x,\tau)}\\
\Pi^b(x,\tau)=\sbraket{\tilde b\dn_R\tilde b\dn_{R'}(\delta,0)\tilde b\dg_{M'}(x+\delta',\tau)\tilde b\dg_{M}(x,\tau)}.
\eea
We can apply Wick's contraction to each of these four-point functions and using Eq.\,\pref{eqG} find
\bw

\bean
\chi&=&\frac{ 2}{{\cal N}^3}\sum_{k_1k_2q_1q_2}\int{\frac{d\omega_1\omega_2\omega'_1\omega'_2}{(2\pi)^4}}\delta_{k_1+k_2+q_1+q_2,0}\frac{e^{\beta(\omega_1+\omega_2)}-e^{\beta(\omega'_1+\omega'_2)}}{\omega_1+\omega_2-\omega'_1-\omega'_2}f(\omega_1)f(\omega_2)n_B(\omega'_1)n_B(\omega'_2)A^f_{MR}(k_1,\omega_1)A^f_{M'R'}(k_2,\omega_2)\\
&& \sum_{\delta\delta'}({\cal O}_\delta\sigma^y)_{RR'}(\sigma^y{\cal O}\dg_{\delta'})_{M'M}[A^b_{MR}(q_1,\omega'_1)A^b_{M'R'}(q_2,\omega_2')e^{i(k_2+q_2)(\delta'-\delta)}+A^b_{MR'}(q_1,\omega'_1)A^b_{M'R}(q_2,\omega'_2)e^{i(k_2+q_2)\delta'-i(k_2+q_1)\delta}]
\eean
Using Eq.\,\pref{eq41} we find 
\bean
\chi&=&\frac{ 2}{{\cal N}^3}\hspace{-.3cm}\sum_{\text{\scalebox{0.8}{$\matn{k_1k_2q_1q_2\\n_1n_2m_1m_2}$}}}\hspace{-.3cm}\delta_{\vec k}\frac{e^{\beta(\eps^f_{n_1k_1}+\eps^f_{n_2k_2})}-e^{\beta(\eps^b_{m_1q_1}+\eps^b_{m_2q_2})}}{\eps_{n_1k_1}+\eps_{n_2k_2}-\eps_{m_1q_1}-\eps_{m_2q_2}}f(\eps^f_{n_1k_1})f(\eps^f_{n_2k_2})n_B(\eps^b_{m_1q_1})n_B(\eps^b_{m_2q_2})\varphi^f_{M}(nk_1)\varphi^{f*}_R(nk_1)\varphi^f_{M'}(n_2k_2)\varphi^{f*}_{R'}(n_2k_2)\\
&&\hspace{-.5cm}[\sigma^y{\cal O}\dg(k_2+q_2)]_{M'M}\varphi^{b*}_{M}(m_1q_1)\varphi^{b*}_{M'}(m_2q_2)\Big([{\cal O}(k_2+q_2)\sigma^y]_{RR'}\varphi^b_R(m_1q_1)\varphi^b_{R'}(m_2q_2)+[{\cal O}(k_2+q_1)\sigma^y]_{RR'}\varphi^b_{R'}(m_1q_1)\varphi^b_{R}(m_2q_2)\Big)
\eean

\ew

A numerical evaluation of this sum is computationally costly. A simplification happens in the coherent regime $n_B(\eps_{mq})\approx{\cal N}{n_m}\delta_{\eps_{mq},0}$ which happens for one of the bosonic bands. In that case, $\varphi_M^b(0)\sim 1$ and we find the same equation as Eq.\,\pref{eqchi} except that the renormalized ${\cal M}$ is given by
\be
{\cal M}_{nm}(\vec k)=n^b_nn^b_m\vert{\varphi^T_{n,-\vec k}\sigma^y{\cal O}(\vec k) \varphi_{m,\vec k}}\vert^2\label{eqM}
\ee
We examine this approximation closely in the single-band Hubbard model.

\section{Single-band Hubbard model}\label{1band}
In the limit of infinite-$U$, the single-band Hubbard is mapped to the $t-J$ model, 
\bea
H&=&\sum_{{ij},\sigma}t_{ij}b\dn_i f\dg_{i\sigma}f\dn_{j\sigma}b\dg_j+\sum_{{ij},\sigma}J_{ij}\vec S_i\cdot\vec S_j\nonumber\\
&&\qqquad+\sum_i\lambda_i(f\dg_{i\sigma}f\dn_{i\sigma}+b\dg_ib\dn_i-1),\label{eqtJ}
\eea
The $J$-term can be decoupled in the singlet channel using Eq.\,\pref{eqS7}  in particle-hole and particle-particle channels. Again, both channels are attractive and can acquire finite expectation value. We find
\bea
J_{ij}\vec S_i\cdot\vec S_j&&\quad\to\quad\frac{\abs{\kappa_{ij}}^2}{J_{ij}}+\sum_\sigma(\kappa_{ij}f\dg_{i\sigma}f\dn_{j\sigma}b_ib\dg_j+h.c.)\nonumber\\
&&+\frac{\abs{\Delta_{ij}}^2}{J_{ij}}+\sum_\sigma(\Delta_{ij}f\dg_{i\sigma}\bar f\dn_{j\sigma}b_ib\dn_j+h.c.).
\eea
The new feature compared to \cite{KotliarLiu1988} is the additional factor of $b_i$ which result from decoupling of hidden charge projectors in \pref{eqtJ}. The two channels $\kappa$ and $\Delta$ behave differently as $\kappa_{ij}$ is being driven by the $t_{ij}$ and just renormalizes $t_{ij}\to\tilde t_{ij}$. So, we can write $H=H_0+H_\Delta$ where
\bea
H_0&=&\sum_{ij,\sigma}\tilde t_{ij}b\dn_i f\dg_{i\sigma}f\dn_{j\sigma}b\dg_j+\sum_i\lambda_i(f\dg_{i\sigma}f\dn_{i\sigma}+b\dg_ib\dn_i-1),\nonumber
\eea
\bea
H_{\Delta}=\frac{\abs{\Delta_{ij}}^2}{J}+\sum_{ij,\sigma}(\Delta_{ij}f\dg_{i\sigma}\bar f\dn_{j\sigma}b\dn_{i}b\dn_{j}+h.c.).
\eea
Note that this Hamiltonian is expressed entirely in terms of infinite-$U$ real electrons $\psi\dn_{i\sigma}=b\dg_if\dn_{i\sigma}$. The plan we follow in this paper is to solve this problem in the normal state using mean-field decoupling of holons and fermions and then compute the pair susceptibility of the real electrons.  A mean-field decoupling $H_0\to H_0^f+H_0^b-\sbraket{H_0^f}$ gives up to a constant shift in energy
\bea
H_0&\to& \sum_{ij,\sigma}f\dg_{i\sigma}(t_{ij}^f-\mu_i^f\delta_{ij})f\dn_{j\sigma}+\sum_{ij}b\dn_i(t_{ij}^b-\mu_i^b\delta_{ij})b\dg_j,\qquad
\eea
where
\be
t_{ij}^f=t_{ij}\sbraket{b_ib\dg_j}\andd t_{ij}^b=t_{ij}\sum_\sigma\sbraket{f\dg_{i\sigma}f\dn_{j\sigma}}.
\ee
At low-T, we find $t_{ij}^f\to t_{ij}q_b$ whereas $t_{ij}^b$ is determined by the average kinetic energy of occupied fermions. A self-consistent solution to these equations represent a fixed point solution (in a statistical mechanical sense) to the interacting problem $H_0$. In terms of these $\kappa^*_{ij}=J_{ij}t^b_{ij}t^t_{ij}/t_{ij}$. Using translational invariance
\be
H_0=\sum_{k\sigma}\eps_k^ff\dg_{k\sigma}f\dn_{k\sigma}+\sum_k\eps_k^bb_k\dg b_k.
\ee
At low temperature and sufficiently large dimension, the bosons will condense. The computation below is done in a finite system size and contains this transition as a crossover. A comment about possible interaction 
\be
H_{\rm int}=\sum_{ij}V_{ij}b\dg_ib\dn_ib\dg_jb\dn_j
\ee
is in order. Clearly $V_{ij}$ can derive various forms of charge-density wave. Within the mean-field theory and as long as translational invariance is assumed, this interaction does not play any role and only changes the relation between chemical potential and the doping. Next, we compute the pair susceptibility for this state, assuming translational invariance $\Delta_{i,i+\delta}=\Delta{\cal O}_\delta$. For the d-wave,
\be
\delta=(\hat x,-\hat x,\hat y,-\hat y), \so {\cal O}_\delta=(1,1,-1,-1).
\ee
It is convenient to define a form-factor
\be
{\cal O}(k)=\sum_\delta {\cal O}_\delta e^{-ik\delta}\quad \to\quad {\cal O}_{\rm d-wave}(k)=2(\cos k_x-\cos k_y).\nonumber
\ee
The pair susceptibility is
\bea
\chi&=&\frac{1}{2}\sum_{\delta\delta'}\int{dx}\intob{d\tau}{\cal O}_\delta{\cal O}_{\delta'}[{\cal G}_F(x,\tau){\cal G}_B(x,\tau)+\tau\to-\tau]\nonumber
\eea
expressed in terms of the fermion/holon bubbles,
\bea
{\cal G}_F(x,\tau)&=&\sum_{\sigma,\sigma'}\tilde\sigma\tilde\sigma'\braket{f\dg_\sigma(x,\tau)f\dg_{\bar\sigma}(x+\delta,\tau)f\dn_{\bar\sigma'}(\delta')f\dn_{\sigma'}(0)},\nonumber\\
{\cal G}_B(x,\tau)&=&\braket{b(x,\tau)b(x+\delta,\tau)b\dg(\delta')b\dg(0)}.
\eea
A straightforward calculation gives a simplified version of the expression from the previous appendix:
\bw

\be
\chi=\frac{2}{{\cal N}^3}\sum_{kq_1q_2}\frac{R(k,q_1,q_2)}{\eps_{k}+\eps_{-k-q_1-q_2}-\eps_{q_1}-\eps_{q_2}}\Big(e^{\beta(\eps_{k}+\eps_{-k-q_1-q_2})}-e^{\beta(\eps_{q_1}+\eps_{q_2})}\Big)f(\eps_{k}^f)f(\eps_{-k-q_1-q_2}^f)n_B(\eps_{q_1}^b)n_B(\eps_{q_2}^b)
\ee
\ew

where
\bea
R(k;q_1,q_2)&=&2{\cal O}(k_2+q_2)[{\cal O}(k_2+q_2)+{\cal O}(k_2+q_1)]\nonumber\\
&=&2{\cal O}(k+q_1)[{\cal O}(k+q_1)+{\cal O}(k+q_2)]
\eea
$400\times 400$.
At low-T, we can approximate $n_b(\eps_q)\approx{\cal N}q_b\delta_{q,0}+e^{-\beta\eps_q}$. The first term gives the usual contribution
\be
\chi_{00}=\frac{8q_b^2}{\cal N}\sum_k{\cal O}^2(k)\frac{f(\tilde\eps_k)-f(-\tilde\eps_{-k})}{\tilde\eps_k+\tilde\eps_{-k}}
\ee
These two functions are shown side-by-side in Fig...

A problem is that both over-estimate the value of the optical doping. We suspect that this is due to the mean-field decoupling.

\begin{figure}[h!]
\includegraphics[width=0.48\linewidth]{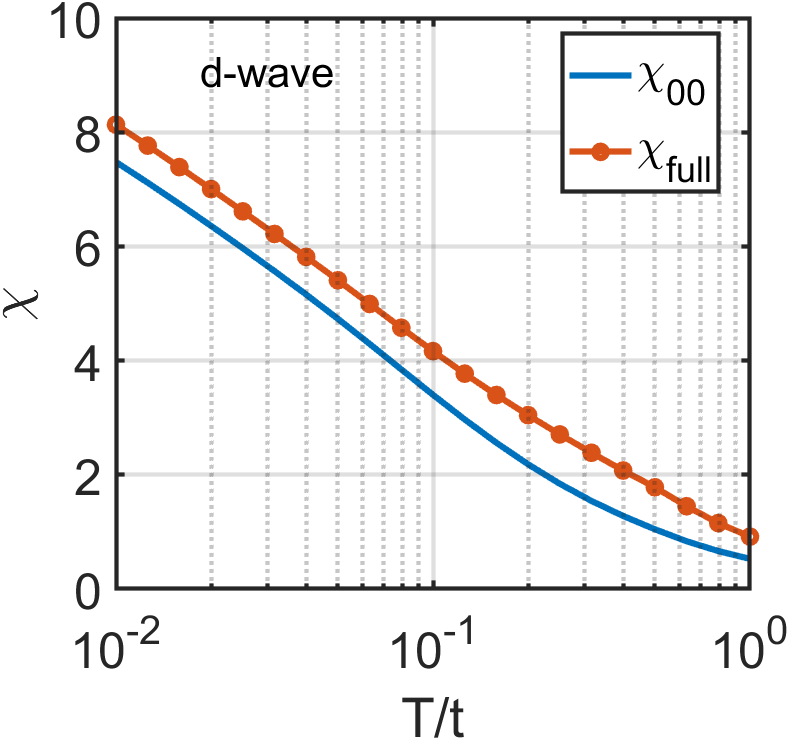}
\includegraphics[width=0.5\linewidth]{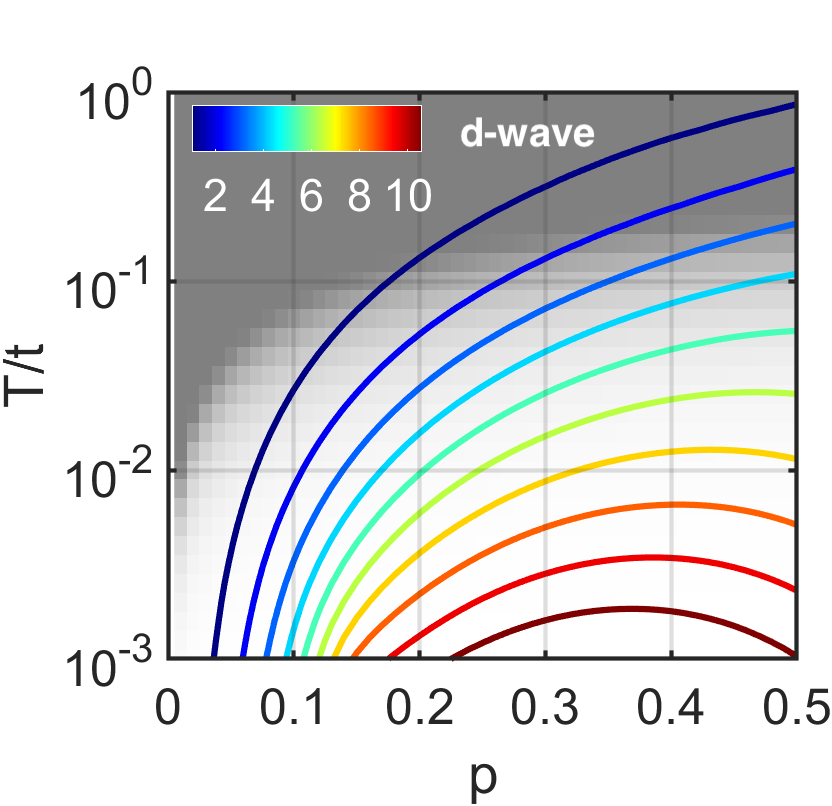}
\caption{\small (a) Pair susceptibility of single-band infinite-$U$ Hubbard model for $t'/t=-0.17$ and 0.37\% hole doping to d-wave pairing. $\chi_{\rm full}$ the full mean-field result vs. the qualitatively similar $\chi_{00}$ is the condensate contribution. (b) The (condensate contribution of) susceptibility to d-wave pairing for the single-band Hubbard model as a function of hole doping $p$ and temperature $T/t$. The gray background shows the condensation fraction of the boson (0\% gray to 100\% white). The value of inverse coupling constant $1/g=\chi$ will determine the transition temperature, indicated in color bar.}
\end{figure}

\begin{figure}[h!]
\includegraphics[width=0.46\linewidth]{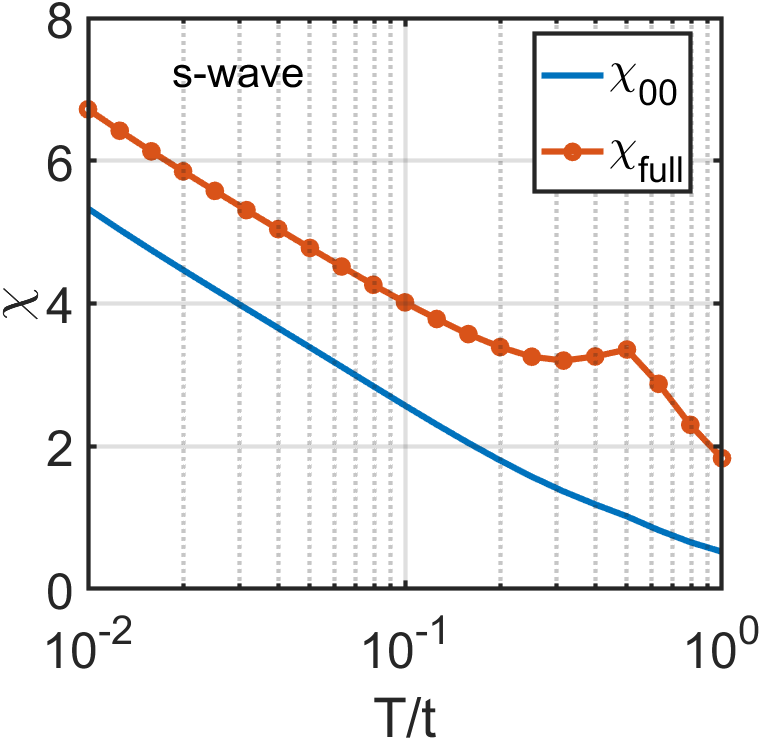}
\caption{\small 
A non-zero result for the latter is due to the failure of the mean-field decoupling. 
}
\end{figure}

\section{Slave-boson mean-field calculation}\label{sec:details}
\subsection*{Diagonalizing Bosonic Hamiltonian}
Since we have used mixed holon-doublon description of the orbitals, the bosonic Hamiltonian will contain bosonic pairing terms. For an $N$-orbital problem we form the $2N\times 2N$ Hamiltonian
\be
H_b=\sum_k\mat{b_k\\ \hline b_k^*}\dg{\cal H}_b(k)\mat{b_k\\ \hline b_k^*}.
\ee
Rotating to $p-x$ form we obtain
\be
{\cal R}= \frac{1}{\sqrt 2}\mat{i\bb 1 & \bb 1 \\ -i\bb1 & \bb1},\quad \mat{p\\\hline x}={\cal R}\mat{b\\\hline b^*}.
\ee
The Hamiltonian becomes
\be
H_b=\sum_k\mat{p_k\\ \hline x_k}^T{\cal R}\dg{\cal H}_b(k){\cal R}\mat{p_k\\ \hline x_k},
\ee
where ${\cal R}\dg{\cal H}_b{\cal R}$ is real, symmetric and positive definite  and can be diagonalized \cite{Son2021} using a symplectic transformation $S\in Sp(2n,\bb c)$.
\be
\mat{p\\\hline x}=S\mat{\breve p \\\hline \breve x}\quad \to\quad S^T_k{\cal R}\dg{\cal H}_b(k){\cal R}S_k=\matc{c|c}{D & 0 \\ \hline 0 & D}\nonumber
\ee
where $D$ is a diagonal matrix containing symplectic eigenvalues. Therefore, the original bosons are related via ${\cal U}_k={\cal R}S_k{\cal R}\dg$ to a new set of canonical boson
\be
\mat{b_k\\\hline b_k^*}={\cal U}_k\mat{\breve b_k \\\hline \breve b_k^*}
\ee
in terms of which the Hamiltonian is diagonal:
\be
H_b=\frac{1}{2}\sum_k\mat{\breve b_k \\\hline \breve b_k^*}\dg\matc{c|c}{D & 0 \\ \hline 0 & D}\mat{\breve b_k \\\hline \breve b_k^*}.
\ee
\subsection*{Details of the procedure}
After a decoupling, we find
\be
H_b=\sum_kB\dg_k
{\cal H}^b(k) B_k
, \quad
H_f=\sum_kF\dg_{k\sigma}{\cal H}^f(k)F_{k\sigma},
\ee
in terms of bosonic $B$ and fermionic $F$ operators
\be
F_\sigma=\mat{\tilde\sigma f\dg_{xz,\sigma} \\ \tilde\sigma f\dg_{yz,\sigma} \\ f_{xy,\sigma}}
,\qquad B=\mat{b_{xz} \\ b_{yz} \\ b\dg_{xy}}.
\ee
We assume that $f$ spinons inherit the symmetry of the electron orbitals, whereas $b$ bosons are invariant under crystal rotations. Therefore, ${\cal H}^f(k)$ has the same form as ${\cal H}^0(k)$ only with renormalized parameters. ${\cal H}^b(k)$ is similar, with the difference that all $i\sin k_\mu$ are replaced with $\cos k_\mu$.

\bibliography{slaveboson}

\begin{thebibliography}{77}%
\makeatletter
\providecommand \@ifxundefined [1]{%
 \@ifx{#1\undefined}
}%
\providecommand \@ifnum [1]{%
 \ifnum #1\expandafter \@firstoftwo
 \else \expandafter \@secondoftwo
 \fi
}%
\providecommand \@ifx [1]{%
 \ifx #1\expandafter \@firstoftwo
 \else \expandafter \@secondoftwo
 \fi
}%
\providecommand \natexlab [1]{#1}%
\providecommand \enquote  [1]{``#1''}%
\providecommand \bibnamefont  [1]{#1}%
\providecommand \bibfnamefont [1]{#1}%
\providecommand \citenamefont [1]{#1}%
\providecommand \href@noop [0]{\@secondoftwo}%
\providecommand \href [0]{\begingroup \@sanitize@url \@href}%
\providecommand \@href[1]{\@@startlink{#1}\@@href}%
\providecommand \@@href[1]{\endgroup#1\@@endlink}%
\providecommand \@sanitize@url [0]{\catcode `\\12\catcode `\$12\catcode
  `\&12\catcode `\#12\catcode `\^12\catcode `\_12\catcode `\%12\relax}%
\providecommand \@@startlink[1]{}%
\providecommand \@@endlink[0]{}%
\providecommand \url  [0]{\begingroup\@sanitize@url \@url }%
\providecommand \@url [1]{\endgroup\@href {#1}{\urlprefix }}%
\providecommand \urlprefix  [0]{URL }%
\providecommand \Eprint [0]{\href }%
\providecommand \doibase [0]{https://doi.org/}%
\providecommand \selectlanguage [0]{\@gobble}%
\providecommand \bibinfo  [0]{\@secondoftwo}%
\providecommand \bibfield  [0]{\@secondoftwo}%
\providecommand \translation [1]{[#1]}%
\providecommand \BibitemOpen [0]{}%
\providecommand \bibitemStop [0]{}%
\providecommand \bibitemNoStop [0]{.\EOS\space}%
\providecommand \EOS [0]{\spacefactor3000\relax}%
\providecommand \BibitemShut  [1]{\csname bibitem#1\endcsname}%
\let\auto@bib@innerbib\@empty
\bibitem [{\citenamefont {Hosono}\ \emph {et~al.}(2018)\citenamefont {Hosono},
  \citenamefont {Yamamoto}, \citenamefont {Hiramatsu},\ and\ \citenamefont
  {Ma}}]{Hosono2018}%
  \BibitemOpen
  \bibfield  {author} {\bibinfo {author} {\bibfnamefont {H.}~\bibnamefont
  {Hosono}}, \bibinfo {author} {\bibfnamefont {A.}~\bibnamefont {Yamamoto}},
  \bibinfo {author} {\bibfnamefont {H.}~\bibnamefont {Hiramatsu}},\ and\
  \bibinfo {author} {\bibfnamefont {Y.}~\bibnamefont {Ma}},\ }\bibfield
  {title} {\bibinfo {title} {Recent advances in iron-based superconductors
  toward applications},\ }\href
  {http://www.sciencedirect.com/science/article/pii/S1369702117306545}
  {\bibfield  {journal} {\bibinfo  {journal} {Materials Today}\ }\textbf
  {\bibinfo {volume} {21}},\ \bibinfo {pages} {278} (\bibinfo {year}
  {2018})}\BibitemShut {NoStop}%
\bibitem [{\citenamefont {Mazin}\ \emph {et~al.}(2008)\citenamefont {Mazin},
  \citenamefont {Singh}, \citenamefont {Johannes},\ and\ \citenamefont
  {Du}}]{Mazin2008}%
  \BibitemOpen
  \bibfield  {author} {\bibinfo {author} {\bibfnamefont {I.~I.}\ \bibnamefont
  {Mazin}}, \bibinfo {author} {\bibfnamefont {D.~J.}\ \bibnamefont {Singh}},
  \bibinfo {author} {\bibfnamefont {M.~D.}\ \bibnamefont {Johannes}},\ and\
  \bibinfo {author} {\bibfnamefont {M.~H.}\ \bibnamefont {Du}},\ }\bibfield
  {title} {\bibinfo {title} {{Unconventional Superconductivity with a Sign
  Reversal in the Order Parameter of
  ${\mathrm{LaFeAsO}}_{1-x}{\mathrm{F}}_{x}$}},\ }\href
  {https://journals.aps.org/prl/abstract/10.1103/PhysRevLett.101.057003}
  {\bibfield  {journal} {\bibinfo  {journal} {Phys. Rev. Lett.}\ }\textbf
  {\bibinfo {volume} {101}},\ \bibinfo {pages} {057003} (\bibinfo {year}
  {2008})}\BibitemShut {NoStop}%
\bibitem [{\citenamefont {Hirschfeld}\ \emph {et~al.}(2011)\citenamefont
  {Hirschfeld}, \citenamefont {Korshunov},\ and\ \citenamefont
  {Mazin}}]{Hirschfeld2011}%
  \BibitemOpen
  \bibfield  {author} {\bibinfo {author} {\bibfnamefont {P.~J.}\ \bibnamefont
  {Hirschfeld}}, \bibinfo {author} {\bibfnamefont {M.~M.}\ \bibnamefont
  {Korshunov}},\ and\ \bibinfo {author} {\bibfnamefont {I.~I.}\ \bibnamefont
  {Mazin}},\ }\bibfield  {title} {\bibinfo {title} {Gap symmetry and structure
  of fe-based superconductors},\ }\href
  {https://doi.org/10.1088/0034-4885/74/12/124508} {\bibfield  {journal}
  {\bibinfo  {journal} {Reports on Progress in Physics}\ }\textbf {\bibinfo
  {volume} {74}},\ \bibinfo {pages} {124508} (\bibinfo {year}
  {2011})}\BibitemShut {NoStop}%
\bibitem [{\citenamefont {Chubukov}(2012)}]{Chubukov2012}%
  \BibitemOpen
  \bibfield  {author} {\bibinfo {author} {\bibfnamefont {A.}~\bibnamefont
  {Chubukov}},\ }\bibfield  {title} {\bibinfo {title} {Pairing mechanism in
  fe-based superconductors},\ }\href
  {https://doi.org/10.1146/annurev-conmatphys-020911-125055} {\bibfield
  {journal} {\bibinfo  {journal} {Annual Review of Condensed Matter Physics}\
  }\textbf {\bibinfo {volume} {3}},\ \bibinfo {pages} {57} (\bibinfo {year}
  {2012})}\BibitemShut {NoStop}%
\bibitem [{\citenamefont {Chubukov}(2015)}]{Chubukov2015a}%
  \BibitemOpen
  \bibfield  {author} {\bibinfo {author} {\bibfnamefont {A.~V.}\ \bibnamefont
  {Chubukov}},\ }\bibfield  {title} {\bibinfo {title} {Itinerant electron
  scenario for {Fe}-based superconductors}\ }\href
  {https://doi.org/10.48550/arxiv.1507.03856} {10.48550/arxiv.1507.03856}
  (\bibinfo {year} {2015})\BibitemShut {NoStop}%
\bibitem [{\citenamefont {Si}\ and\ \citenamefont {Abrahams}(2008)}]{Si08}%
  \BibitemOpen
  \bibfield  {author} {\bibinfo {author} {\bibfnamefont {Q.}~\bibnamefont
  {Si}}\ and\ \bibinfo {author} {\bibfnamefont {E.}~\bibnamefont {Abrahams}},\
  }\bibfield  {title} {\bibinfo {title} {Strong correlations and magnetic
  frustration in the high ${T}_{c}$ iron pnictides},\ }\href
  {https://doi.org/10.1103/PhysRevLett.101.076401} {\bibfield  {journal}
  {\bibinfo  {journal} {Phys. Rev. Lett.}\ }\textbf {\bibinfo {volume} {101}},\
  \bibinfo {pages} {076401} (\bibinfo {year} {2008})}\BibitemShut {NoStop}%
\bibitem [{\citenamefont {Fernandes}\ and\ \citenamefont
  {Chubukov}(2016)}]{Fernandes2016}%
  \BibitemOpen
  \bibfield  {author} {\bibinfo {author} {\bibfnamefont {R.~M.}\ \bibnamefont
  {Fernandes}}\ and\ \bibinfo {author} {\bibfnamefont {A.~V.}\ \bibnamefont
  {Chubukov}},\ }\bibfield  {title} {\bibinfo {title} {Low-energy microscopic
  models for iron-based superconductors: a review},\ }\href@noop {} {\bibfield
  {journal} {\bibinfo  {journal} {Reports on Progress in Physics}\ }\textbf
  {\bibinfo {volume} {80}},\ \bibinfo {pages} {014503} (\bibinfo {year}
  {2016})}\BibitemShut {NoStop}%
\bibitem [{\citenamefont {Xu}\ \emph {et~al.}(2020)\citenamefont {Xu},
  \citenamefont {Dai}, \citenamefont {Li}, \citenamefont {Yin}, \citenamefont
  {Rong}, \citenamefont {Tian}, \citenamefont {Liu}, \citenamefont {Wang},
  \citenamefont {Xing}, \citenamefont {Wei}, \citenamefont {Kajimoto},
  \citenamefont {Ikeuchi}, \citenamefont {Abernathy}, \citenamefont {Wang},
  \citenamefont {Jin}, \citenamefont {Lu}, \citenamefont {Tan},\ and\
  \citenamefont {Dai}}]{Xu2020}%
  \BibitemOpen
  \bibfield  {author} {\bibinfo {author} {\bibfnamefont {Z.}~\bibnamefont
  {Xu}}, \bibinfo {author} {\bibfnamefont {G.}~\bibnamefont {Dai}}, \bibinfo
  {author} {\bibfnamefont {Y.}~\bibnamefont {Li}}, \bibinfo {author}
  {\bibfnamefont {Z.}~\bibnamefont {Yin}}, \bibinfo {author} {\bibfnamefont
  {Y.}~\bibnamefont {Rong}}, \bibinfo {author} {\bibfnamefont {L.}~\bibnamefont
  {Tian}}, \bibinfo {author} {\bibfnamefont {P.}~\bibnamefont {Liu}}, \bibinfo
  {author} {\bibfnamefont {H.}~\bibnamefont {Wang}}, \bibinfo {author}
  {\bibfnamefont {L.}~\bibnamefont {Xing}}, \bibinfo {author} {\bibfnamefont
  {Y.}~\bibnamefont {Wei}}, \bibinfo {author} {\bibfnamefont {R.}~\bibnamefont
  {Kajimoto}}, \bibinfo {author} {\bibfnamefont {K.}~\bibnamefont {Ikeuchi}},
  \bibinfo {author} {\bibfnamefont {D.~L.}\ \bibnamefont {Abernathy}}, \bibinfo
  {author} {\bibfnamefont {X.}~\bibnamefont {Wang}}, \bibinfo {author}
  {\bibfnamefont {C.}~\bibnamefont {Jin}}, \bibinfo {author} {\bibfnamefont
  {X.}~\bibnamefont {Lu}}, \bibinfo {author} {\bibfnamefont {G.}~\bibnamefont
  {Tan}},\ and\ \bibinfo {author} {\bibfnamefont {P.}~\bibnamefont {Dai}},\
  }\bibfield  {title} {\bibinfo {title} {Strong local moment antiferromagnetic
  spin fluctuations in v-doped {LiFeAs}},\ }\bibfield  {journal} {\bibinfo
  {journal} {npj Quantum Materials}\ }\textbf {\bibinfo {volume} {5}},\ \href
  {https://doi.org/10.1038/s41535-020-0212-x} {10.1038/s41535-020-0212-x}
  (\bibinfo {year} {2020})\BibitemShut {NoStop}%
\bibitem [{\citenamefont {Coleman}\ \emph {et~al.}(2020)\citenamefont
  {Coleman}, \citenamefont {Komijani},\ and\ \citenamefont {K\"onig}}]{tRVB}%
  \BibitemOpen
  \bibfield  {author} {\bibinfo {author} {\bibfnamefont {P.}~\bibnamefont
  {Coleman}}, \bibinfo {author} {\bibfnamefont {Y.}~\bibnamefont {Komijani}},\
  and\ \bibinfo {author} {\bibfnamefont {E.~J.}\ \bibnamefont {K\"onig}},\
  }\bibfield  {title} {\bibinfo {title} {Triplet resonating valence bond state
  and superconductivity in hund's metals},\ }\href
  {https://doi.org/10.1103/PhysRevLett.125.077001} {\bibfield  {journal}
  {\bibinfo  {journal} {Phys. Rev. Lett.}\ }\textbf {\bibinfo {volume} {125}},\
  \bibinfo {pages} {077001} (\bibinfo {year} {2020})}\BibitemShut {NoStop}%
\bibitem [{\citenamefont {Vafek}\ and\ \citenamefont
  {Chubukov}(2017)}]{VafekChubukov2017}%
  \BibitemOpen
  \bibfield  {author} {\bibinfo {author} {\bibfnamefont {O.}~\bibnamefont
  {Vafek}}\ and\ \bibinfo {author} {\bibfnamefont {A.~V.}\ \bibnamefont
  {Chubukov}},\ }\bibfield  {title} {\bibinfo {title} {Hund interaction,
  spin-orbit coupling, and the mechanism of superconductivity in strongly
  hole-doped iron pnictides},\ }\href
  {https://link.aps.org/doi/10.1103/PhysRevLett.118.087003} {\bibfield
  {journal} {\bibinfo  {journal} {Phys. Rev. Lett.}\ }\textbf {\bibinfo
  {volume} {118}},\ \bibinfo {pages} {087003} (\bibinfo {year}
  {2017})}\BibitemShut {NoStop}%
\bibitem [{\citenamefont {Lee}\ and\ \citenamefont {Wen}(2008)}]{LeeWen2008}%
  \BibitemOpen
  \bibfield  {author} {\bibinfo {author} {\bibfnamefont {P.~A.}\ \bibnamefont
  {Lee}}\ and\ \bibinfo {author} {\bibfnamefont {X.-G.}\ \bibnamefont {Wen}},\
  }\bibfield  {title} {\bibinfo {title} {Spin-triplet $p$-wave pairing in a
  three-orbital model for iron pnictide superconductors},\ }\href
  {https://doi.org/10.1103/PhysRevB.78.144517} {\bibfield  {journal} {\bibinfo
  {journal} {Phys. Rev. B}\ }\textbf {\bibinfo {volume} {78}},\ \bibinfo
  {pages} {144517} (\bibinfo {year} {2008})}\BibitemShut {NoStop}%
\bibitem [{\citenamefont {Puetter}\ and\ \citenamefont
  {Kee}(2012)}]{PuetterKee2012}%
  \BibitemOpen
  \bibfield  {author} {\bibinfo {author} {\bibfnamefont {C.~M.}\ \bibnamefont
  {Puetter}}\ and\ \bibinfo {author} {\bibfnamefont {H.-Y.}\ \bibnamefont
  {Kee}},\ }\bibfield  {title} {\bibinfo {title} {Identifying spin-triplet
  pairing in spin-orbit coupled multi-band superconductors},\ }\href
  {https://doi.org/10.1209/0295-5075/98/27010} {\bibfield  {journal} {\bibinfo
  {journal} {{EPL} (Europhysics Letters)}\ }\textbf {\bibinfo {volume} {98}},\
  \bibinfo {pages} {27010} (\bibinfo {year} {2012})}\BibitemShut {NoStop}%
\bibitem [{\citenamefont {Hu}(2013)}]{Hu13}%
  \BibitemOpen
  \bibfield  {author} {\bibinfo {author} {\bibfnamefont {J.}~\bibnamefont
  {Hu}},\ }\bibfield  {title} {\bibinfo {title} {Iron-based superconductors as
  odd-parity superconductors},\ }\href
  {https://doi.org/10.1103/PhysRevX.3.031004} {\bibfield  {journal} {\bibinfo
  {journal} {Phys. Rev. X}\ }\textbf {\bibinfo {volume} {3}},\ \bibinfo {pages}
  {031004} (\bibinfo {year} {2013})}\BibitemShut {NoStop}%
\bibitem [{\citenamefont {Hao}\ and\ \citenamefont
  {Hu}(2014{\natexlab{a}})}]{Hao14}%
  \BibitemOpen
  \bibfield  {author} {\bibinfo {author} {\bibfnamefont {N.}~\bibnamefont
  {Hao}}\ and\ \bibinfo {author} {\bibfnamefont {J.}~\bibnamefont {Hu}},\
  }\bibfield  {title} {\bibinfo {title} {Odd parity pairing and nodeless
  antiphase ${s}_\pm$ in iron-based superconductors},\ }\href
  {https://doi.org/10.1103/PhysRevB.89.045144} {\bibfield  {journal} {\bibinfo
  {journal} {Phys. Rev. B}\ }\textbf {\bibinfo {volume} {89}},\ \bibinfo
  {pages} {045144} (\bibinfo {year} {2014}{\natexlab{a}})}\BibitemShut
  {NoStop}%
\bibitem [{\citenamefont {Hao}\ and\ \citenamefont
  {Hu}(2014{\natexlab{b}})}]{Hao14b}%
  \BibitemOpen
  \bibfield  {author} {\bibinfo {author} {\bibfnamefont {N.}~\bibnamefont
  {Hao}}\ and\ \bibinfo {author} {\bibfnamefont {J.}~\bibnamefont {Hu}},\
  }\bibfield  {title} {\bibinfo {title} {Topological phases in the single-layer
  fese},\ }\href {https://doi.org/10.1103/PhysRevX.4.031053} {\bibfield
  {journal} {\bibinfo  {journal} {Phys. Rev. X}\ }\textbf {\bibinfo {volume}
  {4}},\ \bibinfo {pages} {031053} (\bibinfo {year}
  {2014}{\natexlab{b}})}\BibitemShut {NoStop}%
\bibitem [{\citenamefont {Georges}\ \emph {et~al.}(1996)\citenamefont
  {Georges}, \citenamefont {Kotliar}, \citenamefont {Krauth},\ and\
  \citenamefont {Rozenberg}}]{Georges96}%
  \BibitemOpen
  \bibfield  {author} {\bibinfo {author} {\bibfnamefont {A.}~\bibnamefont
  {Georges}}, \bibinfo {author} {\bibfnamefont {G.}~\bibnamefont {Kotliar}},
  \bibinfo {author} {\bibfnamefont {W.}~\bibnamefont {Krauth}},\ and\ \bibinfo
  {author} {\bibfnamefont {M.~J.}\ \bibnamefont {Rozenberg}},\ }\bibfield
  {title} {\bibinfo {title} {Dynamical mean-field theory of strongly correlated
  fermion systems and the limit of infinite dimensions},\ }\href
  {https://doi.org/10.1103/RevModPhys.68.13} {\bibfield  {journal} {\bibinfo
  {journal} {Rev. Mod. Phys.}\ }\textbf {\bibinfo {volume} {68}},\ \bibinfo
  {pages} {13} (\bibinfo {year} {1996})}\BibitemShut {NoStop}%
\bibitem [{\citenamefont {Haule}\ \emph {et~al.}(2008)\citenamefont {Haule},
  \citenamefont {Shim},\ and\ \citenamefont {Kotliar}}]{Haule08}%
  \BibitemOpen
  \bibfield  {author} {\bibinfo {author} {\bibfnamefont {K.}~\bibnamefont
  {Haule}}, \bibinfo {author} {\bibfnamefont {J.~H.}\ \bibnamefont {Shim}},\
  and\ \bibinfo {author} {\bibfnamefont {G.}~\bibnamefont {Kotliar}},\
  }\bibfield  {title} {\bibinfo {title} {Correlated electronic structure of
  ${\mathrm{lao}}_{1\ensuremath{-}x}{\mathrm{f}}_{x}\mathrm{FeAs}$},\ }\href
  {https://doi.org/10.1103/PhysRevLett.100.226402} {\bibfield  {journal}
  {\bibinfo  {journal} {Phys. Rev. Lett.}\ }\textbf {\bibinfo {volume} {100}},\
  \bibinfo {pages} {226402} (\bibinfo {year} {2008})}\BibitemShut {NoStop}%
\bibitem [{\citenamefont {Haule}\ and\ \citenamefont
  {Kotliar}(2009)}]{Haule09}%
  \BibitemOpen
  \bibfield  {author} {\bibinfo {author} {\bibfnamefont {K.}~\bibnamefont
  {Haule}}\ and\ \bibinfo {author} {\bibfnamefont {G.}~\bibnamefont
  {Kotliar}},\ }\bibfield  {title} {\bibinfo {title} {Coherence-incoherence
  crossover in the normal state of iron oxypnictides and importance of hund's
  rule coupling},\ }\href {http://stacks.iop.org/1367-2630/11/i=2/a=025021}
  {\bibfield  {journal} {\bibinfo  {journal} {New Journal of Physics}\ }\textbf
  {\bibinfo {volume} {11}},\ \bibinfo {pages} {025021} (\bibinfo {year}
  {2009})}\BibitemShut {NoStop}%
\bibitem [{\citenamefont {Yin}\ \emph {et~al.}(2011)\citenamefont {Yin},
  \citenamefont {Haule},\ and\ \citenamefont {Kotliar}}]{Yin11}%
  \BibitemOpen
  \bibfield  {author} {\bibinfo {author} {\bibfnamefont {Z.~P.}\ \bibnamefont
  {Yin}}, \bibinfo {author} {\bibfnamefont {K.}~\bibnamefont {Haule}},\ and\
  \bibinfo {author} {\bibfnamefont {G.}~\bibnamefont {Kotliar}},\ }\bibfield
  {title} {\bibinfo {title} {Kinetic frustration and the nature of the magnetic
  and paramagnetic states in iron pnictides and iron chalcogenides},\ }\href
  {http://dx.doi.org/10.1038/nmat3120} {\bibfield  {journal} {\bibinfo
  {journal} {Nature Materials}\ }\textbf {\bibinfo {volume} {10}},\ \bibinfo
  {pages} {932} (\bibinfo {year} {2011})}\BibitemShut {NoStop}%
\bibitem [{\citenamefont {Yin}\ \emph {et~al.}(2012)\citenamefont {Yin},
  \citenamefont {Haule},\ and\ \citenamefont {Kotliar}}]{Yin12}%
  \BibitemOpen
  \bibfield  {author} {\bibinfo {author} {\bibfnamefont {Z.~P.}\ \bibnamefont
  {Yin}}, \bibinfo {author} {\bibfnamefont {K.}~\bibnamefont {Haule}},\ and\
  \bibinfo {author} {\bibfnamefont {G.}~\bibnamefont {Kotliar}},\ }\bibfield
  {title} {\bibinfo {title} {Fractional power-law behavior and its origin in
  iron-chalcogenide and ruthenate superconductors: Insights from
  first-principles calculations},\ }\href
  {https://doi.org/10.1103/PhysRevB.86.195141} {\bibfield  {journal} {\bibinfo
  {journal} {Phys. Rev. B}\ }\textbf {\bibinfo {volume} {86}},\ \bibinfo
  {pages} {195141} (\bibinfo {year} {2012})}\BibitemShut {NoStop}%
\bibitem [{\citenamefont {Georges}\ \emph {et~al.}(2013)\citenamefont
  {Georges}, \citenamefont {Medici},\ and\ \citenamefont
  {Mravlje}}]{Georges13}%
  \BibitemOpen
  \bibfield  {author} {\bibinfo {author} {\bibfnamefont {A.}~\bibnamefont
  {Georges}}, \bibinfo {author} {\bibfnamefont {L.~d.}\ \bibnamefont
  {Medici}},\ and\ \bibinfo {author} {\bibfnamefont {J.}~\bibnamefont
  {Mravlje}},\ }\bibfield  {title} {\bibinfo {title} {Strong correlations from
  hund’s coupling},\ }\href
  {https://doi.org/10.1146/annurev-conmatphys-020911-125045} {\bibfield
  {journal} {\bibinfo  {journal} {Annu. Rev. Condens. Matter Phys.}\ }\textbf
  {\bibinfo {volume} {4}},\ \bibinfo {pages} {137} (\bibinfo {year}
  {2013})}\BibitemShut {NoStop}%
\bibitem [{\citenamefont {Deng}\ \emph {et~al.}(2019)\citenamefont {Deng},
  \citenamefont {Stadler}, \citenamefont {Haule}, \citenamefont {Weichselbaum},
  \citenamefont {von Delft},\ and\ \citenamefont {Kotliar}}]{Deng2019}%
  \BibitemOpen
  \bibfield  {author} {\bibinfo {author} {\bibfnamefont {X.}~\bibnamefont
  {Deng}}, \bibinfo {author} {\bibfnamefont {K.~M.}\ \bibnamefont {Stadler}},
  \bibinfo {author} {\bibfnamefont {K.}~\bibnamefont {Haule}}, \bibinfo
  {author} {\bibfnamefont {A.}~\bibnamefont {Weichselbaum}}, \bibinfo {author}
  {\bibfnamefont {J.}~\bibnamefont {von Delft}},\ and\ \bibinfo {author}
  {\bibfnamefont {G.}~\bibnamefont {Kotliar}},\ }\bibfield  {title} {\bibinfo
  {title} {Signatures of mottness and hundness in archetypal correlated
  metals},\ }\bibfield  {journal} {\bibinfo  {journal} {Nature Communications}\
  }\textbf {\bibinfo {volume} {10}},\ \href
  {https://doi.org/10.1038/s41467-019-10257-2} {10.1038/s41467-019-10257-2}
  (\bibinfo {year} {2019})\BibitemShut {NoStop}%
\bibitem [{\citenamefont {Schrieffer}(1967)}]{Schrieffer1967}%
  \BibitemOpen
  \bibfield  {author} {\bibinfo {author} {\bibfnamefont {J.~R.}\ \bibnamefont
  {Schrieffer}},\ }\bibfield  {title} {\bibinfo {title} {The kondo effect—the
  link between magnetic and nonmagnetic impurities in metals?},\ }\href
  {https://doi.org/10.1063/1.1709517} {\bibfield  {journal} {\bibinfo
  {journal} {Journal of Applied Physics}\ }\textbf {\bibinfo {volume} {38}},\
  \bibinfo {pages} {1143} (\bibinfo {year} {1967})}\BibitemShut {NoStop}%
\bibitem [{\citenamefont {Nevidomskyy}\ and\ \citenamefont
  {Coleman}(2009)}]{Nevidomskyy2009}%
  \BibitemOpen
  \bibfield  {author} {\bibinfo {author} {\bibfnamefont {A.~H.}\ \bibnamefont
  {Nevidomskyy}}\ and\ \bibinfo {author} {\bibfnamefont {P.}~\bibnamefont
  {Coleman}},\ }\bibfield  {title} {\bibinfo {title} {Kondo resonance narrowing
  in$d$-and $f$-electron systems},\ }\href
  {https://doi.org/10.1103/physrevlett.103.147205} {\bibfield  {journal}
  {\bibinfo  {journal} {Physical Review Letters}\ }\textbf {\bibinfo {volume}
  {103}},\ \bibinfo {pages} {147205} (\bibinfo {year} {2009})}\BibitemShut
  {NoStop}%
\bibitem [{\citenamefont {Stadler}\ \emph {et~al.}(2015)\citenamefont
  {Stadler}, \citenamefont {Yin}, \citenamefont {von Delft}, \citenamefont
  {Kotliar},\ and\ \citenamefont {Weichselbaum}}]{Stadler15}%
  \BibitemOpen
  \bibfield  {author} {\bibinfo {author} {\bibfnamefont {K.~M.}\ \bibnamefont
  {Stadler}}, \bibinfo {author} {\bibfnamefont {Z.~P.}\ \bibnamefont {Yin}},
  \bibinfo {author} {\bibfnamefont {J.}~\bibnamefont {von Delft}}, \bibinfo
  {author} {\bibfnamefont {G.}~\bibnamefont {Kotliar}},\ and\ \bibinfo {author}
  {\bibfnamefont {A.}~\bibnamefont {Weichselbaum}},\ }\bibfield  {title}
  {\bibinfo {title} {Dynamical mean-field theory plus numerical
  renormalization-group study of spin-orbital separation in a three-band hund
  metal},\ }\href {https://doi.org/10.1103/PhysRevLett.115.136401} {\bibfield
  {journal} {\bibinfo  {journal} {Phys. Rev. Lett.}\ }\textbf {\bibinfo
  {volume} {115}},\ \bibinfo {pages} {136401} (\bibinfo {year}
  {2015})}\BibitemShut {NoStop}%
\bibitem [{\citenamefont {Drouin-Touchette}\ \emph {et~al.}(2021)\citenamefont
  {Drouin-Touchette}, \citenamefont {K\"onig}, \citenamefont {Komijani},\ and\
  \citenamefont {Coleman}}]{DrouinTouchette2021}%
  \BibitemOpen
  \bibfield  {author} {\bibinfo {author} {\bibfnamefont {V.}~\bibnamefont
  {Drouin-Touchette}}, \bibinfo {author} {\bibfnamefont {E.~J.}\ \bibnamefont
  {K\"onig}}, \bibinfo {author} {\bibfnamefont {Y.}~\bibnamefont {Komijani}},\
  and\ \bibinfo {author} {\bibfnamefont {P.}~\bibnamefont {Coleman}},\
  }\bibfield  {title} {\bibinfo {title} {Emergent moments in a hund's
  impurity},\ }\href {https://doi.org/10.1103/physrevb.103.205147} {\bibfield
  {journal} {\bibinfo  {journal} {Physical Review B}\ }\textbf {\bibinfo
  {volume} {103}},\ \bibinfo {pages} {205147} (\bibinfo {year}
  {2021})}\BibitemShut {NoStop}%
\bibitem [{\citenamefont {Werner}\ \emph {et~al.}(2008)\citenamefont {Werner},
  \citenamefont {Gull}, \citenamefont {Troyer},\ and\ \citenamefont
  {Millis}}]{Werner2008}%
  \BibitemOpen
  \bibfield  {author} {\bibinfo {author} {\bibfnamefont {P.}~\bibnamefont
  {Werner}}, \bibinfo {author} {\bibfnamefont {E.}~\bibnamefont {Gull}},
  \bibinfo {author} {\bibfnamefont {M.}~\bibnamefont {Troyer}},\ and\ \bibinfo
  {author} {\bibfnamefont {A.~J.}\ \bibnamefont {Millis}},\ }\bibfield  {title}
  {\bibinfo {title} {Spin freezing transition and non-fermi-liquid self-energy
  in a three-orbital model},\ }\href
  {https://doi.org/10.1103/physrevlett.101.166405} {\bibfield  {journal}
  {\bibinfo  {journal} {Physical Review Letters}\ }\textbf {\bibinfo {volume}
  {101}},\ \bibinfo {pages} {166405} (\bibinfo {year} {2008})}\BibitemShut
  {NoStop}%
\bibitem [{\citenamefont {Hoshino}\ and\ \citenamefont
  {Werner}(2015)}]{Hoshino15}%
  \BibitemOpen
  \bibfield  {author} {\bibinfo {author} {\bibfnamefont {S.}~\bibnamefont
  {Hoshino}}\ and\ \bibinfo {author} {\bibfnamefont {P.}~\bibnamefont
  {Werner}},\ }\bibfield  {title} {\bibinfo {title} {Superconductivity from
  emerging magnetic moments},\ }\href
  {https://doi.org/10.1103/PhysRevLett.115.247001} {\bibfield  {journal}
  {\bibinfo  {journal} {Phys. Rev. Lett.}\ }\textbf {\bibinfo {volume} {115}},\
  \bibinfo {pages} {247001} (\bibinfo {year} {2015})}\BibitemShut {NoStop}%
\bibitem [{\citenamefont {Miao}\ \emph {et~al.}(2018)\citenamefont {Miao},
  \citenamefont {Brito}, \citenamefont {Yin} \emph {et~al.}}]{MiaoDing2018}%
  \BibitemOpen
  \bibfield  {author} {\bibinfo {author} {\bibfnamefont {H.}~\bibnamefont
  {Miao}}, \bibinfo {author} {\bibfnamefont {W.}~\bibnamefont {Brito}},
  \bibinfo {author} {\bibfnamefont {Z.}~\bibnamefont {Yin}}, \emph {et~al.},\
  }\bibfield  {title} {\bibinfo {title} {Universal $2 \delta_{\rm max}/t_c$
  scaling decoupled from the electronic coherence in iron-based
  superconductors},\ }\href@noop {} {\bibfield  {journal} {\bibinfo  {journal}
  {Physical Review B}\ }\textbf {\bibinfo {volume} {98}},\ \bibinfo {pages}
  {020502(R)} (\bibinfo {year} {2018})}\BibitemShut {NoStop}%
\bibitem [{\citenamefont {Lee}\ \emph {et~al.}(2018)\citenamefont {Lee},
  \citenamefont {Chubukov}, \citenamefont {Miao},\ and\ \citenamefont
  {Kotliar}}]{Lee18}%
  \BibitemOpen
  \bibfield  {author} {\bibinfo {author} {\bibfnamefont {T.-H.}\ \bibnamefont
  {Lee}}, \bibinfo {author} {\bibfnamefont {A.}~\bibnamefont {Chubukov}},
  \bibinfo {author} {\bibfnamefont {H.}~\bibnamefont {Miao}},\ and\ \bibinfo
  {author} {\bibfnamefont {G.}~\bibnamefont {Kotliar}},\ }\bibfield  {title}
  {\bibinfo {title} {Pairing mechanism in {H}und's metal superconductors and
  the universality of the superconducting gap to critical temperature ratio},\
  }\href {https://doi.org/10.1103/PhysRevLett.121.187003} {\bibfield  {journal}
  {\bibinfo  {journal} {Phys. Rev. Lett.}\ }\textbf {\bibinfo {volume} {121}},\
  \bibinfo {pages} {187003} (\bibinfo {year} {2018})}\BibitemShut {NoStop}%
\bibitem [{\citenamefont {Classen}\ and\ \citenamefont
  {Chubukov}(2021)}]{Classen2021}%
  \BibitemOpen
  \bibfield  {author} {\bibinfo {author} {\bibfnamefont {L.}~\bibnamefont
  {Classen}}\ and\ \bibinfo {author} {\bibfnamefont {A.}~\bibnamefont
  {Chubukov}},\ }\bibfield  {title} {\bibinfo {title} {Superconductivity of
  incoherent electrons in the yukawa sachdev-ye-kitaev model},\ }\href
  {https://doi.org/10.1103/PhysRevB.104.125120} {\bibfield  {journal} {\bibinfo
   {journal} {Phys. Rev. B}\ }\textbf {\bibinfo {volume} {104}},\ \bibinfo
  {pages} {125120} (\bibinfo {year} {2021})}\BibitemShut {NoStop}%
\bibitem [{\citenamefont {Inkof}\ \emph {et~al.}(2022)\citenamefont {Inkof},
  \citenamefont {Schlam},\ and\ \citenamefont
  {Schmalian}}]{InkofSchmalian2022}%
  \BibitemOpen
  \bibfield  {author} {\bibinfo {author} {\bibfnamefont {G.-A.}\ \bibnamefont
  {Inkof}}, \bibinfo {author} {\bibfnamefont {K.}~\bibnamefont {Schlam}},\ and\
  \bibinfo {author} {\bibfnamefont {J.}~\bibnamefont {Schmalian}},\ }\href@noop
  {} {\bibfield  {journal} {\bibinfo  {journal} {npj quantum materials}\
  }\textbf {\bibinfo {volume} {7}},\ \bibinfo {pages} {56} (\bibinfo {year}
  {2022})}\BibitemShut {NoStop}%
\bibitem [{Wal(2020)}]{WaltervonDelft2020}%
  \BibitemOpen
  \bibfield  {title} {\bibinfo {title} {Uncovering non-fermi-liquid behavior in
  hund metals: Conformal field theory analysis of an $su(2)\times su(3)$
  spin-orbital kondo model},\ }\href@noop {} {\bibfield  {journal} {\bibinfo
  {journal} {Physical Review X}\ }\textbf {\bibinfo {volume} {10}},\ \bibinfo
  {pages} {031052} (\bibinfo {year} {2020})}\BibitemShut {NoStop}%
\bibitem [{\citenamefont {Drouin-Touchette}\ \emph {et~al.}(2022)\citenamefont
  {Drouin-Touchette}, \citenamefont {König}, \citenamefont {Komijani},\ and\
  \citenamefont {Coleman}}]{DrouinTouchette2022}%
  \BibitemOpen
  \bibfield  {author} {\bibinfo {author} {\bibfnamefont {V.}~\bibnamefont
  {Drouin-Touchette}}, \bibinfo {author} {\bibfnamefont {E.~J.}\ \bibnamefont
  {König}}, \bibinfo {author} {\bibfnamefont {Y.}~\bibnamefont {Komijani}},\
  and\ \bibinfo {author} {\bibfnamefont {P.}~\bibnamefont {Coleman}},\ }\href
  {https://doi.org/10.48550/ARXIV.2203.05172} {\bibinfo {title} {Interplay of
  charge and spin fluctuations in a hund's coupled impurity}} (\bibinfo {year}
  {2022})\BibitemShut {NoStop}%
\bibitem [{\citenamefont {de' Medici}\ \emph {et~al.}(2014)\citenamefont {de'
  Medici}, \citenamefont {Giovannetti},\ and\ \citenamefont
  {Capone}}]{Medici14}%
  \BibitemOpen
  \bibfield  {author} {\bibinfo {author} {\bibfnamefont {L.}~\bibnamefont {de'
  Medici}}, \bibinfo {author} {\bibfnamefont {G.}~\bibnamefont {Giovannetti}},\
  and\ \bibinfo {author} {\bibfnamefont {M.}~\bibnamefont {Capone}},\
  }\bibfield  {title} {\bibinfo {title} {Selective mott physics as a key to
  iron superconductors},\ }\href
  {https://doi.org/10.1103/PhysRevLett.112.177001} {\bibfield  {journal}
  {\bibinfo  {journal} {Phys. Rev. Lett.}\ }\textbf {\bibinfo {volume} {112}},\
  \bibinfo {pages} {177001} (\bibinfo {year} {2014})}\BibitemShut {NoStop}%
\bibitem [{\citenamefont {Sprau}\ \emph {et~al.}(2017)\citenamefont {Sprau},
  \citenamefont {Kostin}, \citenamefont {Kreisel}, \citenamefont {B{\"o}hmer},
  \citenamefont {Taufour}, \citenamefont {Canfield}, \citenamefont {Mukherjee},
  \citenamefont {Hirschfeld}, \citenamefont {Andersen},\ and\ \citenamefont
  {Davis}}]{SprauDavis2017}%
  \BibitemOpen
  \bibfield  {author} {\bibinfo {author} {\bibfnamefont {P.~O.}\ \bibnamefont
  {Sprau}}, \bibinfo {author} {\bibfnamefont {A.}~\bibnamefont {Kostin}},
  \bibinfo {author} {\bibfnamefont {A.}~\bibnamefont {Kreisel}}, \bibinfo
  {author} {\bibfnamefont {A.~E.}\ \bibnamefont {B{\"o}hmer}}, \bibinfo
  {author} {\bibfnamefont {V.}~\bibnamefont {Taufour}}, \bibinfo {author}
  {\bibfnamefont {P.~C.}\ \bibnamefont {Canfield}}, \bibinfo {author}
  {\bibfnamefont {S.}~\bibnamefont {Mukherjee}}, \bibinfo {author}
  {\bibfnamefont {P.~J.}\ \bibnamefont {Hirschfeld}}, \bibinfo {author}
  {\bibfnamefont {B.~M.}\ \bibnamefont {Andersen}},\ and\ \bibinfo {author}
  {\bibfnamefont {J.~C.~S.}\ \bibnamefont {Davis}},\ }\bibfield  {title}
  {\bibinfo {title} {Discovery of orbital-selective {C}ooper pairing in
  {F}e{S}e},\ }\href {https://science.sciencemag.org/content/357/6346/75}
  {\bibfield  {journal} {\bibinfo  {journal} {Science}\ }\textbf {\bibinfo
  {volume} {357}},\ \bibinfo {pages} {75} (\bibinfo {year} {2017})}\BibitemShut
  {NoStop}%
\bibitem [{\citenamefont {Klug}\ \emph {et~al.}(2018)\citenamefont {Klug},
  \citenamefont {Kang}, \citenamefont {Fernandes},\ and\ \citenamefont
  {Schmalian}}]{Klug2018}%
  \BibitemOpen
  \bibfield  {author} {\bibinfo {author} {\bibfnamefont {M.}~\bibnamefont
  {Klug}}, \bibinfo {author} {\bibfnamefont {J.}~\bibnamefont {Kang}}, \bibinfo
  {author} {\bibfnamefont {R.~M.}\ \bibnamefont {Fernandes}},\ and\ \bibinfo
  {author} {\bibfnamefont {J.}~\bibnamefont {Schmalian}},\ }\bibfield  {title}
  {\bibinfo {title} {Orbital loop currents in iron-based superconductors},\
  }\href {https://doi.org/10.1103/physrevb.97.155130} {\bibfield  {journal}
  {\bibinfo  {journal} {Physical Review B}\ }\textbf {\bibinfo {volume} {97}},\
  \bibinfo {pages} {155130} (\bibinfo {year} {2018})}\BibitemShut {NoStop}%
\bibitem [{\citenamefont {Yue}\ and\ \citenamefont {Werner}(2021)}]{Yue2021}%
  \BibitemOpen
  \bibfield  {author} {\bibinfo {author} {\bibfnamefont {C.}~\bibnamefont
  {Yue}}\ and\ \bibinfo {author} {\bibfnamefont {P.}~\bibnamefont {Werner}},\
  }\bibfield  {title} {\bibinfo {title} {Pairing enhanced by local orbital
  fluctuations in a model for monolayer {FeSe}},\ }\href
  {https://doi.org/10.1103/physrevb.104.184507} {\bibfield  {journal} {\bibinfo
   {journal} {Physical Review B}\ }\textbf {\bibinfo {volume} {104}},\ \bibinfo
  {pages} {184507} (\bibinfo {year} {2021})}\BibitemShut {NoStop}%
\bibitem [{\citenamefont {Yi}\ \emph {et~al.}(2013)\citenamefont {Yi},
  \citenamefont {Lu}, \citenamefont {Yu}, \citenamefont {Riggs}, \citenamefont
  {Chu}, \citenamefont {Lv}, \citenamefont {Liu}, \citenamefont {Lu},
  \citenamefont {Cui}, \citenamefont {Hashimoto}, \citenamefont {Mo},
  \citenamefont {Hussain}, \citenamefont {Chu}, \citenamefont {Fisher},
  \citenamefont {Si},\ and\ \citenamefont {Shen}}]{Yi13}%
  \BibitemOpen
  \bibfield  {author} {\bibinfo {author} {\bibfnamefont {M.}~\bibnamefont
  {Yi}}, \bibinfo {author} {\bibfnamefont {D.~H.}\ \bibnamefont {Lu}}, \bibinfo
  {author} {\bibfnamefont {R.}~\bibnamefont {Yu}}, \bibinfo {author}
  {\bibfnamefont {S.~C.}\ \bibnamefont {Riggs}}, \bibinfo {author}
  {\bibfnamefont {J.-H.}\ \bibnamefont {Chu}}, \bibinfo {author} {\bibfnamefont
  {B.}~\bibnamefont {Lv}}, \bibinfo {author} {\bibfnamefont {Z.~K.}\
  \bibnamefont {Liu}}, \bibinfo {author} {\bibfnamefont {M.}~\bibnamefont
  {Lu}}, \bibinfo {author} {\bibfnamefont {Y.-T.}\ \bibnamefont {Cui}},
  \bibinfo {author} {\bibfnamefont {M.}~\bibnamefont {Hashimoto}}, \bibinfo
  {author} {\bibfnamefont {S.-K.}\ \bibnamefont {Mo}}, \bibinfo {author}
  {\bibfnamefont {Z.}~\bibnamefont {Hussain}}, \bibinfo {author} {\bibfnamefont
  {C.~W.}\ \bibnamefont {Chu}}, \bibinfo {author} {\bibfnamefont {I.~R.}\
  \bibnamefont {Fisher}}, \bibinfo {author} {\bibfnamefont {Q.}~\bibnamefont
  {Si}},\ and\ \bibinfo {author} {\bibfnamefont {Z.-X.}\ \bibnamefont {Shen}},\
  }\bibfield  {title} {\bibinfo {title} {Observation of temperature-induced
  crossover to an orbital-selective mott phase in
  ${\mathrm{a}}_{x}{\mathrm{fe}}_{2\mathrm{\text{\ensuremath{-}}}y}{\mathrm{se}}_{2}$
  ($a\mathbf{=}\mathrm{K}$, rb) superconductors},\ }\href
  {https://doi.org/10.1103/PhysRevLett.110.067003} {\bibfield  {journal}
  {\bibinfo  {journal} {Phys. Rev. Lett.}\ }\textbf {\bibinfo {volume} {110}},\
  \bibinfo {pages} {067003} (\bibinfo {year} {2013})}\BibitemShut {NoStop}%
\bibitem [{\citenamefont {Yi}\ \emph {et~al.}(2017)\citenamefont {Yi},
  \citenamefont {Zhang}, \citenamefont {Shen},\ and\ \citenamefont
  {Lu}}]{Yi17}%
  \BibitemOpen
  \bibfield  {author} {\bibinfo {author} {\bibfnamefont {M.}~\bibnamefont
  {Yi}}, \bibinfo {author} {\bibfnamefont {Y.}~\bibnamefont {Zhang}}, \bibinfo
  {author} {\bibfnamefont {Z.-X.}\ \bibnamefont {Shen}},\ and\ \bibinfo
  {author} {\bibfnamefont {D.}~\bibnamefont {Lu}},\ }\bibfield  {title}
  {\bibinfo {title} {Role of the orbital degree of freedom in iron-based
  superconductors},\ }\bibfield  {journal} {\bibinfo  {journal} {npj Quantum
  Materials}\ }\textbf {\bibinfo {volume} {2}},\ \href
  {https://doi.org/10.1038/s41535-017-0059-y} {10.1038/s41535-017-0059-y}
  (\bibinfo {year} {2017})\BibitemShut {NoStop}%
\bibitem [{\citenamefont {Bogolyubov}\ \emph {et~al.}(1958)\citenamefont
  {Bogolyubov}, \citenamefont {Tolmachev},\ and\ \citenamefont
  {Shirkov}}]{BogolyubovShirkov1959}%
  \BibitemOpen
  \bibfield  {author} {\bibinfo {author} {\bibfnamefont {N.}~\bibnamefont
  {Bogolyubov}}, \bibinfo {author} {\bibfnamefont {V.}~\bibnamefont
  {Tolmachev}},\ and\ \bibinfo {author} {\bibfnamefont {D.}~\bibnamefont
  {Shirkov}},\ }\href@noop {} {\emph {\bibinfo {title} {Noviy metod v teorii
  sverkhprovodimosti}}}\ (\bibinfo  {publisher} {Izdatel'stvo akademii nauk
  SSSR, Moscow},\ \bibinfo {year} {1958})\ \bibinfo {note} {[Engl. transl. "A
  new method in the theory of superconductivity", Consultants Bureau, New York,
  1959.]}\BibitemShut {NoStop}%
\bibitem [{\citenamefont {McMillan}(1968)}]{McMillan1968}%
  \BibitemOpen
  \bibfield  {author} {\bibinfo {author} {\bibfnamefont {W.~L.}\ \bibnamefont
  {McMillan}},\ }\bibfield  {title} {\bibinfo {title} {Transition temperature
  of strong-coupled superconductors},\ }\href@noop {} {\bibfield  {journal}
  {\bibinfo  {journal} {Phys. Rev.}\ }\textbf {\bibinfo {volume} {167}},\
  \bibinfo {pages} {331} (\bibinfo {year} {1968})}\BibitemShut {NoStop}%
\bibitem [{\citenamefont {Anderson}\ and\ \citenamefont
  {Morel}(1961)}]{AndersonMorel}%
  \BibitemOpen
  \bibfield  {author} {\bibinfo {author} {\bibfnamefont {P.~W.}\ \bibnamefont
  {Anderson}}\ and\ \bibinfo {author} {\bibfnamefont {P.}~\bibnamefont
  {Morel}},\ }\bibfield  {title} {\bibinfo {title} {Generalized
  bardeen-cooper-schrieffer states and the proposed low-temperature phase of
  liquid ${\mathrm{he}}^{3}$},\ }\href
  {https://doi.org/10.1103/PhysRev.123.1911} {\bibfield  {journal} {\bibinfo
  {journal} {Phys. Rev.}\ }\textbf {\bibinfo {volume} {123}},\ \bibinfo {pages}
  {1911} (\bibinfo {year} {1961})}\BibitemShut {NoStop}%
\bibitem [{\citenamefont {Coleman}(2015)}]{Coleman2015}%
  \BibitemOpen
  \bibfield  {author} {\bibinfo {author} {\bibfnamefont {P.}~\bibnamefont
  {Coleman}},\ }\href {https://doi.org/10.1017/CBO9781139020916} {\emph
  {\bibinfo {title} {Introduction to Many-Body Physics}}}\ (\bibinfo
  {publisher} {Cambridge University Press},\ \bibinfo {year}
  {2015})\BibitemShut {NoStop}%
\bibitem [{\citenamefont {K\"onig}\ and\ \citenamefont
  {Coleman}(2019)}]{KoenigColeman2019}%
  \BibitemOpen
  \bibfield  {author} {\bibinfo {author} {\bibfnamefont {E.~J.}\ \bibnamefont
  {K\"onig}}\ and\ \bibinfo {author} {\bibfnamefont {P.}~\bibnamefont
  {Coleman}},\ }\bibfield  {title} {\bibinfo {title} {Coulomb problem in
  iron-based superconductors},\ }\href
  {https://link.aps.org/doi/10.1103/PhysRevB.99.144522} {\bibfield  {journal}
  {\bibinfo  {journal} {Phys. Rev. B}\ }\textbf {\bibinfo {volume} {99}},\
  \bibinfo {pages} {144522} (\bibinfo {year} {2019})}\BibitemShut {NoStop}%
\bibitem [{\citenamefont {Anderson}(1987)}]{Anderson87}%
  \BibitemOpen
  \bibfield  {author} {\bibinfo {author} {\bibfnamefont {P.~W.}\ \bibnamefont
  {Anderson}},\ }\href {https://doi.org/10.1126/science.235.4793.1196}
  {\bibfield  {journal} {\bibinfo  {journal} {Science}\ }\textbf {\bibinfo
  {volume} {235}},\ \bibinfo {pages} {1196} (\bibinfo {year}
  {1987})}\BibitemShut {NoStop}%
\bibitem [{\citenamefont {Anderson}\ \emph {et~al.}(1987)\citenamefont
  {Anderson}, \citenamefont {Baskaran}, \citenamefont {Zou},\ and\
  \citenamefont {Hsu}}]{Anderson1987}%
  \BibitemOpen
  \bibfield  {author} {\bibinfo {author} {\bibfnamefont {P.~W.}\ \bibnamefont
  {Anderson}}, \bibinfo {author} {\bibfnamefont {G.}~\bibnamefont {Baskaran}},
  \bibinfo {author} {\bibfnamefont {Z.}~\bibnamefont {Zou}},\ and\ \bibinfo
  {author} {\bibfnamefont {T.}~\bibnamefont {Hsu}},\ }\bibfield  {title}
  {\bibinfo {title} {Resonating-valence-bond theory of phase transitions and
  superconductivity in la$_2$cuo$_4$-based compounds},\ }\href
  {https://doi.org/10.1103/physrevlett.58.2790} {\bibfield  {journal} {\bibinfo
   {journal} {Physical Review Letters}\ }\textbf {\bibinfo {volume} {58}},\
  \bibinfo {pages} {2790} (\bibinfo {year} {1987})}\BibitemShut {NoStop}%
\bibitem [{\citenamefont {Baskaran}\ \emph {et~al.}(1987)\citenamefont
  {Baskaran}, \citenamefont {Zou},\ and\ \citenamefont
  {Anderson}}]{Baskaran87}%
  \BibitemOpen
  \bibfield  {author} {\bibinfo {author} {\bibfnamefont {G.}~\bibnamefont
  {Baskaran}}, \bibinfo {author} {\bibfnamefont {Z.}~\bibnamefont {Zou}},\ and\
  \bibinfo {author} {\bibfnamefont {P.}~\bibnamefont {Anderson}},\ }\bibfield
  {title} {\bibinfo {title} {The resonating valence bond state and high-tc
  superconductivity — a mean field theory},\ }\href
  {https://doi.org/https://doi.org/10.1016/0038-1098(87)90642-9} {\bibfield
  {journal} {\bibinfo  {journal} {Solid State Communications}\ }\textbf
  {\bibinfo {volume} {63}},\ \bibinfo {pages} {973 } (\bibinfo {year}
  {1987})}\BibitemShut {NoStop}%
\bibitem [{\citenamefont {Lee}\ \emph {et~al.}(2006)\citenamefont {Lee},
  \citenamefont {Nagaosa},\ and\ \citenamefont {Wen}}]{Lee06}%
  \BibitemOpen
  \bibfield  {author} {\bibinfo {author} {\bibfnamefont {P.~A.}\ \bibnamefont
  {Lee}}, \bibinfo {author} {\bibfnamefont {N.}~\bibnamefont {Nagaosa}},\ and\
  \bibinfo {author} {\bibfnamefont {X.-G.}\ \bibnamefont {Wen}},\ }\bibfield
  {title} {\bibinfo {title} {Doping a {M}ott insulator: Physics of
  high-temperature superconductivity},\ }\href
  {https://link.aps.org/doi/10.1103/RevModPhys.78.17} {\bibfield  {journal}
  {\bibinfo  {journal} {Rev. Mod. Phys.}\ }\textbf {\bibinfo {volume} {78}},\
  \bibinfo {pages} {17} (\bibinfo {year} {2006})}\BibitemShut {NoStop}%
\bibitem [{\citenamefont {Anderson}(1984{\natexlab{a}})}]{Anderson84a}%
  \BibitemOpen
  \bibfield  {author} {\bibinfo {author} {\bibfnamefont {P.~W.}\ \bibnamefont
  {Anderson}},\ }\bibfield  {title} {\bibinfo {title} {Heavy-electron
  superconductors, spin fluctuations, and triplet pairing},\ }\href
  {https://doi.org/10.1103/PhysRevB.30.1549} {\bibfield  {journal} {\bibinfo
  {journal} {Phys. Rev. B}\ }\textbf {\bibinfo {volume} {30}},\ \bibinfo
  {pages} {1549} (\bibinfo {year} {1984}{\natexlab{a}})}\BibitemShut {NoStop}%
\bibitem [{\citenamefont {Anderson}(1984{\natexlab{b}})}]{Anderson84b}%
  \BibitemOpen
  \bibfield  {author} {\bibinfo {author} {\bibfnamefont {P.~W.}\ \bibnamefont
  {Anderson}},\ }\bibfield  {title} {\bibinfo {title} {Structure of "triplet"
  superconducting energy gaps},\ }\href
  {https://doi.org/10.1103/PhysRevB.30.4000} {\bibfield  {journal} {\bibinfo
  {journal} {Phys. Rev. B}\ }\textbf {\bibinfo {volume} {30}},\ \bibinfo
  {pages} {4000} (\bibinfo {year} {1984}{\natexlab{b}})}\BibitemShut {NoStop}%
\bibitem [{\citenamefont {Anderson}(1985)}]{Anderson85}%
  \BibitemOpen
  \bibfield  {author} {\bibinfo {author} {\bibfnamefont {P.~W.}\ \bibnamefont
  {Anderson}},\ }\bibfield  {title} {\bibinfo {title} {Further consequences of
  symmetry in heavy-electron superconductors},\ }\href
  {https://doi.org/10.1103/PhysRevB.32.499} {\bibfield  {journal} {\bibinfo
  {journal} {Phys. Rev. B}\ }\textbf {\bibinfo {volume} {32}},\ \bibinfo
  {pages} {499} (\bibinfo {year} {1985})}\BibitemShut {NoStop}%
\bibitem [{\citenamefont {Shen}\ \emph {et~al.}(2020)\citenamefont {Shen},
  \citenamefont {Zhang}, \citenamefont {Komijani}, \citenamefont {Nicklas},
  \citenamefont {Borth}, \citenamefont {Wang}, \citenamefont {Chen},
  \citenamefont {Nie}, \citenamefont {Li}, \citenamefont {Lu} \emph
  {et~al.}}]{Cerge}%
  \BibitemOpen
  \bibfield  {author} {\bibinfo {author} {\bibfnamefont {B.}~\bibnamefont
  {Shen}}, \bibinfo {author} {\bibfnamefont {Y.}~\bibnamefont {Zhang}},
  \bibinfo {author} {\bibfnamefont {Y.}~\bibnamefont {Komijani}}, \bibinfo
  {author} {\bibfnamefont {M.}~\bibnamefont {Nicklas}}, \bibinfo {author}
  {\bibfnamefont {R.}~\bibnamefont {Borth}}, \bibinfo {author} {\bibfnamefont
  {A.}~\bibnamefont {Wang}}, \bibinfo {author} {\bibfnamefont {Y.}~\bibnamefont
  {Chen}}, \bibinfo {author} {\bibfnamefont {Z.}~\bibnamefont {Nie}}, \bibinfo
  {author} {\bibfnamefont {R.}~\bibnamefont {Li}}, \bibinfo {author}
  {\bibfnamefont {X.}~\bibnamefont {Lu}}, \emph {et~al.},\ }\bibfield  {title}
  {\bibinfo {title} {Strange-metal behaviour in a pure ferromagnetic kondo
  lattice},\ }\href {https://www.nature.com/articles/s41586-020-2052-z}
  {\bibfield  {journal} {\bibinfo  {journal} {Nature}\ }\textbf {\bibinfo
  {volume} {579}},\ \bibinfo {pages} {51} (\bibinfo {year} {2020})}\BibitemShut
  {NoStop}%
\bibitem [{\citenamefont {König}\ \emph {et~al.}(2022)\citenamefont {König},
  \citenamefont {Komijani},\ and\ \citenamefont {Coleman}}]{Koenig2022}%
  \BibitemOpen
  \bibfield  {author} {\bibinfo {author} {\bibfnamefont {E.~J.}\ \bibnamefont
  {König}}, \bibinfo {author} {\bibfnamefont {Y.}~\bibnamefont {Komijani}},\
  and\ \bibinfo {author} {\bibfnamefont {P.}~\bibnamefont {Coleman}},\
  }\bibfield  {title} {\bibinfo {title} {Triplet resonating valence bond theory
  and transition metal chalcogenides},\ }\href
  {https://doi.org/10.1103/physrevb.105.075142} {\bibfield  {journal} {\bibinfo
   {journal} {Physical Review B}\ }\textbf {\bibinfo {volume} {105}},\ \bibinfo
  {pages} {075142} (\bibinfo {year} {2022})}\BibitemShut {NoStop}%
\bibitem [{\citenamefont {Lopez}\ \emph {et~al.}(2022)\citenamefont {Lopez},
  \citenamefont {Powell},\ and\ \citenamefont
  {Merino}}]{LopezPowellMerino2022}%
  \BibitemOpen
  \bibfield  {author} {\bibinfo {author} {\bibfnamefont {M.}~\bibnamefont
  {Lopez}}, \bibinfo {author} {\bibfnamefont {B.}~\bibnamefont {Powell}},\ and\
  \bibinfo {author} {\bibfnamefont {J.}~\bibnamefont {Merino}},\ }\bibfield
  {title} {\bibinfo {title} {Topological superconductivity from doping a
  triplet quantum spin liquid in a flat band system},\ }\href@noop {} {\
  (\bibinfo {year} {2022})},\ \Eprint {https://arxiv.org/abs/arXiv:2210.05275}
  {arXiv:2210.05275} \BibitemShut {NoStop}%
\bibitem [{\citenamefont {Carretta}\ and\ \citenamefont
  {Prando}(2020)}]{Carretta2020}%
  \BibitemOpen
  \bibfield  {author} {\bibinfo {author} {\bibfnamefont {P.}~\bibnamefont
  {Carretta}}\ and\ \bibinfo {author} {\bibfnamefont {G.}~\bibnamefont
  {Prando}},\ }\bibfield  {title} {\bibinfo {title} {Iron-based
  superconductors: tales from the nuclei},\ }\href
  {https://doi.org/10.1007/s40766-019-0001-1} {\bibfield  {journal} {\bibinfo
  {journal} {La Rivista del Nuovo Cimento}\ }\textbf {\bibinfo {volume} {43}},\
  \bibinfo {pages} {1} (\bibinfo {year} {2020})}\BibitemShut {NoStop}%
\bibitem [{\citenamefont {Hanaguri}\ \emph {et~al.}(2010)\citenamefont
  {Hanaguri}, \citenamefont {Niitaka}, \citenamefont {Kuroki},\ and\
  \citenamefont {Takagi}}]{HanaguriTakagi2010}%
  \BibitemOpen
  \bibfield  {author} {\bibinfo {author} {\bibfnamefont {T.}~\bibnamefont
  {Hanaguri}}, \bibinfo {author} {\bibfnamefont {S.}~\bibnamefont {Niitaka}},
  \bibinfo {author} {\bibfnamefont {K.}~\bibnamefont {Kuroki}},\ and\ \bibinfo
  {author} {\bibfnamefont {H.}~\bibnamefont {Takagi}},\ }\bibfield  {title}
  {\bibinfo {title} {Unconventional s-wave superconductivity in {F}e ({S}e,
  {T}e)},\ }\href
  {https://science.sciencemag.org/content/328/5977/474.figures-only} {\bibfield
   {journal} {\bibinfo  {journal} {Science}\ }\textbf {\bibinfo {volume}
  {328}},\ \bibinfo {pages} {474} (\bibinfo {year} {2010})}\BibitemShut
  {NoStop}%
\bibitem [{\citenamefont {Chi}\ \emph {et~al.}(2014)\citenamefont {Chi},
  \citenamefont {Johnston}, \citenamefont {Levy}, \citenamefont {Grothe},
  \citenamefont {Szedlak}, \citenamefont {Ludbrook}, \citenamefont {Liang},
  \citenamefont {Dosanjh}, \citenamefont {Burke}, \citenamefont {Damascelli},
  \citenamefont {Bonn}, \citenamefont {Hardy},\ and\ \citenamefont
  {Pennec}}]{ChiPennec2014}%
  \BibitemOpen
  \bibfield  {author} {\bibinfo {author} {\bibfnamefont {S.}~\bibnamefont
  {Chi}}, \bibinfo {author} {\bibfnamefont {S.}~\bibnamefont {Johnston}},
  \bibinfo {author} {\bibfnamefont {G.}~\bibnamefont {Levy}}, \bibinfo {author}
  {\bibfnamefont {S.}~\bibnamefont {Grothe}}, \bibinfo {author} {\bibfnamefont
  {R.}~\bibnamefont {Szedlak}}, \bibinfo {author} {\bibfnamefont
  {B.}~\bibnamefont {Ludbrook}}, \bibinfo {author} {\bibfnamefont
  {R.}~\bibnamefont {Liang}}, \bibinfo {author} {\bibfnamefont
  {P.}~\bibnamefont {Dosanjh}}, \bibinfo {author} {\bibfnamefont {S.~A.}\
  \bibnamefont {Burke}}, \bibinfo {author} {\bibfnamefont {A.}~\bibnamefont
  {Damascelli}}, \bibinfo {author} {\bibfnamefont {D.~A.}\ \bibnamefont
  {Bonn}}, \bibinfo {author} {\bibfnamefont {W.~N.}\ \bibnamefont {Hardy}},\
  and\ \bibinfo {author} {\bibfnamefont {Y.}~\bibnamefont {Pennec}},\
  }\bibfield  {title} {\bibinfo {title} {Sign inversion in the superconducting
  order parameter of {L}i{F}e{A}s inferred from bogoliubov quasiparticle
  interference},\ }\href {https://doi.org/10.1103/PhysRevB.89.104522}
  {\bibfield  {journal} {\bibinfo  {journal} {Phys. Rev. B}\ }\textbf {\bibinfo
  {volume} {89}},\ \bibinfo {pages} {104522} (\bibinfo {year}
  {2014})}\BibitemShut {NoStop}%
\bibitem [{\citenamefont {Barnes}(1976)}]{Barnes1976}%
  \BibitemOpen
  \bibfield  {author} {\bibinfo {author} {\bibfnamefont {S.~E.}\ \bibnamefont
  {Barnes}},\ }\bibfield  {title} {\bibinfo {title} {New method for the
  anderson model},\ }\href {https://doi.org/10.1088/0305-4608/6/7/018}
  {\bibfield  {journal} {\bibinfo  {journal} {Journal of Physics F: Metal
  Physics}\ }\textbf {\bibinfo {volume} {6}},\ \bibinfo {pages} {1375}
  (\bibinfo {year} {1976})}\BibitemShut {NoStop}%
\bibitem [{\citenamefont {Coleman}(1984)}]{Coleman1984}%
  \BibitemOpen
  \bibfield  {author} {\bibinfo {author} {\bibfnamefont {P.}~\bibnamefont
  {Coleman}},\ }\bibfield  {title} {\bibinfo {title} {New approach to the
  mixed-valence problem},\ }\href {https://doi.org/10.1103/physrevb.29.3035}
  {\bibfield  {journal} {\bibinfo  {journal} {Physical Review B}\ }\textbf
  {\bibinfo {volume} {29}},\ \bibinfo {pages} {3035} (\bibinfo {year}
  {1984})}\BibitemShut {NoStop}%
\bibitem [{\citenamefont {Ruckenstein}\ \emph {et~al.}(1987)\citenamefont
  {Ruckenstein}, \citenamefont {Hirschfeld},\ and\ \citenamefont
  {Appel}}]{Ruckenstein87}%
  \BibitemOpen
  \bibfield  {author} {\bibinfo {author} {\bibfnamefont {A.~E.}\ \bibnamefont
  {Ruckenstein}}, \bibinfo {author} {\bibfnamefont {P.~J.}\ \bibnamefont
  {Hirschfeld}},\ and\ \bibinfo {author} {\bibfnamefont {J.}~\bibnamefont
  {Appel}},\ }\bibfield  {title} {\bibinfo {title} {Mean-field theory of
  high-${T}_{c}$ superconductivity: The superexchange mechanism},\ }\href
  {https://doi.org/10.1103/PhysRevB.36.857} {\bibfield  {journal} {\bibinfo
  {journal} {Phys. Rev. B}\ }\textbf {\bibinfo {volume} {36}},\ \bibinfo
  {pages} {857} (\bibinfo {year} {1987})}\BibitemShut {NoStop}%
\bibitem [{\citenamefont {Kotliar}\ and\ \citenamefont
  {Liu}(1988)}]{KotliarLiu1988}%
  \BibitemOpen
  \bibfield  {author} {\bibinfo {author} {\bibfnamefont {G.}~\bibnamefont
  {Kotliar}}\ and\ \bibinfo {author} {\bibfnamefont {J.}~\bibnamefont {Liu}},\
  }\bibfield  {title} {\bibinfo {title} {Superexchange mechanism and d-wave
  superconductivity},\ }\href
  {https://link.aps.org/doi/10.1103/PhysRevB.38.5142} {\bibfield  {journal}
  {\bibinfo  {journal} {Phys. Rev. B}\ }\textbf {\bibinfo {volume} {38}},\
  \bibinfo {pages} {5142} (\bibinfo {year} {1988})}\BibitemShut {NoStop}%
\bibitem [{\citenamefont {Yu}\ and\ \citenamefont {Si}(2013)}]{Yu13}%
  \BibitemOpen
  \bibfield  {author} {\bibinfo {author} {\bibfnamefont {R.}~\bibnamefont
  {Yu}}\ and\ \bibinfo {author} {\bibfnamefont {Q.}~\bibnamefont {Si}},\
  }\bibfield  {title} {\bibinfo {title} {Orbital-selective mott phase in
  multiorbital models for alkaline iron selenides
  ${\mathbf{k}}_{1\ensuremath{-}x}{\mathrm{fe}}_{2\ensuremath{-}y}{\mathrm{se}}_{2}$},\
  }\href {https://doi.org/10.1103/PhysRevLett.110.146402} {\bibfield  {journal}
  {\bibinfo  {journal} {Phys. Rev. Lett.}\ }\textbf {\bibinfo {volume} {110}},\
  \bibinfo {pages} {146402} (\bibinfo {year} {2013})}\BibitemShut {NoStop}%
\bibitem [{\citenamefont {Daghofer}\ \emph {et~al.}(2010)\citenamefont
  {Daghofer}, \citenamefont {Nicholson}, \citenamefont {Moreo},\ and\
  \citenamefont {Dagotto}}]{DaghoferDagotto2010}%
  \BibitemOpen
  \bibfield  {author} {\bibinfo {author} {\bibfnamefont {M.}~\bibnamefont
  {Daghofer}}, \bibinfo {author} {\bibfnamefont {A.}~\bibnamefont {Nicholson}},
  \bibinfo {author} {\bibfnamefont {A.}~\bibnamefont {Moreo}},\ and\ \bibinfo
  {author} {\bibfnamefont {E.}~\bibnamefont {Dagotto}},\ }\bibfield  {title}
  {\bibinfo {title} {Three orbital model for the iron-based superconductors},\
  }\href {https://link.aps.org/doi/10.1103/PhysRevB.81.014511} {\bibfield
  {journal} {\bibinfo  {journal} {Phys. Rev. B}\ }\textbf {\bibinfo {volume}
  {81}},\ \bibinfo {pages} {014511} (\bibinfo {year} {2010})}\BibitemShut
  {NoStop}%
\bibitem [{\citenamefont {Albuquerque}\ \emph {et~al.}(2012)\citenamefont
  {Albuquerque}, \citenamefont {Alet},\ and\ \citenamefont
  {Moessner}}]{Albuquerque12}%
  \BibitemOpen
  \bibfield  {author} {\bibinfo {author} {\bibfnamefont {A.~F.}\ \bibnamefont
  {Albuquerque}}, \bibinfo {author} {\bibfnamefont {F.}~\bibnamefont {Alet}},\
  and\ \bibinfo {author} {\bibfnamefont {R.}~\bibnamefont {Moessner}},\
  }\bibfield  {title} {\bibinfo {title} {Coexistence of long-range and
  algebraic correlations for short-range valence-bond wave functions in three
  dimensions},\ }\href {https://doi.org/10.1103/PhysRevLett.109.147204}
  {\bibfield  {journal} {\bibinfo  {journal} {Phys. Rev. Lett.}\ }\textbf
  {\bibinfo {volume} {109}},\ \bibinfo {pages} {147204} (\bibinfo {year}
  {2012})}\BibitemShut {NoStop}%
\bibitem [{\citenamefont {Eschrig}\ and\ \citenamefont
  {Koepernik}(2009{\natexlab{a}})}]{Eschrig2009}%
  \BibitemOpen
  \bibfield  {author} {\bibinfo {author} {\bibfnamefont {H.}~\bibnamefont
  {Eschrig}}\ and\ \bibinfo {author} {\bibfnamefont {K.}~\bibnamefont
  {Koepernik}},\ }\bibfield  {title} {\bibinfo {title} {Tight-binding models
  for the iron-based superconductors},\ }\href
  {https://doi.org/10.1103/PhysRevB.80.104503} {\bibfield  {journal} {\bibinfo
  {journal} {Phys. Rev. B}\ }\textbf {\bibinfo {volume} {80}},\ \bibinfo
  {pages} {104503} (\bibinfo {year} {2009}{\natexlab{a}})}\BibitemShut
  {NoStop}%
\bibitem [{\citenamefont {Yu}\ and\ \citenamefont {Si}(2017)}]{Yu17}%
  \BibitemOpen
  \bibfield  {author} {\bibinfo {author} {\bibfnamefont {R.}~\bibnamefont
  {Yu}}\ and\ \bibinfo {author} {\bibfnamefont {Q.}~\bibnamefont {Si}},\
  }\bibfield  {title} {\bibinfo {title} {Orbital-selective mott phase in
  multiorbital models for iron pnictides and chalcogenides},\ }\href
  {https://doi.org/10.1103/PhysRevB.96.125110} {\bibfield  {journal} {\bibinfo
  {journal} {Phys. Rev. B}\ }\textbf {\bibinfo {volume} {96}},\ \bibinfo
  {pages} {125110} (\bibinfo {year} {2017})}\BibitemShut {NoStop}%
\bibitem [{\citenamefont {Komijani}\ and\ \citenamefont
  {Kotliar}(2017)}]{Komijani17}%
  \BibitemOpen
  \bibfield  {author} {\bibinfo {author} {\bibfnamefont {Y.}~\bibnamefont
  {Komijani}}\ and\ \bibinfo {author} {\bibfnamefont {G.}~\bibnamefont
  {Kotliar}},\ }\bibfield  {title} {\bibinfo {title} {Analytical slave-spin
  mean-field approach to orbital selective mott insulators},\ }\href
  {https://doi.org/10.1103/PhysRevB.96.125111} {\bibfield  {journal} {\bibinfo
  {journal} {Phys. Rev. B}\ }\textbf {\bibinfo {volume} {96}},\ \bibinfo
  {pages} {125111} (\bibinfo {year} {2017})}\BibitemShut {NoStop}%
\bibitem [{\citenamefont {Komijani}\ \emph {et~al.}(2019)\citenamefont
  {Komijani}, \citenamefont {Hallberg},\ and\ \citenamefont
  {Kotliar}}]{Komijani2019}%
  \BibitemOpen
  \bibfield  {author} {\bibinfo {author} {\bibfnamefont {Y.}~\bibnamefont
  {Komijani}}, \bibinfo {author} {\bibfnamefont {K.}~\bibnamefont {Hallberg}},\
  and\ \bibinfo {author} {\bibfnamefont {G.}~\bibnamefont {Kotliar}},\
  }\bibfield  {title} {\bibinfo {title} {Renormalized dispersing multiplets in
  the spectrum of nearly mott localized systems},\ }\href
  {https://doi.org/10.1103/physrevb.99.125150} {\bibfield  {journal} {\bibinfo
  {journal} {Physical Review B}\ }\textbf {\bibinfo {volume} {99}},\ \bibinfo
  {pages} {125150} (\bibinfo {year} {2019})}\BibitemShut {NoStop}%
\bibitem [{\citenamefont {Si}\ \emph {et~al.}(2016)\citenamefont {Si},
  \citenamefont {Yu},\ and\ \citenamefont {Abrahams}}]{Si2016}%
  \BibitemOpen
  \bibfield  {author} {\bibinfo {author} {\bibfnamefont {Q.}~\bibnamefont
  {Si}}, \bibinfo {author} {\bibfnamefont {R.}~\bibnamefont {Yu}},\ and\
  \bibinfo {author} {\bibfnamefont {E.}~\bibnamefont {Abrahams}},\ }\bibfield
  {title} {\bibinfo {title} {High-temperature superconductivity in iron
  pnictides and chalcogenides},\ }\href
  {https://doi.org/10.1038/natrevmats.2016.17} {\bibfield  {journal} {\bibinfo
  {journal} {Nature Reviews Materials}\ }\textbf {\bibinfo {volume} {1}},\
  \bibinfo {pages} {16017} (\bibinfo {year} {2016})}\BibitemShut {NoStop}%
\bibitem [{\citenamefont {Liu}\ \emph {et~al.}(2015)\citenamefont {Liu},
  \citenamefont {Yi}, \citenamefont {Zhang}, \citenamefont {Hu}, \citenamefont
  {Yu}, \citenamefont {Zhu}, \citenamefont {He}, \citenamefont {Chen},
  \citenamefont {Hashimoto}, \citenamefont {Moore}, \citenamefont {Mo},
  \citenamefont {Hussain}, \citenamefont {Si}, \citenamefont {Mao},
  \citenamefont {Lu},\ and\ \citenamefont {Shen}}]{Liu15}%
  \BibitemOpen
  \bibfield  {author} {\bibinfo {author} {\bibfnamefont {Z.~K.}\ \bibnamefont
  {Liu}}, \bibinfo {author} {\bibfnamefont {M.}~\bibnamefont {Yi}}, \bibinfo
  {author} {\bibfnamefont {Y.}~\bibnamefont {Zhang}}, \bibinfo {author}
  {\bibfnamefont {J.}~\bibnamefont {Hu}}, \bibinfo {author} {\bibfnamefont
  {R.}~\bibnamefont {Yu}}, \bibinfo {author} {\bibfnamefont {J.-X.}\
  \bibnamefont {Zhu}}, \bibinfo {author} {\bibfnamefont {R.-H.}\ \bibnamefont
  {He}}, \bibinfo {author} {\bibfnamefont {Y.~L.}\ \bibnamefont {Chen}},
  \bibinfo {author} {\bibfnamefont {M.}~\bibnamefont {Hashimoto}}, \bibinfo
  {author} {\bibfnamefont {R.~G.}\ \bibnamefont {Moore}}, \bibinfo {author}
  {\bibfnamefont {S.-K.}\ \bibnamefont {Mo}}, \bibinfo {author} {\bibfnamefont
  {Z.}~\bibnamefont {Hussain}}, \bibinfo {author} {\bibfnamefont
  {Q.}~\bibnamefont {Si}}, \bibinfo {author} {\bibfnamefont {Z.~Q.}\
  \bibnamefont {Mao}}, \bibinfo {author} {\bibfnamefont {D.~H.}\ \bibnamefont
  {Lu}},\ and\ \bibinfo {author} {\bibfnamefont {Z.-X.}\ \bibnamefont {Shen}},\
  }\bibfield  {title} {\bibinfo {title} {Experimental observation of
  incoherent-coherent crossover and orbital-dependent band renormalization in
  iron chalcogenide superconductors},\ }\href
  {https://doi.org/10.1103/PhysRevB.92.235138} {\bibfield  {journal} {\bibinfo
  {journal} {Phys. Rev. B}\ }\textbf {\bibinfo {volume} {92}},\ \bibinfo
  {pages} {235138} (\bibinfo {year} {2015})}\BibitemShut {NoStop}%
\bibitem [{\citenamefont {Huang}\ \emph {et~al.}(2022)\citenamefont {Huang},
  \citenamefont {Yu}, \citenamefont {Xu}, \citenamefont {Zhu}, \citenamefont
  {Oh}, \citenamefont {Jiang}, \citenamefont {Wang}, \citenamefont {Wu},
  \citenamefont {Chen}, \citenamefont {Denlinger}, \citenamefont {Mo},
  \citenamefont {Hashimoto}, \citenamefont {Michiardi}, \citenamefont
  {Pedersen}, \citenamefont {Gorovikov}, \citenamefont {Zhdanovich},
  \citenamefont {Damascelli}, \citenamefont {Gu}, \citenamefont {Dai},
  \citenamefont {Chu}, \citenamefont {Lu}, \citenamefont {Si}, \citenamefont
  {Birgeneau},\ and\ \citenamefont {Yi}}]{Huang2022}%
  \BibitemOpen
  \bibfield  {author} {\bibinfo {author} {\bibfnamefont {J.}~\bibnamefont
  {Huang}}, \bibinfo {author} {\bibfnamefont {R.}~\bibnamefont {Yu}}, \bibinfo
  {author} {\bibfnamefont {Z.}~\bibnamefont {Xu}}, \bibinfo {author}
  {\bibfnamefont {J.-X.}\ \bibnamefont {Zhu}}, \bibinfo {author} {\bibfnamefont
  {J.~S.}\ \bibnamefont {Oh}}, \bibinfo {author} {\bibfnamefont
  {Q.}~\bibnamefont {Jiang}}, \bibinfo {author} {\bibfnamefont
  {M.}~\bibnamefont {Wang}}, \bibinfo {author} {\bibfnamefont {H.}~\bibnamefont
  {Wu}}, \bibinfo {author} {\bibfnamefont {T.}~\bibnamefont {Chen}}, \bibinfo
  {author} {\bibfnamefont {J.~D.}\ \bibnamefont {Denlinger}}, \bibinfo {author}
  {\bibfnamefont {S.-K.}\ \bibnamefont {Mo}}, \bibinfo {author} {\bibfnamefont
  {M.}~\bibnamefont {Hashimoto}}, \bibinfo {author} {\bibfnamefont
  {M.}~\bibnamefont {Michiardi}}, \bibinfo {author} {\bibfnamefont {T.~M.}\
  \bibnamefont {Pedersen}}, \bibinfo {author} {\bibfnamefont {S.}~\bibnamefont
  {Gorovikov}}, \bibinfo {author} {\bibfnamefont {S.}~\bibnamefont
  {Zhdanovich}}, \bibinfo {author} {\bibfnamefont {A.}~\bibnamefont
  {Damascelli}}, \bibinfo {author} {\bibfnamefont {G.}~\bibnamefont {Gu}},
  \bibinfo {author} {\bibfnamefont {P.}~\bibnamefont {Dai}}, \bibinfo {author}
  {\bibfnamefont {J.-H.}\ \bibnamefont {Chu}}, \bibinfo {author} {\bibfnamefont
  {D.}~\bibnamefont {Lu}}, \bibinfo {author} {\bibfnamefont {Q.}~\bibnamefont
  {Si}}, \bibinfo {author} {\bibfnamefont {R.~J.}\ \bibnamefont {Birgeneau}},\
  and\ \bibinfo {author} {\bibfnamefont {M.}~\bibnamefont {Yi}},\ }\bibfield
  {title} {\bibinfo {title} {Correlation-driven electronic reconstruction in
  {FeTe}1-{xSex}},\ }\bibfield  {journal} {\bibinfo  {journal} {Communications
  Physics}\ }\textbf {\bibinfo {volume} {5}},\ \href
  {https://doi.org/10.1038/s42005-022-00805-6} {10.1038/s42005-022-00805-6}
  (\bibinfo {year} {2022})\BibitemShut {NoStop}%
\bibitem [{\citenamefont {Graser}\ \emph {et~al.}(2009)\citenamefont {Graser},
  \citenamefont {Maier}, \citenamefont {Hirschfeld},\ and\ \citenamefont
  {Scalapino}}]{Graser2009}%
  \BibitemOpen
  \bibfield  {author} {\bibinfo {author} {\bibfnamefont {S.}~\bibnamefont
  {Graser}}, \bibinfo {author} {\bibfnamefont {T.~A.}\ \bibnamefont {Maier}},
  \bibinfo {author} {\bibfnamefont {P.~J.}\ \bibnamefont {Hirschfeld}},\ and\
  \bibinfo {author} {\bibfnamefont {D.~J.}\ \bibnamefont {Scalapino}},\
  }\bibfield  {title} {\bibinfo {title} {Near-degeneracy of several pairing
  channels in multiorbital models for the fe pnictides},\ }\href@noop {}
  {\bibfield  {journal} {\bibinfo  {journal} {New Journal of Physics}\ }\textbf
  {\bibinfo {volume} {11}},\ \bibinfo {pages} {025016} (\bibinfo {year}
  {2009})}\BibitemShut {NoStop}%
\bibitem [{\citenamefont {Eschrig}\ and\ \citenamefont
  {Koepernik}(2009{\natexlab{b}})}]{Eschrig09}%
  \BibitemOpen
  \bibfield  {author} {\bibinfo {author} {\bibfnamefont {H.}~\bibnamefont
  {Eschrig}}\ and\ \bibinfo {author} {\bibfnamefont {K.}~\bibnamefont
  {Koepernik}},\ }\bibfield  {title} {\bibinfo {title} {Tight-binding models
  for the iron-based superconductors},\ }\href
  {https://doi.org/10.1103/PhysRevB.80.104503} {\bibfield  {journal} {\bibinfo
  {journal} {Phys. Rev. B}\ }\textbf {\bibinfo {volume} {80}},\ \bibinfo
  {pages} {104503} (\bibinfo {year} {2009}{\natexlab{b}})}\BibitemShut
  {NoStop}%
\bibitem [{\citenamefont {Borisenko}\ \emph {et~al.}(2016)\citenamefont
  {Borisenko}, \citenamefont {Evtushinsky}, \citenamefont {Liu}, \citenamefont
  {Morozov}, \citenamefont {Kappenberger}, \citenamefont {Wurmehl},
  \citenamefont {B{\"u}chner}, \citenamefont {Yaresko}, \citenamefont {Kim},
  \citenamefont {Hoesch} \emph {et~al.}}]{BorisenkoZhigadlo2016}%
  \BibitemOpen
  \bibfield  {author} {\bibinfo {author} {\bibfnamefont {S.}~\bibnamefont
  {Borisenko}}, \bibinfo {author} {\bibfnamefont {D.}~\bibnamefont
  {Evtushinsky}}, \bibinfo {author} {\bibfnamefont {Z.-H.}\ \bibnamefont
  {Liu}}, \bibinfo {author} {\bibfnamefont {I.}~\bibnamefont {Morozov}},
  \bibinfo {author} {\bibfnamefont {R.}~\bibnamefont {Kappenberger}}, \bibinfo
  {author} {\bibfnamefont {S.}~\bibnamefont {Wurmehl}}, \bibinfo {author}
  {\bibfnamefont {B.}~\bibnamefont {B{\"u}chner}}, \bibinfo {author}
  {\bibfnamefont {A.}~\bibnamefont {Yaresko}}, \bibinfo {author} {\bibfnamefont
  {T.}~\bibnamefont {Kim}}, \bibinfo {author} {\bibfnamefont {M.}~\bibnamefont
  {Hoesch}}, \emph {et~al.},\ }\bibfield  {title} {\bibinfo {title} {Direct
  observation of spin--orbit coupling in iron-based superconductors},\ }\href
  {https://www.nature.com/articles/nphys3594} {\bibfield  {journal} {\bibinfo
  {journal} {Nature Physics}\ }\textbf {\bibinfo {volume} {12}},\ \bibinfo
  {pages} {311} (\bibinfo {year} {2016})}\BibitemShut {NoStop}%
\bibitem [{\citenamefont {Wang}\ and\ \citenamefont {Zhang}(2012)}]{Wang2012}%
  \BibitemOpen
  \bibfield  {author} {\bibinfo {author} {\bibfnamefont {Z.}~\bibnamefont
  {Wang}}\ and\ \bibinfo {author} {\bibfnamefont {S.-C.}\ \bibnamefont
  {Zhang}},\ }\bibfield  {title} {\bibinfo {title} {Simplified topological
  invariants for interacting insulators},\ }\href
  {https://doi.org/10.1103/physrevx.2.031008} {\bibfield  {journal} {\bibinfo
  {journal} {Physical Review X}\ }\textbf {\bibinfo {volume} {2}},\ \bibinfo
  {pages} {031008} (\bibinfo {year} {2012})}\BibitemShut {NoStop}%
\bibitem [{\citenamefont {Son}\ \emph {et~al.}(2021)\citenamefont {Son},
  \citenamefont {Absil}, \citenamefont {Gao},\ and\ \citenamefont
  {Stykel}}]{Son2021}%
  \BibitemOpen
  \bibfield  {author} {\bibinfo {author} {\bibfnamefont {N.~T.}\ \bibnamefont
  {Son}}, \bibinfo {author} {\bibfnamefont {P.-A.}\ \bibnamefont {Absil}},
  \bibinfo {author} {\bibfnamefont {B.}~\bibnamefont {Gao}},\ and\ \bibinfo
  {author} {\bibfnamefont {T.}~\bibnamefont {Stykel}},\ }\bibfield  {title}
  {\bibinfo {title} {Computing symplectic eigenpairs of symmetric
  positive-definite matrices via trace minimization and riemannian
  optimization},\ }\href {https://doi.org/10.1137/21m1390621} {\bibfield
  {journal} {\bibinfo  {journal} {{SIAM} Journal on Matrix Analysis and
  Applications}\ }\textbf {\bibinfo {volume} {42}},\ \bibinfo {pages} {1732}
  (\bibinfo {year} {2021})}\BibitemShut {NoStop}%
\end{thebibliography}%
\end{document}